\documentclass[traditabstract]{aa}

\usepackage[utf8]{inputenc}
\usepackage[dvipsnames]{xcolor}
\usepackage[T1]{fontenc}
\usepackage{lmodern}
\usepackage{graphicx}
\usepackage{hyperref}
\usepackage{placeins}
\usepackage{float}
\usepackage{subfigure}
\usepackage{stfloats}
\usepackage[normalem]{ulem}
\usepackage{amsmath}
\usepackage{physics}
\usepackage{bm}

\definecolor{elizabra1}{RGB}{0,180,180}
\definecolor{elizabra2}{RGB}{200,160,255}
\definecolor{elizabra}{RGB}{50,180,180}
\definecolor{newgreen}{RGB}{0,140,70}
\definecolor{todoplaceholder}{RGB}{200,30,30}

\hypersetup{
  colorlinks=true,
  linkcolor=elizabra2,
  citecolor=elizabra1,
  urlcolor=elizabra2
}

\usepackage{txfonts}

\numberwithin{subsection}{section}

\begin{document}

\title{Composition-dependent acceleration of UHE Cosmic Rays in AGN electric fields}

\titlerunning{UHECR acceleration in AGN}
\authorrunning{Koutsoumpou et al.}

\author{E. Koutsoumpou\inst{1}\fnmsep\thanks{\email{evkoutso@phys.uoa.gr}},
I. Contopoulos\inst{2}, 
A. Nathanail\inst{2},
A. Loules \inst{1},
\and D. Kazanas\inst{3}
}
\institute{Section of Astrophysics, Astronomy and Mechanics, Department of Physics, National and Kapodistrian University of Athens, University Campus, Zografos GR-15784, Athens, Greece\\
\and
Research Center for Astronomy and Applied Mathematics, Academy of Athens, Soranou Efessiou 4, GR-11527 Athens, Greece
\and
Astrophysics Science Division, NASA Goddard Space Flight Center, Greenbelt, MD 20771, USA
}

%

\abstract{
The origin of ultra-high-energy cosmic rays (UHECRs) remains elusive,
as proposed acceleration mechanisms are rarely connected directly to
composition-sensitive air-shower observables. 
We establish such a connection for voltage-drop
acceleration in the near-horizon current sheets of magnetically
arrested disks around supermassive black holes. 
Fragmentation of the
acceleration region is represented by a log-normal distribution of
effective potential-sampling factors $\eta$, with
$\mu_\eta=\langle\ln\eta\rangle$,
$\sigma_\eta=[\mathrm{Var}(\ln\eta)]^{1/2}$, and
median $\widetilde{\eta}=e^{\mu_\eta}$, while the particle energy gain
scales with nuclear charge $Z$ and with the maximum potential
$\mathcal{V}_{\max}$ available at the source. The observed UHECR
composition can therefore constrain both the characteristic
acceleration scale and its particle-to-particle dispersion.
We propagate the accelerated particles through continuous
Bethe--Heitler and adiabatic losses, catastrophic photopion attenuation
of protons, and a mass-evolving photodisintegration cascade. The
resulting populations are converted into predictions for the mean and
standard deviation of the atmospheric depth of shower maximum,
$\langle X_{\max}\rangle$ and $\sigma(X_{\max})$. We perform independent
Bayesian fits of $(\mu_\eta,\sigma_\eta,\log\mathcal{V}_{\max})$ to the
Pierre Auger Observatory 2025 deep-neural-network measurements for two
fixed injection templates and propagation distances of 10, 30, and
100\,Mpc. 
An Auger-inspired, proton-poor injection reproduces both
shower moments at 10 and 30\,Mpc, with reduced chi-square values
$\chi^2/\nu=0.33$ and $0.47$, respectively, where $\nu=27$ is the
number of degrees of freedom.
At 30\,Mpc, the inferred logarithmic
width is $\sigma_\eta=0.59^{+0.13}_{-0.10}$, corresponding to a
multiplicative dispersion of approximately 1.8 in the sampled
potential. The median sampling factor and the maximum available
potential remain strongly degenerate, with their product
$\widetilde{\eta}\mathcal{V}_{\max}$ more directly constrained.
A metal-enriched injection containing $\sim53\,\%$ hydrogen is less
favoured, with $\chi^2$ larger by $17.9$ and $25.9$ at 10 and
30\,Mpc, respectively, under the adopted joint likelihood for
$\langle X_{\max}\rangle$ and $\sigma(X_{\max})$. Both templates remain viable at 10 and 30\,Mpc but give poor fits at 100\,Mpc.
Within the scenarios tested, the results favour a relatively
proton-poor accelerated population and demonstrate that the evolution
of the shower-width distribution can constrain the stochastic
structure of electric-field acceleration in AGN. The required
conditions are attainable in nearby massive, strongly magnetised
systems such as M87, although the analysis identifies a source class
rather than a unique object.
}


\maketitle


\section{Introduction}\label{sec:intro}

Ultra-high-energy cosmic rays (UHECRs) with $E\gtrsim10^{18}$\,eV remain a
major puzzle in astroparticle physics. At lower energies, Galactic sources,
particularly supernova remnants (SNRs), are leading candidates for cosmic-ray
acceleration, with diffusive shock acceleration potentially reaching the
cosmic-ray knee at a few PeV \citep{Bell2013}. The transition from Galactic
to extragalactic cosmic rays is generally associated with the ankle region
near $5\times10^{18}$\,eV, although its precise interpretation remains
model dependent \citep{Globus}. Among candidate extragalactic sources,
active galactic nuclei (AGN) are particularly well motivated because their
jets and radio lobes can satisfy the Hillas confinement criterion
\citep{Hillas1984,Aabetal}. Early hadronic models of AGN cores showed that
shock-accelerated relativistic protons can undergo pion-producing
interactions, generating secondary electrons, gamma rays, and neutrinos
\citep{Kazanas1983,Kazanas1986}. Because UHECRs are charged particles,
their arrival directions are deflected by Galactic and extragalactic
magnetic fields, and inference of their source directions therefore depends on magnetic-field and composition models \citep{Globus}. 
Claimed associations have been reported with nearby
galaxies such as M82 and Centaurus~A \citep{Aabetal,Globus}, while no
corresponding excess has been found from the Virgo cluster \citep{Globus}.

Large-scale poloidal magnetic-field structures and associated electric-current
systems can be supported by rotating magnetized plasma configurations,
including accretion flows and rotating magnetospheres
\citep{ContopoulosKazanas1998,Contopoulos2005}.
In accreting black-hole systems, such magnetic flux can be advected inward
and accumulate toward the central black hole, where it may become dynamically
important \citep{Narayan2003,Igumenshchev2008,Tchekhovskoy2011}.
Theoretical and numerical studies show that when the magnetic pressure
associated with the accumulated flux becomes comparable to the ram pressure
of the inflowing plasma, the accretion flow enters a magnetically arrested
disk (MAD) state \citep{Narayan2003,Igumenshchev2008,Tchekhovskoy2011},
characterized by a strongly magnetized inner flow and intermittent,
non-axisymmetric accretion
\citep{Igumenshchev2008,Tchekhovskoy2011,Chatterjee2022}.
A robust outcome of MAD simulations is the formation of extended current
sheets and quasi-periodic magnetic-flux eruptions near the black-hole
horizon, associated with magnetic reconnection and interchange instabilities
\citep{Tchekhovskoy2011,Dexter2020,Ripperda2022,Chatterjee2022,
Vos2024,Loules2025}. These eruptions produce localized reconnection-driven dissipation and rapid variability \citep{Dexter2020,Ripperda2022,antonopoulou2025}, while
reconnection X-points host non-ideal electric fields that efficiently
accelerate particles \citep{Vos2024}. Observationally, Event Horizon Telescope
polarimetry of M87* and Sgr~A* favors general-relativistic
magnetohydrodynamic (GRMHD) models with organized, dynamically important
magnetic fields, particularly MAD configurations \citep{ehtc2021,ehtc2024}, supporting the astrophysical relevance of MAD environments near supermassive black holes.

While MADs offer a promising environment for particle acceleration
\citep{Vos2024}, alternative AGN-driven scenarios have also been considered.
Notably, wind-termination shocks of ultra-fast outflows (UFOs) have been
proposed as potential acceleration sites, although intense AGN photon fields
strongly attenuate escaping nuclei through photodisintegration, with only rare
low-emission systems allowing nuclei to escape up to $\sim10^{19.8}$\,eV
\citep{Ehlert2025}. Recent GRMHD simulations by \citet{nathanail2025} have shown that
incorporating modest physical resistivity does not suppress the MAD state,
with flux eruptions and the associated variability remaining robust and
close to the ideal-MHD behavior. Complementary semi-analytic studies of resistive relativistic jets show that electric-potential gradients along poloidal magnetic-field lines generate parallel electric fields, providing potential sites for charged-particle
acceleration \citep{Loules2025a}.

Observationally, the UHECR spectrum exhibits a strong suppression above
$\sim50$\,EeV, consistent with energy losses during propagation through
background photon fields \citep{Kampert2012,Aab2020,Aab2020b,Supanitsky2022}.
Composition-sensitive measurements further indicate a trend toward heavier
nuclei above the ankle \citep{Kampert2012,Anchordoqui2019,Supanitsky2022}.
At lower energies, between $10^{17}$ and $10^{17.5}$\,eV, measurements of
the depth of air-shower maxima ($X_{\max}$) obtained via radio detection
with the LOw-Frequency ARray (LOFAR) instead reveal a large light component
of protons and helium nuclei \citep{Buitink2016}. This finding is difficult
to reconcile with a simple rigidity-limited Galactic population and has
prompted interpretations involving either a second Galactic cosmic-ray
component beyond the canonical supernova-remnant population or a revision
of the conventional view of the ankle as the Galactic--extragalactic
transition point \citep{Buitink2016,Taylor2016}. In either case, these results underscore the
importance of composition-sensitive modelling capable of reconciling observed
spectral features and $X_{\max}$ distributions across the full UHECR energy
range \citep{Kampert2012,Supanitsky2022}.

Models of UHECR acceleration in AGN are usually assessed mainly by whether a source can reach the required energies, whereas composition-sensitive observables provide an additional test of the acceleration process. Because the energy gain in an electric potential scales with nuclear charge, the propagated mass distribution is imprinted on the mean and width of the shower-maximum distribution. Rather than introducing a new charge-scaling prescription, we test whether stochastic variations in near-horizon electric-field acceleration can leave a measurable signature in the observed UHECR composition after propagation. Section~\ref{sec:acceleration} outlines the available near-horizon potential in a MAD SMBH, Section~\ref{sec:mcmc} introduces the stochastic acceleration model and source compositions, and Section~\ref{sec:losses} describes propagation. We then compare the predicted shower-maximum moments with the Auger DNN 2025 measurements using fixed-distance Bayesian fits. Results are presented in Section~\ref{sec:results}, discussed in Section~\ref{sec:discussion}, and summarized in Section~\ref{sec:summary}.

\section{Maximum electric potential and particle acceleration near a supermassive black hole}\label{sec:acceleration}

The capacity of a supermassive black hole (SMBH) to accelerate nuclei is fundamentally set by the maximum electric potential difference that can be established in its immediate environment, in the presence of a strong magnetic field anchored by an accretion flow~\citep{Blandford1977, Tchekhovskoy2015, Nathanail2022}.

In MAD states, which are thought to represent the most efficient jet-producing accretion flows, the black hole is threaded by dynamically significant magnetic flux. The strength of the poloidal magnetic field near the event horizon is determined by the balance between the accretion ram pressure and the magnetic pressure, and for rapidly spinning SMBHs accreting near the Eddington limit, a characteristic field strength can be approximated as~\citep{Tchekhovskoy2015}:

\begin{equation}
B_{\rm MAD} \approx 2 \times 10^4~\mathrm{G}
\left( \frac{L}{0.1\,L_{\mathrm{Edd}}} \right)^{1/2}
\left( \frac{M}{10^9\,M_\odot} \right)^{-1/2}
\left[
\frac{(v_r/c)(h/r)}{0.05}
\right]^{-1/2},
\label{eq:Bfield}
\end{equation}
where $L$ is the bolometric luminosity, $L_{\mathrm{Edd}}$ is the Eddington luminosity, $M$ is the black hole mass, $v_r$ is the radial inflow velocity, and $h/r$ is the disk aspect ratio. For a MAD, close to the horizon $v_r/c \sim 1$ and $h/r \sim 0.05$ \citep{Tchekhovskoy2015}.

In the relativistic regime near a maximally rotating black hole, the induced electric field $\mathcal{E}$ on the black-hole horizon is comparable in magnitude to $B$, namely $\mathcal{E} \sim B$
\citep{Blandford1977, Tchekhovskoy2015}.

The resulting near-horizon potential difference across the accelerating region of size $l = k\, r_g$ (with $r_g = GM/c^2$ and $k$ a geometric factor satisfying $k \lesssim 1$)
is a fundamental property of the AGN central engine, set by the black-hole mass and the magnetic field strength near the event horizon:
\begin{equation}
\mathcal{V}_{\mathrm{max}} \sim \mathcal{E} \cdot l \sim B \cdot k\,\frac{GM}{c^2}.
\label{eq:Vmax_def}
\end{equation}
Substituting the expression for $B$ into Eq.~\eqref{eq:Vmax_def}, we obtain
\begin{align}
\mathcal{V}_{\mathrm{max}} &\sim 8.9\times 10^{20}~\mathrm{V} \cdot k \left( \frac{L}{0.1\,L_{\mathrm{Edd}}} \right)^{1/2}
\left( \frac{M}{10^9\,M_\odot} \right)^{1/2}
\left[
\frac{(v_r/c)(h/r)}{0.05}
\right]^{-1/2}.
\label{eq:Vmax}
\end{align}
The available potential therefore scales with the square root of the black-hole mass as $\mathcal{V}_{\mathrm{max}} \propto M^{1/2}$. The maximum energy attainable by a nucleus is determined by its atomic number $Z$  and the available potential,
\begin{equation}
E_{\mathrm{max}} = Z \mathrm{e}\mathcal{V}_{\mathrm{max}}.
\label{eq:Emax}
\end{equation}
Our scaling indicates that the largest electric potentials are favoured in
massive, strongly magnetised SMBHs, particularly those capable of supporting
MAD-like conditions, and can reach values sufficient for UHECR acceleration.

A physically motivated acceleration site is the extended current sheet
that forms along the jet--disk boundary and can reach the near-horizon
region during magnetic-flux eruption events \citep[e.g.][]{Ripperda2022,Vos2024}. 
Such current sheets are transient and structured by reconnection instabilities and plasmoids, motivating a stochastic description in which different particles may sample different effective fractions of the available potential. We do not attempt to derive the distribution of current-sheet structures or particle trajectories microscopically. 
Instead, we adopt a log-normal distribution as a minimal phenomenological model for variability in the effective potential sampled by different particles. Log-normal behaviour and multiplicative variability occur in several AGN contexts, while log-normal distributions can also emerge from fluctuations in particle-acceleration processes \citep{Rieger_2010,Sinha_2018,Matthews_2021}. These studies, however, do not predict a unique distribution for the effective potential sampled by particles in near-horizon current-sheet acceleration. The inferred standard deviation of $\ln \eta$, $\sigma_\eta$, should therefore be interpreted as a constraint on the adopted stochastic description rather than as a direct measurement of a particular current-sheet size distribution.

\section{Monte Carlo source modeling}\label{sec:mcmc}

We model the energy and composition of UHECRs accelerated in AGN environments with a stochastic Monte Carlo source model. Each simulated
particle is assigned a nuclear species and an effective potential-sampling
factor, while the energy gained during acceleration is determined by the
nuclear charge and the electric-potential scale available in the source
region.

For each realization, the source model samples $(Z,A,\eta)$, where $(Z,A)$
are the charge and mass number of the injected nucleus and $\eta$ is the
effective potential-sampling factor. The energy after acceleration is
\begin{equation}
  E_0 \simeq \eta\,Z\,\mathrm{e}\,\mathcal{V}_{\max}.
  \label{eq:electric_potential_energy}
\end{equation}
In the numerical implementation $E_0$ is the total energy, which also includes
the rest-mass energy $A\,m_pc^2$ and a small initial kinetic energy $E_{\rm init}$
(Gamma-distributed with mean $30\,\mathrm{MeV}$). For all particles retained in
the analysis ($E\geq10^{18}$~eV) both terms are negligible 
and have no effect on the results. The generated particles are then propagated through
the energy-loss model described in Section~\ref{sec:losses}.

\subsection{Injected chemical composition}\label{sec:species}

The injected chemical composition is a central ingredient of the model
because the energy gained from the electric potential scales with nuclear
charge. For the same available potential, nuclei with higher charge $Z$ can
therefore reach higher energies than protons. The composition also affects
the subsequent propagation, because nuclei undergo photodisintegration and
produce daughter nuclei with different masses and charges. Consequently, the
assumed source composition directly influences the predicted energy spectrum
and mass-composition observables at Earth.

Since the true composition of UHECR sources is not known a priori, we consider
two alternative injection scenarios. Scenario~A adopts an empirically
motivated composition inferred from the combined spectrum-and-composition
analysis of the Pierre Auger Observatory. Scenario~B adopts a phenomenological
metal-enriched abundance template intended to represent an AGN circumnuclear
environment. The two scenarios are fitted independently but are treated using
the same acceleration and energy-loss pipeline.

In both scenarios, each nuclear species $i$ is assigned a non-negative
composition weight $w_i$. These weights are converted into discrete injection
probabilities according to
\begin{equation}
  P_i =
  \frac{w_i}{\sum_j w_j},
  \label{eq:species_prob}
\end{equation}
where the sum extends over all species included in the corresponding
composition scenario. In Scenario~A, $w_i$ is the Auger-inspired source
fraction, whereas in Scenario~B it is the adopted injection
weight derived from the solar abundance pattern and modified as described
below. Each Monte Carlo particle is then assigned a nuclear
species by sampling from the discrete probability distribution defined by
equation~\eqref{eq:species_prob}.

\subsubsection{Scenario A: Auger-inspired composition}\label{scenarioA}

As an observationally motivated baseline, we adopt the injected source
fractions reported by the Pierre Auger Observatory in its combined fit of the
UHECR energy spectrum and mass-composition observables. The
systematics-inclusive reference fit represents the injected composition by four nuclear species, with weights
$w_{\rm H}=0.064$,
$w_{\rm He}=0.467$,
$w_{\rm N}=0.375$, and
$w_{\rm Si}=0.094$ \citep[table~3]{Aab2017JCAP}, which sum to unity and therefore correspond
directly to the sampling probabilities in
equation~\eqref{eq:species_prob}. Iron is absent from this four-component fit
because the preceding Auger reference fits yielded a best-fit iron fraction of
zero. These fractions were inferred by propagating an assumed source population
and fitting the resulting energy spectrum and $X_{\max}$ distributions above
$5\times10^{18}\,\mathrm{eV}$; they therefore depend on the adopted source
spectrum, propagation calculation, photodisintegration model, and
hadronic-interaction model, and are not model-independent measurements of the
elemental composition at the sources. These are the four representative nuclei used in the Auger combined fit: H, He, N as a representative intermediate-mass nucleus, and Si as the heavier component. We adopt them as such, and not as a literal claim that UHECR sources contain only these four nuclei, providing an empirically motivated reference case
for comparison with the metal-enriched template described below.

Scenario A is therefore used as an observationally motivated benchmark rather than as an independent prediction of the source composition. Its agreement with the shower observables should not be interpreted as an independent recovery of the Auger-inferred source fractions. Instead, conditional on this injection template, we test whether the adopted charge-dependent acceleration model can reproduce the observed shower moments and constrain the acceleration parameters.

\subsubsection{Scenario B: hydrogen-reduced $3\times$ solar selected-metal
template}
\label{scenarioB}

The second scenario explores a metal-enriched AGN source environment
with an injected UHECR composition that is not assumed to follow the ambient
elemental abundances in a strictly abundance-proportional way. Two independent
considerations point in the same direction. Exploratory
calculations with a solar-like injection mixture retain too large a light
component to reproduce the observed evolution of the shower-width distribution,
which favours a reduced hydrogen contribution. Physically, such a departure
from a solar-like composition is expected in AGN environments. Stars embedded in AGN disks can process hydrogen into helium and enrich the surrounding disk material through their subsequent evolution and mass loss
\citep{Jermyn2022,AliDib2023,Xu2026}. Observationally, He/H enhancements are
inferred in some AGN, while broad-line-region metallicities and
$\alpha$-element abundances are commonly found to be super-solar
\citep{Huang2023,Nagao2006b}.

Guided by both considerations, we construct the following prescription from the
solar photospheric abundances compiled by \citet{Palme2014}: the numerical
weights of the metals are multiplied by a uniform factor of three relative to
their solar values, the helium weight is retained at its solar value, and the
hydrogen injection weight is reduced by a factor of ten, so that
$w_{\rm H}=0.1$,
$w_{\rm He}=10^{A({\rm He})-12}$, and
$w_X=3\,10^{A(X)-12}$ for $X>{\rm He}$,
where $A(X)=\log(n_X/n_{\mathrm{H}})+12$ is the logarithmic solar photospheric
abundance of element $X$. For the selected species H, He, C, N, O, Ne, Mg, Al,
Si, S, Ca, and Fe, the adopted weights are
$w_{\rm H}=1.00\times10^{-1}$,
$w_{\rm He}=8.41\times10^{-2}$,
$w_{\rm C}=9.48\times10^{-4}$,
$w_{\rm N}=2.17\times10^{-4}$,
$w_{\rm O}=1.61\times10^{-3}$,
$w_{\rm Ne}=3.36\times10^{-4}$,
$w_{\rm Mg}=1.04\times10^{-4}$,
$w_{\rm Al}=8.85\times10^{-6}$,
$w_{\rm Si}=9.93\times10^{-5}$,
$w_{\rm S}=4.35\times10^{-5}$,
$w_{\rm Ca}=6.42\times10^{-6}$, and
$w_{\rm Fe}=9.06\times10^{-5}$,
normalized through equation~\eqref{eq:species_prob} before the species are
sampled. Within the resulting mixture the helium-to-hydrogen ratio is
$n_{\rm He}/n_{\rm H}=0.84$, ten times the solar value, which is the composition signature expected if a substantial fraction of the hydrogen has been processed into helium. The reduced hydrogen weight should not be interpreted as a rigidity-dependent suppression acting exclusively on protons; it is a prescribed source-injection weight, while the subsequent acceleration
remains charge dependent for all nuclear species through
Eq.~\eqref{eq:electric_potential_energy}.

Non-solar and heavy-enriched source compositions have independent precedent
in models of UHECR acceleration by radio galaxies. A supersolar abundance of
UHE nuclei is obtained through the injection and shear reacceleration of
pre-existing Galactic CRs, demonstrating that the accelerated composition need
not directly trace the ambient elemental abundances \citep{Kimura2018}. More
recently, solar and Wolf--Rayet source compositions were compared in
source-resolved models of nearby radio galaxies; the solar-composition
scenarios failed to reproduce the measured UHECR spectrum and yielded an
excessively light mass evolution, whereas the heavier Wolf--Rayet composition
provided viable alternatives \citep{Colaco2026}. These results provide
complementary precedent for considering non-solar, heavy-enriched UHECR
injection compositions in radio-galaxy environments.

\subsection{Stochastic potential sampling}\label{sec:eta}

Particles of the same nuclear species are not assumed to experience an
identical effective potential drop. Instead, the acceleration term is
modulated by a positive dimensionless stochastic factor $\eta$, such that
the post-acceleration energy is given by
Eq.~\eqref{eq:electric_potential_energy}. Here, $\eta$ is interpreted as an
effective potential-sampling factor rather than as a conventional
acceleration efficiency. Its stochastic variation represents
particle-to-particle differences in the acceleration histories produced by
the fragmented and time-dependent current-sheet environment.

We model $\eta$ as a log-normal random variable. Equivalently, $\ln\eta$ is
normally distributed,
\begin{equation}
  p(\ln\eta\mid\mu_\eta,\sigma_\eta)
  =
  \mathcal{N}(\ln\eta;\mu_\eta,\sigma_\eta^2)
  =
  \frac{1}{\sigma_\eta\sqrt{2\pi}}
  \exp\!\left[
    -\frac{(\ln\eta-\mu_\eta)^2}{2\sigma_\eta^2}
  \right],
  \label{eq:lneta_normal}
\end{equation}
and, after the change of variables
$\left|\mathrm{d}\ln\eta/\mathrm{d}\eta\right|=1/\eta$,
\begin{equation}
  p(\eta\mid\mu_\eta,\sigma_\eta)
  =
  \frac{1}{\eta\,\sigma_\eta\sqrt{2\pi}}
  \exp\!\left[
    -\frac{(\ln\eta-\mu_\eta)^2}{2\sigma_\eta^2}
  \right],
  \qquad \eta>0.
  \label{eq:eta_lognormal}
\end{equation}

The parameter $\mu_\eta=\langle\ln\eta\rangle$ is the location parameter of
the distribution and is not itself an efficiency. It sets the median
potential-sampling factor,
\begin{equation}
  \widetilde{\eta}=\exp(\mu_\eta),
  \label{eq:eta_median}
\end{equation}
while $\sigma_\eta=[\operatorname{Var}(\ln\eta)]^{1/2}$ sets its logarithmic
width: a one-standard-deviation change corresponds to a multiplicative
factor $e^{\sigma_\eta}$ in $\eta$. Because the energy gain in
Eq.~\eqref{eq:electric_potential_energy} depends on the product
$\eta\,\mathcal{V}_{\max}$, the shower-maximum moments constrain
$\widetilde{\eta}\,\mathcal{V}_{\max}$ more directly than $\mu_\eta$ or
$\mathcal{V}_{\max}$ separately; this degeneracy is quantified in
Sect.~\ref{sec:results_mcmc}.

The log-normal distribution is not explicitly truncated at $\eta=1$.
Accordingly, $\eta$ is not interpreted as a literal fraction of a strict
maximum potential, but as an effective potential-sampling factor describing
the net electric work experienced by each particle in the fragmented
acceleration region. Values $\eta>1$ may therefore represent effective
acceleration histories involving multiple local energisation episodes rather
than a single potential drop. This interpretation is phenomenological, since
the present model does not follow individual particle trajectories through
the current-sheet structure.

The log-normal form is adopted as a phenomenological representation of a
fragmented acceleration environment in which different particles experience
different effective acceleration histories. Multiplicative stochastic
processes can arise in turbulent or reconnecting astrophysical plasmas when
the final energy gain is accumulated through a sequence of local acceleration
episodes \citep{Rieger_2010,Gaspari_2017,Sinha_2018}. Log-normal variability
is also commonly observed in AGN light curves \citep{Matthews_2021}, making
it a natural phenomenological choice for a positive efficiency-like
stochastic parameter. Related UHECR source models also invoke sub-unity
acceleration efficiencies
\citep{Kotera2011,Fang2012,Fang2014,Guepin2018}.

For each Monte Carlo particle, a nuclear species $(Z,A)$ is first drawn from
the fixed injection probabilities of either Scenario~A or Scenario~B,
defined in Sections~\ref{scenarioA} and~\ref{scenarioB}, respectively. An independent value of $\eta$ is then drawn from equation~\eqref{eq:eta_lognormal}, and the accelerated energy is
computed using equation~\eqref{eq:electric_potential_energy}. 
\subsection{Fixed-distance Bayesian inference}\label{sec:bayes}

We constrain the acceleration parameters by fitting the Auger DNN
measurements of $\langle X_{\max}\rangle$ and $\sigma(X_{\max})$
\citep{AbdulHalim2025}. Independent Bayesian fits are performed for
Scenarios~A and B at fixed propagation distances of $d=10$, 30, and
$100\,\mathrm{Mpc}$. Fixing $d$ allows us to examine how propagation
affects the inferred acceleration parameters without interpreting the
$X_{\max}$ moments as a direct distance measurement.

For each fit, the sampled parameter vector is
\begin{equation}
  \boldsymbol{\theta}
  =
  \left(\mu_\eta,\sigma_\eta,\ell_V\right),
  \qquad
  \ell_V
  =
  \log\!\left(
    \frac{\mathcal{V}_{\max}}{\mathrm{V}}
  \right).
  \label{eq:theta_fixed_d}
\end{equation}
The injection probabilities $P_i$ are fixed to those of the corresponding
composition scenario defined in Section~\ref{sec:species}.

\subsubsection{Likelihood and priors}

The likelihood is evaluated over
\begin{equation}
  18.5
  \leq
  \log\!\left(
    \frac{E_i}{\mathrm{eV}}
  \right)
  \leq
  20.0.
  \label{eq:fit_energy_window}
\end{equation}

For compactness, we define
\begin{equation}
  \overline{X}_i
  \equiv
  \langle X_{\max}\rangle_i,
  \qquad
  s_i
  \equiv
  \sigma(X_{\max})_i .
  \label{eq:xmax_moment_notation}
\end{equation}
Both moments contribute to the goodness-of-fit statistic:
\begin{equation}
\begin{aligned}
  \chi^2(\boldsymbol{\theta};d)
  &=
  \sum_{i\in W}
  \left(
    \frac{
      \overline{X}^{\,\mathrm{mod}}_i
      -
      \overline{X}^{\,\mathrm{obs}}_i
    }{
      \delta\overline{X}_i
    }
  \right)^2
  +
  \sum_{i\in W}
  \left(
    \frac{
      s^{\mathrm{mod}}_i
      -
      s^{\mathrm{obs}}_i
    }{
      \delta s_i
    }
  \right)^2 ,
\end{aligned}
\label{eq:chi2}
\end{equation}
where $W$ denotes the fitted energy bins and the model quantities depend
on $\boldsymbol{\theta}$ and $d$. For each observable, the statistical
uncertainty is combined in quadrature with the symmetrised systematic
uncertainty. Correlations between energy bins and between the two moments
are not included. The corresponding log-likelihood is
\begin{equation}
  \ln\mathcal{L}(\boldsymbol{\theta};d)
  =
  -\frac{1}{2}\chi^2(\boldsymbol{\theta};d)
  + \mathrm{const}.
  \label{eq:log_likelihood}
\end{equation}

The 15 fitted energy bins provide 30 measurements. With three free
parameters, the nominal count under the diagonal likelihood is
$\nu=27$ degrees of freedom.

Uniform priors are adopted for the log-normal parameters,
$  -5 < \mu_\eta < 0,\, 0.05 < \sigma_\eta < 2, \label{eq:eta_priors}$
while the prior on the accelerating potential is
\begin{equation}
  p(\ell_V)
  \propto
  \exp\!\left[
    -\frac{(\ell_V-20)^2}{2(1.5)^2}
  \right],
  \qquad
  18.5 < \ell_V < 22.9.
  \label{eq:potential_prior}
\end{equation}

\subsubsection{Posterior sampling}

For the production runs reported here, the posterior distributions are
sampled with \textsc{emcee} \citep{emcee}, using 48 walkers evolved for 3000 steps. The first 1000 steps, corresponding to one third of each chain, are discarded as burn-in. 
Sampling uses a weighted
mixture of differential-evolution \citep{terBraak} and
differential-evolution snooker \citep{terBraak2008} proposals, with
weights of 0.8 and 0.2, respectively.

Each likelihood evaluation uses $N_{\mathrm{mc}}=10^6$ particles. A frozen
Monte Carlo particle bank is reused throughout each fit, making the forward
model and log-posterior deterministic for a fixed random seed. Determinism
is verified through repeated evaluations at the same parameter vector and
seed. The residual dependence on the finite Monte Carlo realisation is
assessed before sampling by evaluating the log-posterior at the initial
parameter vector using five different particle-bank seeds. Convergence is
assessed from the chain traces, walker acceptance fractions, and integrated
autocorrelation times.

\section{Propagation framework}\label{sec:losses}

After acceleration, the simulated particles are propagated over a fixed
source distance $d$. We distinguish between continuous energy losses
(CEL), which reduce the particle energy while preserving its nuclear
identity, and stochastic processes that either remove the particle or
change its mass number. Bethe--Heitler pair production and adiabatic
expansion are treated as continuous losses. Photopion production is
implemented as a catastrophic loss for protons, while
photodisintegration is followed through an explicit mass-evolving
nuclear cascade.

For each fixed distance, we assign the corresponding redshift $z(d)$
using the Planck cosmological parameters \citep{Planck2020}. Since all
source distances considered here are in the nearby Universe, the
resulting redshifts are well approximated by the low-redshift Hubble
relation $z\simeq H_0d/c$; the detailed cosmological model therefore has
a negligible effect on the propagation results. The calculation includes
interactions with the cosmic microwave background (CMB). The
extragalactic background light is not included, and its possible impact
is discussed in Section~\ref{sec:limitations}.

The key composition-changing component of the calculation is an
effective cascade along a reduced Puget--Stecker--Bredekamp (PSB)
nuclear chain \citep{PugetSteckerBredekamp1976}. In contrast to a
fixed-$A$ survival treatment, this approach follows the distribution of
residual daughter nuclei produced during propagation.

\subsection{Continuous energy losses}\label{sec:losses_cel}

\subsubsection{Bethe--Heitler pair production}\label{sec:losses_BH}

Bethe--Heitler pair production,
\begin{equation}
A+\gamma \rightarrow A+e^{+}+e^{-},
\end{equation}
has a small inelasticity per interaction and can therefore be treated
as a continuous energy-loss process
\citep{Blumenthal1970,Chodorowski1992,Rachen1996}. For a nucleus of
total energy $E$, mass number $A$, and charge $Z$, the proton loss
length is evaluated at the same Lorentz factor as the nucleus. We
therefore define the per-nucleon energy
\begin{equation}
E_{\rm N}=\frac{E}{A},
\qquad
E_{{\rm N},18}=\frac{E_{\rm N}}{10^{18}\,{\rm eV}}.
\label{eq:per_nucleon_energy}
\end{equation}

Following \citet{Dermer2009}, the proton-equivalent Bethe--Heitler loss
length on the CMB is
\begin{equation}
\lambda_{\rm BH,p}(E_{\rm N},z)=
\frac{1.02\,E_{{\rm N},18}^{1/2}}
     {(1+z)^{5/2}F_{\phi e}(E_{\rm N},z)}
\,{\rm Gpc},
\label{eq:lambda_BH_proton}
\end{equation}
where
\begin{align}
F_{\phi e}(E_{\rm N},z)={}&
I_{\ell}(b)
+I_{3/2}(b)
\ln\!\left[
\frac{E_{{\rm N},18}(1+z)}
     {2k_{\phi e}}
\right]
\nonumber\\
&+
0.69\,I_0(b)
\left[
\frac{k_{\phi e}}
     {(1+z)E_{{\rm N},18}}
\right]^{3/2},
\label{eq:F_phi_e}
\end{align}
with $k_{\phi e}=6.70$. The dimensionless parameter $b$ is
\begin{equation}
b=\frac{1}{\gamma\Theta(z)},
\qquad
\gamma=\frac{E_{\rm N}}{m_pc^2},
\qquad
\Theta(z)=\frac{k_{\rm B}T_{\rm CMB}(z)}{m_ec^2}.
\label{eq:BH_b_parameter}
\end{equation}

The functions entering Eq.~\eqref{eq:F_phi_e} are
\begin{align}
I_{\ell}(b)
&=
\int_b^\infty
\frac{x^{3/2}\ln x}{\exp(x)-1}\,dx,
\\
I_{3/2}(b)
&=
\int_b^\infty
\frac{x^{3/2}}{\exp(x)-1}\,dx,
\\
I_0(b)
&=
\int_b^\infty
\frac{dx}{\exp(x)-1}
=
-\ln\!\left[1-\exp(-b)\right].
\end{align}
For computational efficiency, these functions are precomputed on a
logarithmic grid in $b$ and evaluated by interpolation during the
Monte Carlo calculation.

The Bethe--Heitler loss length of a nucleus is obtained from the
proton-equivalent expression through
\begin{equation}
\lambda_{\rm BH}(E,Z,A,z)
=
\frac{A}{Z^2}
\lambda_{\rm BH,p}\!\left(\frac{E}{A},z\right).
\label{eq:lambda_BH_nucleus}
\end{equation}
The corresponding fractional energy-loss rate is
\begin{equation}
\left(\frac{b}{E}\right)_{\rm BH}
=
\lambda_{\rm BH}^{-1}.
\label{eq:BH_rate}
\end{equation}

\subsubsection{Adiabatic losses}\label{sec:losses_adiab}

Cosmological expansion produces an additional continuous loss,
\begin{equation}
\left(\frac{b}{E}\right)_{\rm ad}
=
\frac{H(z)}{c},
\label{eq:adiabatic_rate}
\end{equation}
where $H(z)$ is the Hubble rate. For the distances considered here this
term is small, but it is retained for consistency.

The Bethe--Heitler and adiabatic rates are evaluated at the
post-acceleration energy $E_0$ and applied over the full propagation
distance:
\begin{equation}
E_{\rm CEL}
=
E_0
\exp\!\left\{
-d\left[
\left(\frac{b}{E}\right)_{\rm BH}
+
\left(\frac{b}{E}\right)_{\rm ad}
\right]
\right\}.
\label{eq:combined_CEL}
\end{equation}
This treatment approximates the loss rates as constant over the
trajectory rather than updating them continuously as the energy
decreases.

\subsection{Photopion production}\label{sec:losses_pion}

For protons above the Greisen--Zatsepin--Kuzmin threshold, photopion
production proceeds mainly through the $\Delta^{+}$ resonance,
\begin{equation}
p+\gamma\rightarrow\Delta^{+}\rightarrow
\begin{cases}
p+\pi^{0},\\
n+\pi^{+}.
\end{cases}
\label{eq:pion_channels}
\end{equation}

Following the semi-analytic approximation of \citet{Dermer2009}, the
proton interaction length is

\begin{equation}
\lambda_{\pi}(E,z)\approx
\frac{\mathrm{Mpc}}{(1+z)^3}
\begin{cases}
\dfrac{13.7\,e^{\omega}}{1+\omega},
& x<4,
\\[8pt]
\dfrac{27.4}{\mathcal{D}(\omega)},
& x\geq 4,
\end{cases}
\label{eq:lambda_pion}
\end{equation}
where
\begin{equation}
x\equiv E_{20}(1+z),
\qquad
\omega\equiv\frac{4}{x},
\qquad
E_{20}\equiv\frac{E}{10^{20}\,\mathrm{eV}},
\end{equation}
and
\begin{equation}
\mathcal{D}(\omega)\equiv
\Gamma(3)\zeta(3)
+\omega^2
\left[
\ln\!\left(1-e^{-\omega}\right)-\frac{1}{2}
\right].
\end{equation}

Photopion production is treated as a catastrophic process. After the
continuous-loss step, each proton is retained with probability
\begin{equation}
P_{\pi}(E_{\rm CEL},d)
=
\exp\!\left[
-\frac{d}{\lambda_{\pi}(E_{\rm CEL},z)}
\right].
\label{eq:pion_survival}
\end{equation}
A failed interaction removes the proton from the propagated sample.
This provides a first-order representation of the high-energy
suppression but does not follow the lower-energy nucleon produced after
an individual interaction. Photopion interactions of nuclei with
$A>1$ are neglected in the present model.

\subsection{Photodisintegration cascade}\label{sec:losses_dis}

Photodisintegration changes the nuclear identity through the emission
of one or more nucleons,
\begin{equation}
A+\gamma\rightarrow (A-n)+n\,{\rm nucleons},
\qquad n\geq1.
\label{eq:photodisintegration}
\end{equation}
Because the Lorentz factor is approximately conserved, the energy of a
daughter nucleus of mass $A'$ produced from a parent of mass
$A_{\rm inj}$ is
\begin{equation}
E_{A'}
=
\frac{A'}{A_{\rm inj}}E_{\rm inj}.
\label{eq:daughter_energy}
\end{equation}

We follow the residual nucleus through a reduced PSB chain containing
one representative nuclide for each retained mass number
\citep{PugetSteckerBredekamp1976}. The unstable gap nuclei $A=5$ and
$A=8$ are skipped. Successive cascade steps can therefore populate the
retained intermediate mass numbers below $A_{\rm inj}$, with transitions
across the omitted $A=5$ and $A=8$ gaps treated within the reduced chain.
The corresponding charge $Z(A)$ is assigned from the adopted PSB mapping.

\subsubsection{Cross sections and effective loss lengths}
\label{sec:losses_cross_sections}

We adopt the giant-dipole-resonance parametrisation of
\citet{SteckerSalamon1999}, separating single-nucleon, two-nucleon, and
multi-nucleon plateau channels.  The energy-integrated strength of the
giant-dipole resonance is normalised using the Thomas--Reiche--Kuhn
sum rule,
\begin{equation}
\Sigma_d
\simeq
60\frac{(A-Z)Z}{A}
\ {\rm mbarn\,MeV}.
\label{eq:TRK_sum}
\end{equation}

The single- and two-nucleon components are represented by Gaussian
profiles centred at approximately $18$ and $25$ MeV, respectively,
with widths of approximately $8$ MeV. The plateau component extends
from approximately $30$ to $150$ MeV and species-dependent thresholds are
taken from \citet{SteckerSalamon1999}.

For every retained mass number, the channel-dependent interaction
lengths are calculated by integrating the corresponding cross sections
over the present-day CMB photon distribution following
\citet{HooperSarkarTaylor2008}. The resulting rates are tabulated over
$17.5\leq \log(E/{\rm eV})\leq22.5$ and interpolated during propagation.

The channel rates are combined into an effective nucleon-loss rate,
\begin{equation}
\frac{1}{L_A(E)}
=
\frac{1}{\lambda_A^{1{\rm N}}(E)}
+
\frac{2}{\lambda_A^{2{\rm N}}(E)}
+
\frac{\bar n_{\rm pl}}
     {\lambda_A^{\rm pl}(E)},
\label{eq:effective_nucleon_loss}
\end{equation}
where $\bar n_{\rm pl}=4$ is the adopted mean multiplicity of the
plateau component. The multiplicity factors in
Eq.~\eqref{eq:effective_nucleon_loss} convert the channel interaction
rates into an effective rate of mass-number reduction.

\subsubsection{Cascade evolution}\label{sec:losses_cascade}

The multiplicity-weighted rates drive an effective sequential cascade
along the PSB chain. The residual-nucleus populations satisfy
\begin{equation}
\frac{dN_A}{dL}
=
-\frac{N_A}{L_A(E_A)}
+
\frac{N_{A+1}}{L_{A+1}(E_{A+1})},
\label{eq:PSB_cascade}
\end{equation}
following \citet{HooperSarkarTaylor2008,AhlersTaylor2010}. The
two-nucleon and plateau channels are therefore not treated as separate
direct transitions. Instead, their multiplicities enter through the
effective rate in Eq.~\eqref{eq:effective_nucleon_loss}.

For a nucleus injected with $(A_{\rm inj},E_{\rm inj})$, the daughter
energies entering the interaction matrix are evaluated as
\begin{equation}
E_A
=
\frac{A}{A_{\rm inj}}E_{\rm inj}.
\end{equation}
Writing Eq.~\eqref{eq:PSB_cascade} in matrix form,
\begin{equation}
\frac{d\boldsymbol{N}}{dL}
=
\mathbf{M}\boldsymbol{N},
\end{equation}
the population after propagation over distance $d$ is
\begin{equation}
\boldsymbol{N}(d)
=
\exp(\mathbf{M}d)\boldsymbol{N}(0).
\label{eq:matrix_exponential}
\end{equation}

For each Monte Carlo nucleus, a final residual daughter mass is sampled
from the population fractions in Eq.~\eqref{eq:matrix_exponential}.
The calculation follows the residual nuclear fragment; nucleons emitted
during photodisintegration are not added as separate particles to the
arriving population. Under the approximately conserved-Lorentz-factor
approximation, an emitted nucleon would carry an energy of order the
parent energy per nucleon, $E_A/A=E_{\rm inj}/A_{\rm inj}$, and is
therefore not assumed to be energetically negligible. The omission of these secondary nucleons is a simplifying approximation.
Its impact, together with the other approximations in our semi-analytic
propagation treatment, is assessed through comparison with CRPropa in
Appendix~\ref{app:validation}.

\subsection{Combined Monte Carlo pipeline}\label{sec:losses_combined}

Each accelerated particle is propagated as follows. A source species
$(Z,A_{\rm inj})$ and potential-sampling factor $\eta$ are sampled, giving the
post-acceleration energy $E_0$ of Eq.~\eqref{eq:electric_potential_energy}.
Bethe--Heitler and
adiabatic losses are then applied through Eq.~\eqref{eq:combined_CEL}, and
particles with $E_{\rm CEL}<10^{17}$\,eV are discarded. For injected protons,
photopion survival is sampled using Eq.~\eqref{eq:pion_survival}; for nuclei
with $A_{\rm inj}>1$, the PSB cascade is evolved over the distance $d$, and a
residual daughter $A'$ is sampled and assigned the energy
$E'=(A'/A_{\rm inj})E_{\rm CEL}$. Particles with $E'\geq10^{18}$\,eV are
retained and converted into shower observables using the generalised Gumbel
parametrisation of Section~\ref{sec:xmax}. The resulting energy and mass
distributions are binned in $\log(E/{\rm eV})$ and used to calculate
$\langle X_{\max}\rangle$ and $\sigma(X_{\max})$ for comparison with the Auger
measurements.

\subsection{Shower-maximum observables}
\label{sec:xmax}

After propagation, each simulated particle is characterised by its
energy $E$ and residual mass number $A$. To compare the propagated
population with composition-sensitive air-shower measurements, we
convert these quantities into predictions for the atmospheric depth
of shower maximum, $X_{\max}$.

The first two moments of the $X_{\max}$ distribution provide
complementary information about the primary mass composition. At fixed
energy, lighter primaries generally produce showers that develop
deeper in the atmosphere and exhibit larger event-to-event
fluctuations, whereas heavier nuclei produce shallower and less
fluctuating showers. Consequently, $\langle X_{\max}\rangle$ is
primarily sensitive to the mean logarithmic mass, while
$\sigma(X_{\max})$ provides additional information about the spread
of masses in a mixed composition
\citep{2013Domenico,AbdulHalim2025}.

We describe the conditional distribution
$p(X_{\max}\mid E,A)$ using the generalised Gumbel formalism introduced
by \citet{2013Domenico}. For the numerical implementation, we adopt
the EPOS-LHC coefficients reported by \citet{AbdulHalim2023}. These
coefficients were obtained by fitting generalised Gumbel distributions
to CONEX simulations of H-, He-, N-, Si-, and Fe-initiated showers
with energies between $10^{17}$ and $10^{20}\,\mathrm{eV}$. CONEX
combines an explicit Monte Carlo treatment of the high-energy part of
the shower cascade with a numerical solution of cascade equations for
the lower-energy sub-showers \citep{Bergmann2007}. The mapping from the propagated particle properties $(E,A)$ to the $X_{\max}$ distribution is based on generalised-Gumbel fits to CONEX
air-shower simulations performed with the EPOS-LHC
hadronic-interaction model \citep{2013Pierog}.

In our Monte Carlo implementation, the generalised Gumbel distribution
is sampled using the mathematically equivalent representation
\begin{equation}
  X_{\max}
  =
  \mu(E,A)
  -
  \sigma_G(E,A)\ln G,
  \label{eq:xmax_gumbel}
\end{equation}
where
\begin{equation}
  G
  \sim
  \Gamma\!\left(
    \lambda(E,A),
    \frac{1}{\lambda(E,A)}
  \right),
  \label{eq:xmax_gamma}
\end{equation}
and the two arguments of the Gamma distribution denote its shape and
scale parameters, respectively.
Here, $\mu$, $\sigma_G$, and $\lambda$ are the location, scale, and shape
parameters of the generalised Gumbel distribution, respectively; they should
not be identified directly with the mean and standard deviation of
$X_{\max}$.

Their dependence on primary energy and mass follows the parametrisation of
\citet{2013Domenico}. Defining
\begin{equation}
  \ell_E
  \equiv
  \log\left(\frac{E}{\mathrm{eV}}\right)-19,
\end{equation}
each Gumbel parameter $q\in\{\mu,\sigma_G,\lambda\}$ is expressed as
\begin{equation}
q(E,A)
=
\sum_{k=0}^{n_q}
p_k^{(q)}(A)\,\ell_E^k,
\qquad
n_q=
\begin{cases}
2, & q=\mu,\\
1, & q=\sigma_G,\lambda.
\end{cases}
\label{eq:gumbel_energy_dependence}
\end{equation}

The mass-dependent coefficients are quadratic functions of $\ln A$:
\begin{equation}
  p_k^{(q)}(A)
  =
  a_{k,0}^{(q)}
  +
  a_{k,1}^{(q)}\ln A
  +
  a_{k,2}^{(q)}(\ln A)^2.
  \label{eq:gumbel_mass_dependence}
\end{equation}
The numerical EPOS-LHC coefficients are taken from
\citet[Appendix~A, table~5]{AbdulHalim2023}.

For reference, the conditional mean and variance of this distribution
at fixed $E$ and $A$ are
\begin{align}
  \langle X_{\max}\rangle_{E,A}
  &=
  \mu
  +
  \sigma_G
  \left[
    \ln\lambda-\psi(\lambda)
  \right],
  \label{eq:gumbel_mean}\\
  \operatorname{Var}(X_{\max}\mid E,A)
  &=
  \sigma_G^2\psi_1(\lambda),
  \label{eq:gumbel_variance}
\end{align}
where $\psi$ and $\psi_1$ are the digamma and trigamma functions,
respectively \citep{2013Domenico}.

In our forward-model implementation we evaluate the two bin moments directly from
the conditional generalised-Gumbel moments of
Eqs.~\eqref{eq:gumbel_mean}--\eqref{eq:gumbel_variance}, rather than by drawing a
sampled $X_{\max}$ for each particle. 
For the $N_{\rm bin}$ propagated particles in a given
$\log(E/\mathrm{eV})$ bin,
with conditional means $m_i$ and variances $v_i$
evaluated at each particle's $(E_i,A_i)$, the predicted moments follow from the law
of total variance,
\begin{equation}
\begin{aligned}
  \langle X_{\max}\rangle
    &= \frac{1}{N_{\rm bin}}\sum_i m_i, \\
  \sigma^2(X_{\max})
    &= \frac{1}{N_{\rm bin}}\sum_i v_i
    + \frac{1}{N_{\rm bin}}\sum_i
      \left(m_i-\langle X_{\max}\rangle\right)^2,
\end{aligned}
\label{eq:total_variance}
\end{equation}
where the first term is the mean intrinsic single-shower variance and the second is
the dispersion of mean depths across the coexisting energies and nuclear masses in the
bin. This analytic evaluation is exact and removes the Monte Carlo sampling noise that
a single $X_{\max}$ draw per particle would otherwise introduce into
$\sigma(X_{\max})$, the observable that carries most of the composition information.
Finally, we test this treatment against CRPropa~3 \citep{AlvesBatista2016,AlvesBatista2022} in Appendix~\ref{app:validation}.

\section{Results}
\label{sec:results}

\subsection{Posterior-predictive shower-maximum moments}
\label{sec:xmax_results}

We first examine whether the propagated populations reproduce the
composition-sensitive air-shower measurements of the Pierre Auger
Observatory. Figures~\ref{fig:combined_xmax}
and~\ref{fig:combined_sigma} show the posterior-predictive mean and
standard deviation of the $X_{\max}$ distribution for Scenarios~A
and B at fixed propagation distances of $d=10$, 30, and 100\,Mpc.
The predictions are compared with the DNN-reconstructed Auger
measurements over the fitted interval
$18.5\leq\log(E/\mathrm{eV})\leq20.0$
\citep{AbdulHalim2025}. The posterior-predictive bands are obtained by
evaluating the complete acceleration, propagation, and air-shower
model over samples from the inferred posterior distributions of the
acceleration parameters.

\subsubsection{Mean depth of shower maximum}

\begin{figure*}[!!!!!!!!!!!!!!htbt]
    \centering
    \subfigure[]{%
        \includegraphics[width=0.32\textwidth]
        {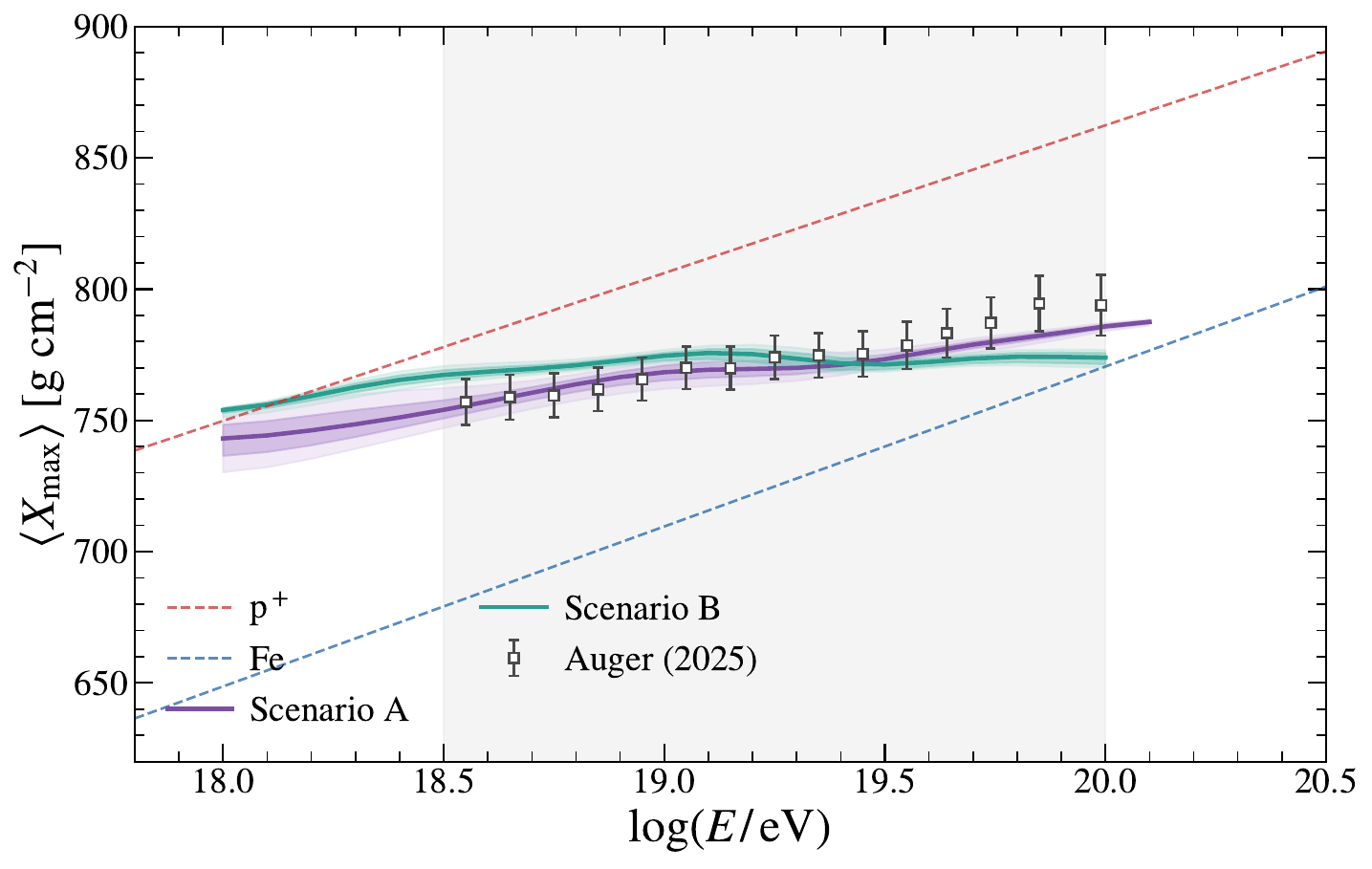}}%
    \hfill
    \subfigure[]{%
        \includegraphics[width=0.32\textwidth]
        {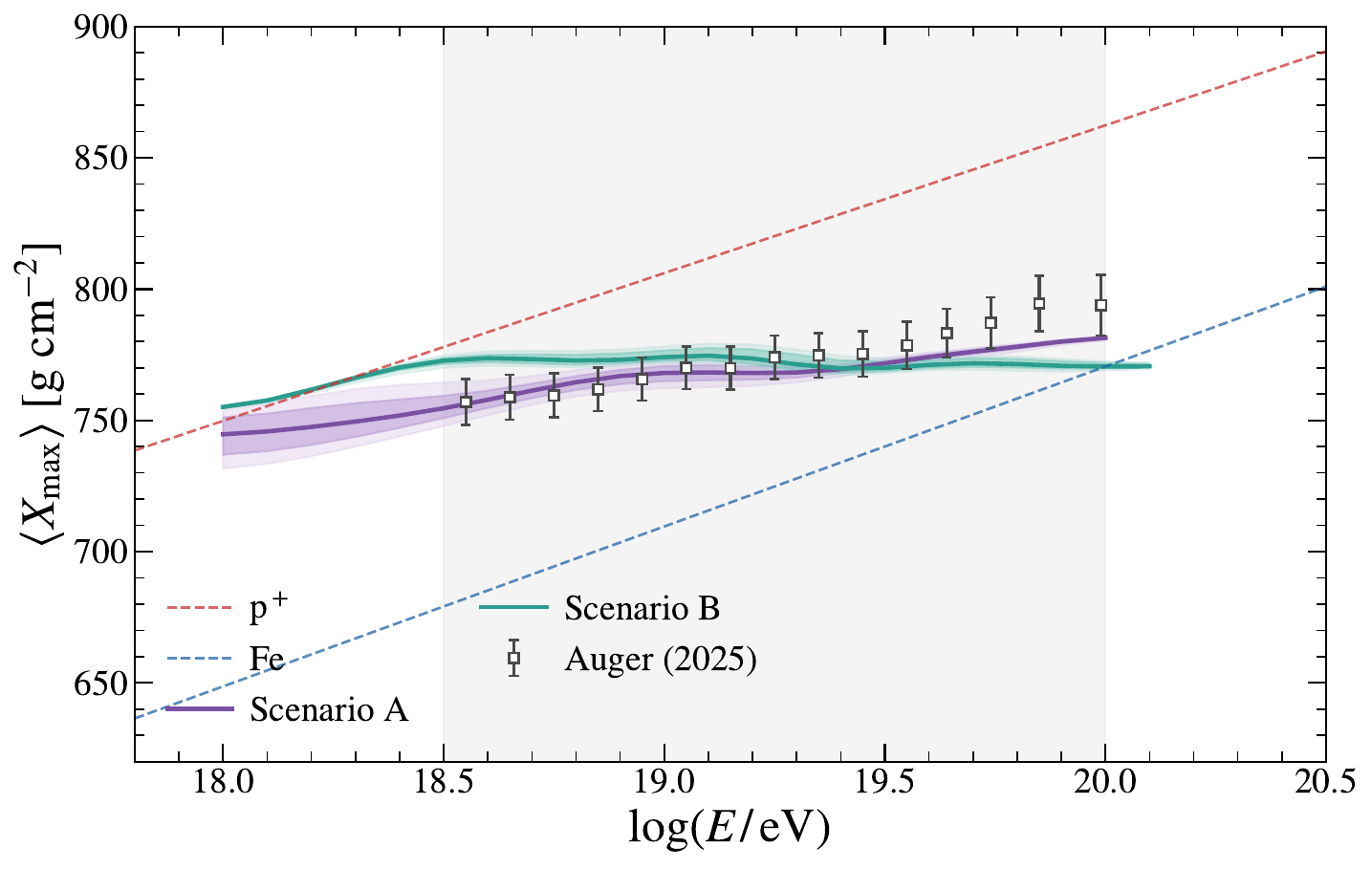}}%
    \hfill
    \subfigure[]{%
        \includegraphics[width=0.32\textwidth]
        {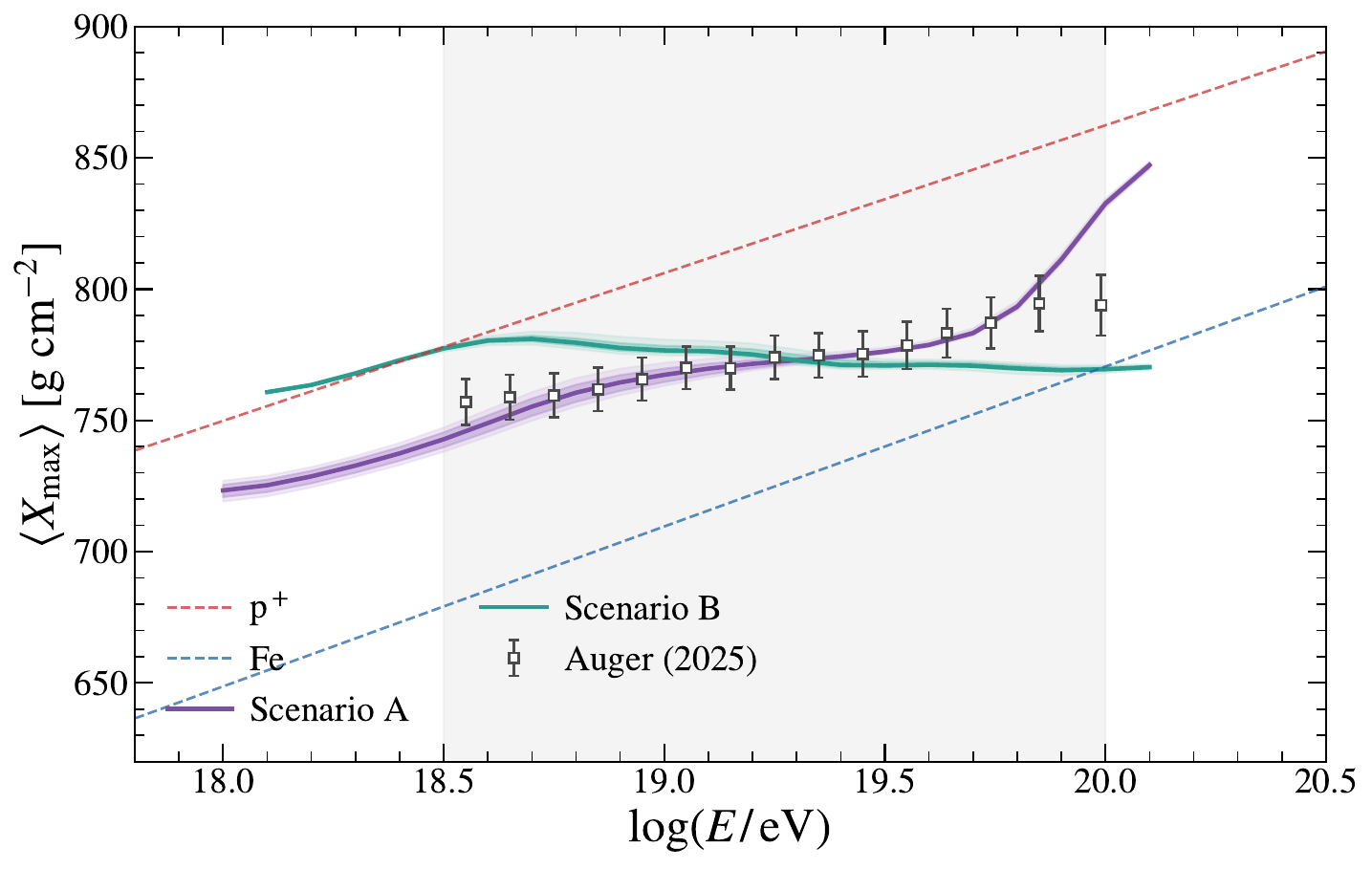}}%
    \caption{
    Posterior-predictive mean depth of shower maximum,
    $\langle X_{\max}\rangle$, at propagation distances of
    $d=10$, 30, and 100\,Mpc (panels a--c, respectively).
    Scenario~A corresponds to Auger 2017 above-ankle composition,
    while Scenario~B corresponds to the hydrogen-reduced
    $3\times$ solar selected-metal template containing approximately
    53\,\% hydrogen after normalisation. Dark and light bands denote
    the 68\,\% and 95\,\% posterior-predictive intervals,
    respectively. Dashed curves indicate the pure-proton and
    pure-iron reference expectations, and square markers show the
    Auger DNN 2025 measurements \citep{AbdulHalim2025}. The grey
    shaded region marks the energy interval included in the
    likelihood. For visualisation, the posterior-predictive median and interval
boundaries have been smoothed along the energy axis with a Gaussian
filter of standard deviation 1.5 bins ($0.15$ dex); the unsmoothed
predictions are used in the likelihood and parameter inference. }
    \label{fig:combined_xmax}
\end{figure*}

Figure~\ref{fig:combined_xmax} presents the predicted mean depth of
shower maximum, $\langle X_{\max}\rangle$. At $d=10$ and 30\,Mpc, both source-composition scenarios reproduce the
overall scale and increasing evolution of the measured mean shower depth,
although their posterior-predictive intervals overlap substantially but not uniformly
across the fitted energy range. The differences are most apparent toward
the lower and upper ends of the fitted interval.

Differences between the scenarios become more apparent toward the
upper end of the fitted interval. Scenario~A retains a rising
$\langle X_{\max}\rangle$ with increasing energy and remains closer
to the highest-energy measurements. Scenario~B reaches a broad
maximum at intermediate energies and subsequently flattens or
decreases slightly, placing its median prediction below the final
Auger measurements. Nevertheless, the overlap of the predictive
intervals indicates that $\langle X_{\max}\rangle$ alone provides
only limited discrimination between the two injection scenarios at
$d=10$--30\,Mpc.

At $d=100$\,Mpc, the two scenarios already differ at the low-energy end
of the fitted interval, approach one another at intermediate energies, and
separate again above $\log(E/\mathrm{eV})\simeq19.6$. Scenario~A remains
comparatively close to the measured $\langle X_{\max}\rangle$ over much of
the intermediate-energy range, but develops a pronounced upward curvature
at the highest energies and eventually exceeds the Auger measurements.
Scenario~B instead lies above the data at the low-energy end and then
flattens or decreases toward the highest energies, where it falls below the
measurements. Thus, neither scenario reproduces the complete
$\langle X_{\max}\rangle$ evolution at $100\,\mathrm{Mpc}$, although the
two templates depart from the observations in different energy ranges.

\subsubsection{Fluctuations of the shower maximum}

\begin{figure*}[!!!!!!!!!!!!htb]
    \centering
    \subfigure[]{%
        \includegraphics[width=0.32\textwidth]
        {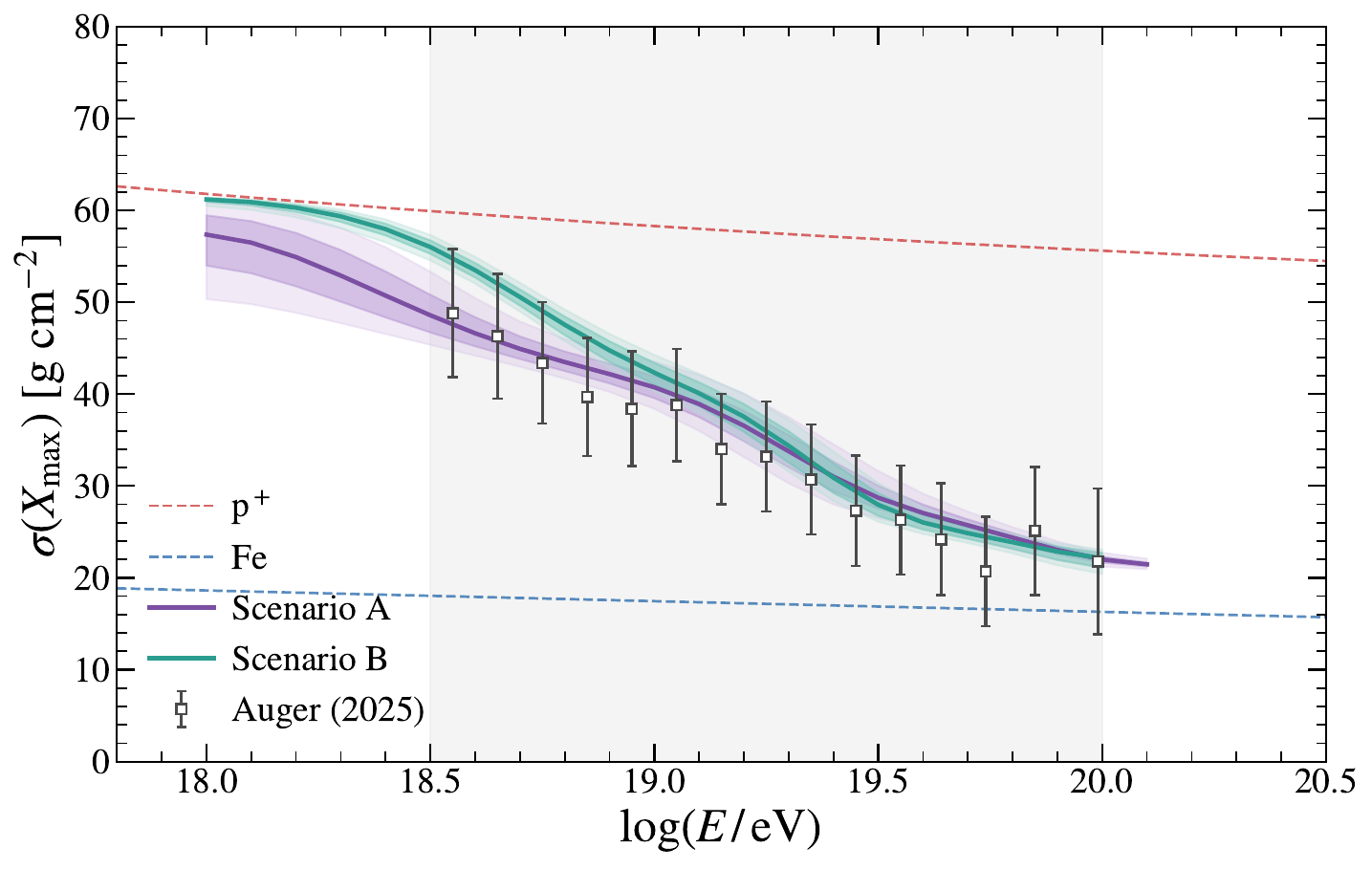}}%
    \hfill
    \subfigure[]{%
        \includegraphics[width=0.32\textwidth]
        {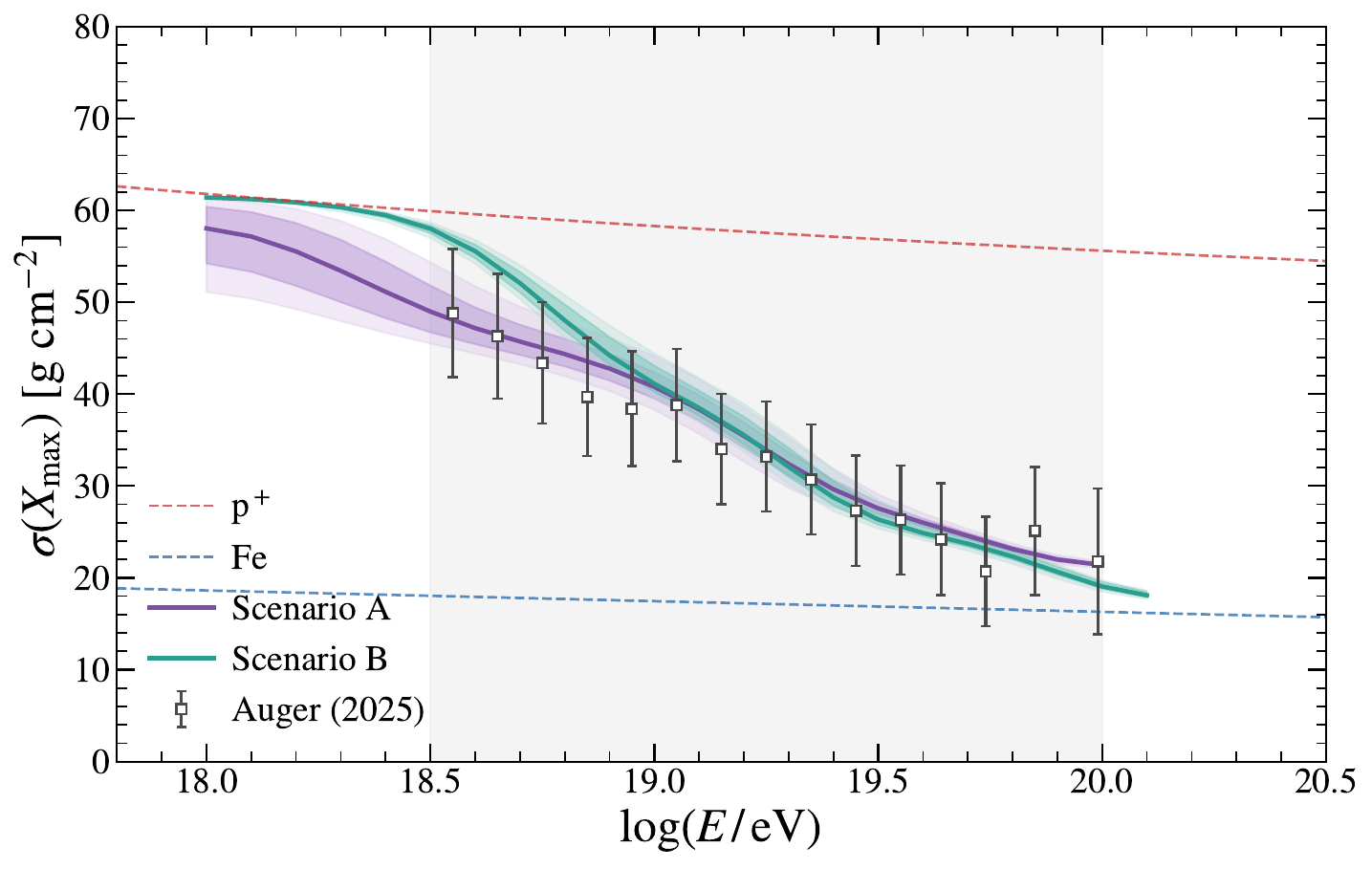}}%
    \hfill
    \subfigure[]{%
        \includegraphics[width=0.32\textwidth]
        {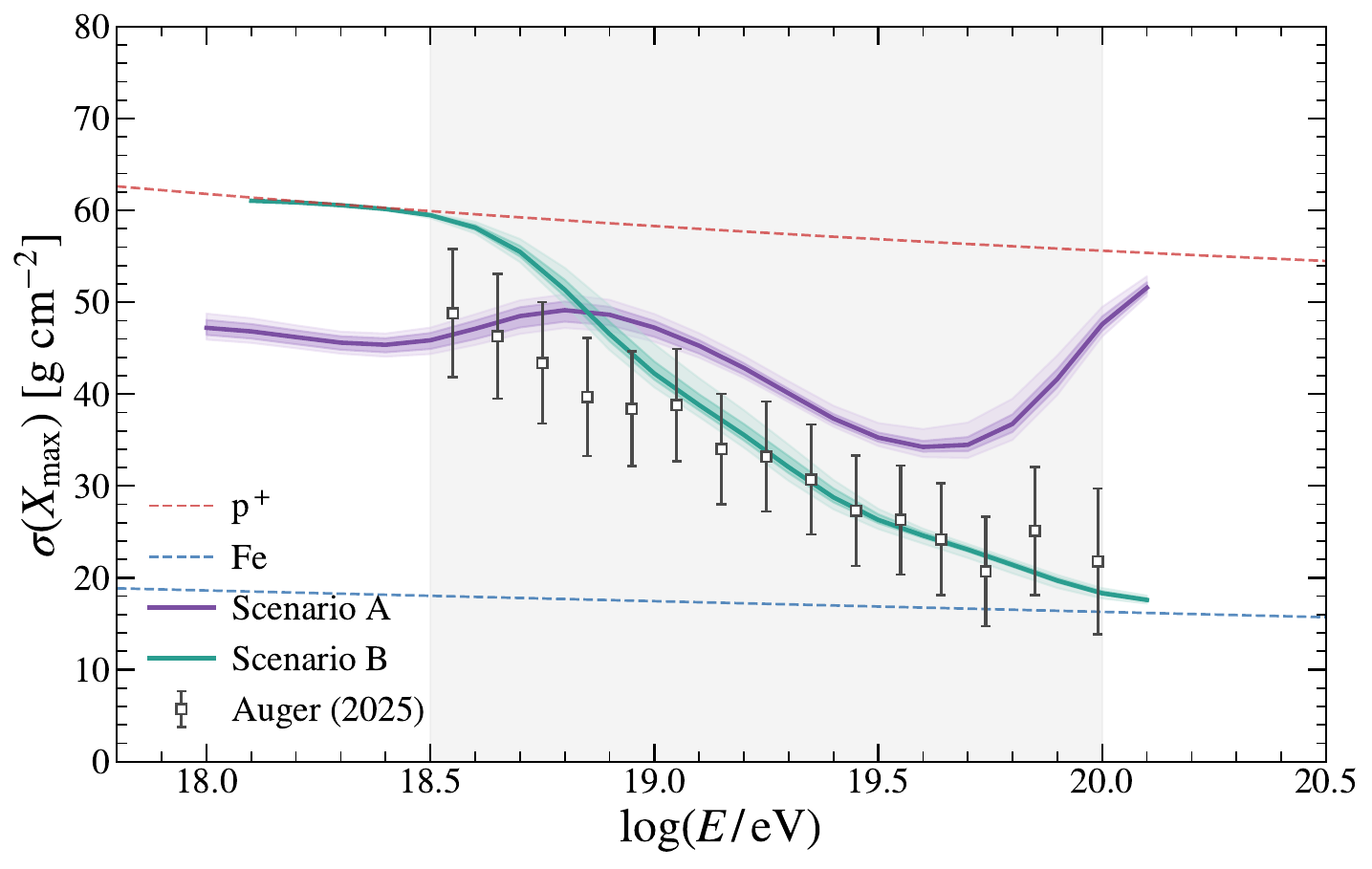}}%
    \caption{
    Posterior-predictive standard deviation of the depth of shower
    maximum, $\sigma(X_{\max})$, at propagation distances of
    $d=10$, 30, and 100\,Mpc (panels a--c, respectively).
    The scenarios, posterior-predictive intervals, reference curves,
    and observational markers follow the definitions in
    Fig.~\ref{fig:combined_xmax}. The grey shaded region marks the
    energy interval included in the likelihood.
    }
    \label{fig:combined_sigma}
\end{figure*}

Figure~\ref{fig:combined_sigma} shows the corresponding standard
deviation, $\sigma(X_{\max})$. This observable is more sensitive than
the mean alone to the width of the propagated mass distribution,
because it contains contributions both from intrinsic air-shower
fluctuations and from the coexistence of different nuclear masses
within the same energy bin.

At $d=10$ and 30\,Mpc, both scenarios reproduce the principal
observed trend: $\sigma(X_{\max})$ decreases progressively with
increasing energy. The model predictions evolve from broad
distributions at the lower boundary of the fitted interval toward
values of approximately $20$--$25\,\mathrm{g\,cm^{-2}}$ in the
highest-energy bins. Scenario~B generally predicts somewhat broader
fluctuations over the lower and intermediate energy range, while the
two scenarios converge toward similarly narrow values at high energy.

At $d=100$\,Mpc, the predictions differ much more strongly.
Scenario~B retains a smooth decline of $\sigma(X_{\max})$: it is broader
than the measurements at the low-energy end of the fitted interval but
approaches the narrow distributions measured at the highest energies.
Scenario~A remains broader than the measurements over much of the fitted
range, reaches a minimum near
$\log(E/\mathrm{eV})\simeq19.5$--19.7, and subsequently develops a
pronounced high-energy broadening, rising toward
$\sigma(X_{\max})\sim50\,\mathrm{g\,cm^{-2}}$.

The high-energy broadening of Scenario~A at 100\,Mpc occurs together
with the upturn in $\langle X_{\max}\rangle$ seen in
Fig.~\ref{fig:combined_xmax}. Taken together, these features suggest
the emergence of a lighter and more heterogeneous high-energy
population after propagation. The long-distance behaviour is therefore sensitive to the interplay between the injected composition, the
charge-dependent acceleration process, and propagation-induced
composition changes.

Overall, the joint comparison of the two shower moments favours the
behaviour obtained at $d=10$ and 30\,Mpc over that obtained at
$d=100$\,Mpc. At the shorter distances, both
$\langle X_{\max}\rangle$ and $\sigma(X_{\max})$ contribute to the
differences between the two source-composition scenarios under the
adopted joint likelihood. In particular, it exposes a clear tension
between the 100\,Mpc Scenario~A prediction and the observed
high-energy shower fluctuations. Scenario~B predicts
somewhat broader fluctuations over parts of the fitted interval,
consistent with its larger injected hydrogen fraction, but its width
also decreases toward the narrower values observed at the highest
energies. Thus, the approximately 53\,\% hydrogen fraction at
injection does not translate directly into a proton-dominated
high-energy population, because the charge-dependent acceleration
term preferentially promotes nuclei with larger $Z$ into the highest
energy bins.

Finally, for visual clarity, the posterior-predictive median and the boundaries of the 68\,\% and 95\,\% intervals shown in
Figs.~\ref{fig:combined_xmax} and~\ref{fig:combined_sigma} are smoothed
along the energy axis using a one-dimensional Gaussian filter with a
standard deviation of 1.5 energy bins, corresponding to
$0.15$ in $\log(E/\mathrm{eV})$. The smoothing is applied only
after the pointwise posterior-predictive quantiles have been computed
and is used exclusively for visualisation; it does not affect the
forward-model predictions, the likelihood, the posterior samples, or
the reported goodness-of-fit values.

\subsection{Bayesian parameter constraints}\label{sec:results_mcmc}

Table~\ref{tab:posterior_summary} summarises the posterior constraints from
the six fixed-distance fits described in Section~\ref{sec:bayes}. The full
marginalised posteriors and chain-trace diagnostics are presented in
Appendix~\ref{app:posteriors}.

\begin{table}[!htb]
\caption{Posterior summary for the Bayesian MCMC fit to the Auger DNN
2025 $X_{\max}$ moments at three fixed propagation distances and two
injection compositions. Central values are posterior medians;
uncertainties denote 16th--84th percentile intervals. The number of
degrees of freedom is $\nu=27$ for all fits.}
\label{tab:posterior_summary}
\centering
\setlength{\tabcolsep}{1.5pt}
\renewcommand{\arraystretch}{0.95}
\begin{tabular}{c|c|c|c|c|c|c}
\hline\hline
$d$ & Scenario & $\chi^2_{\min}$ & $\chi^2/\nu$ &
$\mu_\eta$ & $\sigma_\eta$ & $\log\mathcal{V}_{\max}$ \\
{[Mpc]} & & & & & & {[V]} \\
\hline
10  & A & 8.9  & 0.33
    & $-2.697^{+1.698}_{-1.532}$
    & $0.623^{+0.110}_{-0.096}$
    & $19.741^{+0.670}_{-0.738}$ \\[1pt]
30  & A & 12.7 & 0.47
    & $-2.709^{+1.697}_{-1.513}$
    & $0.590^{+0.129}_{-0.097}$
    & $19.739^{+0.654}_{-0.738}$ \\[1pt]
100 & A & 85.0 & 3.15
    & $-2.894^{+1.641}_{-1.393}$
    & $1.237^{+0.076}_{-0.068}$
    & $19.652^{+0.606}_{-0.723}$ \\[2pt]
10  & B & 26.8 & 0.99
    & $-2.793^{+1.697}_{-1.469}$
    & $0.488^{+0.067}_{-0.057}$
    & $19.657^{+0.639}_{-0.728}$ \\[1pt]
30  & B & 38.6 & 1.43
    & $-2.743^{+1.735}_{-1.487}$
    & $0.359^{+0.068}_{-0.029}$
    & $19.739^{+0.648}_{-0.756}$ \\[1pt]
100 & B & 66.1 & 2.45
    & $-2.640^{+1.668}_{-1.578}$
    & $0.273^{+0.053}_{-0.030}$
    & $19.833^{+0.677}_{-0.730}$ \\
\hline
\end{tabular}
\end{table}

For Scenario~A, the fits at $d=10$ and $30\,\mathrm{Mpc}$ give very similar
constraints and low reduced chi-square values
$\chi^2/\nu=0.33$ and $0.47$, respectively. In both cases,
$\sigma_\eta\simeq0.6$ is comparatively well constrained.
At $d=100\,\mathrm{Mpc}$, the fit deteriorates to
$\chi^2/\nu=3.15$ and the required spread increases to
$\sigma_\eta=1.24^{+0.08}_{-0.07}$. By contrast, the individual
marginal constraints on $\mu_\eta$ and
$\ell_V\equiv\log(\mathcal{V}_{\max}/\mathrm{V})$ are broad and partly
prior dominated because of their strong degeneracy. Their similar median
values across the fixed-distance fits should therefore not be interpreted
as evidence for an unchanged intrinsic $\mathcal{V}_{\max}$.

The corner plots show a strong anticorrelation between $\mu_\eta$ and
$\ell_V$, because both parameters control the characteristic acceleration
scale. Their individual marginal distributions are consequently broad and
partly prior dominated, whereas $\sigma_\eta$ is much more localised.
For the representative $d=30\,\mathrm{Mpc}$ Scenario~A fit, the constrained
quantity is approximately the combination
\begin{equation}
  \log\!\left(
    \frac{\widetilde{\eta}\mathcal{V}_{\max}}{\mathrm{V}}
  \right)
  =
  \frac{\mu_\eta}{\ln 10}+\ell_V,
  \qquad
  \widetilde{\eta}\equiv e^{\mu_\eta},
  \label{eq:com_V_eta}
\end{equation}
rather than either $\mu_\eta$ or $\ell_V$ separately. The median values
correspond to $\widetilde{\eta}\simeq0.07$, although this quantity should
not be interpreted independently of the correlated potential.

Scenario~B exhibits the same $\mu_\eta$--$\ell_V$ degeneracy but favours a
progressively narrower efficiency distribution,
$\sigma_\eta=0.49$, $0.36$, and $0.27$ at $d=10$, 30, and
$100\,\mathrm{Mpc}$, respectively. Its fit quality simultaneously degrades
from $\chi^2/\nu=0.99$ to $1.43$ and $2.45$. Thus, neither composition
provides an acceptable description at $100\,\mathrm{Mpc}$, although they
respond to the increasing propagation distance through opposite trends in
$\sigma_\eta$.

\subsection{Composition contrast and proton content}
\label{sec:proton_suppression}

The two injection templates differ by a factor of approximately eight in
their proton fractions: $6.4\%$ for Scenario~A and approximately $53\%$
for Scenario~B. At $d=10$ and $30\,\mathrm{Mpc}$, Scenario~A gives
$\chi^2/\nu=0.33$ and $0.47$, compared with $0.99$ and $1.43$ for
Scenario~B. The corresponding differences,
$\Delta\chi^2\equiv\chi^2_{\min,\mathrm{B}}-\chi^2_{\min,\mathrm{A}}$,
are $17.9$ and $25.9$, indicating a strong relative preference for
Scenario~A under the adopted likelihood. Scenario~B remains compatible
with the likelihood at $10\,\mathrm{Mpc}$ ($p=0.474$), but shows moderate
tension at $30\,\mathrm{Mpc}$ ($p=0.069$). At $100\,\mathrm{Mpc}$, both
templates fit poorly, so their relative ordering is not physically
informative.

The two shower moments provide complementary discrimination between the
injection templates (Figs.~\ref{fig:combined_xmax}
and~\ref{fig:combined_sigma}), with differences appearing in both
$\langle X_{\max}\rangle$ and $\sigma(X_{\max})$ over the fitted
energy range at $d=10$--$30\,\mathrm{Mpc}$. Scenario~B generally produces broader
fluctuations at lower and intermediate energies. Its large injected
hydrogen fraction does not necessarily imply a proton-dominated
highest-energy population, because the acceleration energy scales as
$E_0\simeq\eta Z e\mathcal{V}_{\max}$ and therefore favours higher-$Z$
nuclei.

The fitted acceleration widths also evolve differently with distance.
For Scenario~B, $\sigma_\eta$ decreases from $0.49$ to $0.36$ and $0.27$,
whereas for Scenario~A it remains stable at $d=10$--$30\,\mathrm{Mpc}$
($0.62$ and $0.59$) before increasing to $1.24$ at
$d=100\,\mathrm{Mpc}$. This indicates that the two injection templates
require different potential-sampling distributions as propagation
losses increase, although the trend should not be interpreted as a direct
measurement of the arriving composition.

Overall, the fits favour the proton-poor Scenario~A template at
$d=10$--$30\,\mathrm{Mpc}$. This supports a relatively proton-poor accelerated
population within the two templates tested, but does not provide a
model-independent constraint on the source proton fraction.

\section{Discussion}\label{sec:discussion}

\subsection{Physical implications}
\label{sec:physical_interpretation}

The strong anticorrelation between $\mu_\eta$ and $\ell_V$ follows directly
from Eq.~\eqref{eq:electric_potential_energy}, because the characteristic
energy gain depends on the product
$\eta\mathcal{V}_{\max}$. The shower-maximum moments therefore constrain
$\widetilde{\eta}\mathcal{V}_{\max}$ more directly than either parameter
separately. Consequently, the broad marginal posterior of $\ell_V$ should
not be interpreted as a measurement of the intrinsic potential of an
individual black hole. For the representative $30\,\mathrm{Mpc}$
Scenario~A fit, the posterior medians give
$\widetilde{\eta}\simeq0.07$, but this value cannot be interpreted
independently of the correlated potential.

By contrast, $\sigma_\eta$ is comparatively well constrained. Scenario~A
gives $\sigma_\eta\simeq0.6$ at both $10$ and $30\,\mathrm{Mpc}$,
corresponding to a multiplicative spread of approximately $1.8$ in the
potential-sampling factor. This is consistent with a fragmented
acceleration region in which different particles experience different
effective potential drops, although the adopted log-normal form does not
uniquely specify the underlying microscopic process
\citep{Rieger_2010,Gaspari_2017,Sinha_2018}. The much larger value required
at $100\,\mathrm{Mpc}$ accompanies a poor fit and is more naturally
interpreted as a compensating response to stronger propagation effects.
The inferred $\sigma_\eta$ should therefore be regarded as an effective
width within the present source description; source-to-source variation
in $\mathcal{V}_{\max}$ could contribute to the same observed dispersion.

Importantly, the $10$, $30$, and $100\,\mathrm{Mpc}$ cases are independent
fits: $\mu_\eta$, $\sigma_\eta$, and $\mathcal{V}_{\max}$ are re-inferred
separately at each fixed distance. The similarity of the constraints at
$10$ and $30\,\mathrm{Mpc}$ is therefore an outcome of the fits rather than
an imposed assumption that the sources are identical. The substantially
poorer fits in both scenarios at $100\,\mathrm{Mpc}$ indicate that, within the fixed-distance grid explored here, the observed shower moments are more naturally
reproduced at the shorter propagation distances.

The two prescribed injection scenarios play complementary roles rather than
constituting first-principles predictions of the source composition.
Scenario~A provides an observationally motivated baseline, while
Scenario~B tests an AGN-motivated, hydrogen-richer and metal-enriched
composition using the same acceleration and propagation framework.
Although Scenario~B gives a poorer absolute fit, both templates show the
same qualitative preference for shorter propagation distances over
$100\,\mathrm{Mpc}$. Their different proton fractions also do not map
directly onto the composition at a fixed observed energy, because
$E_0\simeq\eta Z e\mathcal{V}_{\max}$ preferentially selects higher-$Z$
nuclei into the highest-energy bins. Thus, within the two compositions
tested, the data favour a relatively proton-poor accelerated population,
while $\sigma(X_{\max})$ provides a direct constraint on the spread of
charges selected by the stochastic acceleration process rather than a
model-independent measurement of the ambient AGN abundances or of the
source proton fraction.

\subsection{Candidate AGN and source implications}
\label{sec:source_constraints}

The fixed-distance fits provide a basis for comparing the inferred
acceleration conditions with the candidate systems listed in
Table~\ref{tab:source_table}. Both injection scenarios perform
substantially worse at $100\,\mathrm{Mpc}$ than at the shorter distances
within the present fixed-distance propagation framework, indicating that
the observed shower moments are more naturally reproduced at the shorter
distances explored here. The three adopted distances are representative
tests rather than samples from a continuous distance posterior and therefore
cannot identify a unique source or determine the relative contributions of
individual objects. The tabulated potentials should be regarded as
illustrative source capacities, scaling as
$\propto k\,(L/L_{\rm Edd})^{1/2}M_{\rm BH}^{1/2}$. We adopt $k=1$ as a
horizon-scale limiting normalization; more localized acceleration regions
with $k<1$ reduce the listed potentials linearly.

Among the nearby candidates, M87 and NGC~1407 stand out because of their
large black-hole masses. For the illustrative low-accretion case
$L/L_{\rm Edd}=10^{-6}$ and $k=1$, their estimated potentials are
$7.18\times10^{18}\,\mathrm{V}$ and
$5.97\times10^{18}\,\mathrm{V}$, respectively; these scale as
$(L/L_{\rm Edd})^{1/2}$ according to Eq.~\eqref{eq:Vmax} and decrease
linearly for $k<1$. M87 provides the clearest benchmark in the present
sample, combining an exceptionally massive black hole, a prominent
relativistic jet, a nearby distance, and independent evidence for a strongly
magnetised near-horizon environment. NGC~1407 is energetically comparable,
although its viability depends more strongly on whether its central engine
attains the magnetisation, spin, and current-sheet conditions assumed here.
NGC~4261, at $\simeq32\,\mathrm{Mpc}$, has an illustrative potential of
$3.64\times10^{18}\,\mathrm{V}$ for $L/L_{\rm Edd}=10^{-6}$ and $k=1$
and lies close to the 30\,Mpc propagation case.

Because $\mu_\eta$ and $\mathcal{V}_{\max}$ are strongly degenerate,
the candidate-source potentials should not be compared directly with
the marginal posterior median of $\mathcal{V}_{\max}$ alone. The
acceleration model primarily constrains the characteristic sampled
potential $e^{\mu_\eta}\mathcal{V}_{\max}$. For the representative
$30\,\mathrm{Mpc}$ Scenario~A fit, the posterior medians give
$e^{\mu_\eta}\mathcal{V}_{\max}\simeq3.7\times10^{18}\,\mathrm{V}$,
corresponding to characteristic sampling factors of approximately
$0.52$, $0.62$, and $1.02$ for M87, NGC~1407, and NGC~4261,
respectively. Thus these systems can reproduce the same characteristic
acceleration scale without requiring $k>1$.

NGC~1068 provides a complementary nearby case. High-sensitivity
observations imply a compact central mass of
$M_{\rm BH}\simeq1.7\times10^7\,M_\odot$
\citep{GallimoreImpellizzeri2023}. At $D\simeq14.4$\,Mpc, it lies between
the 10 and 30\,Mpc propagation cases considered here. Its smaller
black-hole mass gives a comparatively low potential in the low-accretion
benchmark, whereas the illustrative Eddington-limited case reaches
$\mathcal{V}_{\max}\simeq3.7\times10^{20}$\,V. Its viability therefore
depends more strongly on the accretion state than for the most massive
black holes in Table~\ref{tab:source_table}. At the same time, NGC~1068
is notable for the evidence of high-energy neutrino emission
\citep{IceCube2022NGC1068}, providing independent evidence for hadronic
activity and supporting its candidacy as a nearby UHECR source in the
present context.

The remaining nearby systems -- Centaurus~A, NGC~1399, and Fornax~A --
have smaller illustrative low-Eddington-ratio potentials, but their proximity
favours the survival of accelerated nuclei, and larger intrinsic potentials
would be obtained for stronger near-horizon magnetic fields and/or larger
$L/L_{\rm Edd}$. Centaurus~A lies closer than the smallest fitted distance,
so its compatibility cannot be assessed directly from the
$10\,\mathrm{Mpc}$ calculation. IC~4296 and NGC~1275 lie between the
$30$ and $100\,\mathrm{Mpc}$ cases and would require dedicated fits at their
actual distances, since the propagated composition cannot be reliably inferred
by interpolation between the fixed-distance calculations. None of these
systems should be excluded on the basis of the illustrative low-Eddington-ratio
estimates alone, particularly because $\mathcal{V}_{\max}$ remains strongly
degenerate with the potential-sampling factor in the present likelihood.
Cygnus~A presents the opposite case: its large black-hole mass allows a
substantial intrinsic potential, but its distance of $\simeq232\,\mathrm{Mpc}$
lies well beyond the explored range, where both injection scenarios already
fit substantially worse at $100\,\mathrm{Mpc}$, making a dominant contribution
from such large distances difficult to reconcile with the observed
$X_{\max}$ moments in the present single-distance framework.

Taken together, the comparison points to a plausible source population rather
than a unique object: nearby, massive, strongly magnetised SMBHs capable of
supporting MAD-like conditions. M87 is the best-characterised benchmark in
the present sample, while NGC~1407 and NGC~4261 illustrate other energetically
plausible systems under the adopted assumptions. More generally, the results
are compatible with a local population of several nearby accelerators
contributing to the observed UHECR flux, rather than requiring a single
dominant source. Table~\ref{tab:source_table} should therefore be viewed as an
illustrative map of nearby source parameter space, not as a ranking or
exclusion of individual AGN.

\subsection{Limitations and future directions}
\label{sec:limitations}

The present calculation is a tractable test of composition-dependent
acceleration in AGN electric potentials, and the inferred constraints remain
conditional on the adopted assumptions. In particular, the injection
composition is restricted to two prescribed templates rather than inferred
from the data. The results therefore support a relatively
proton-poor accelerated population and
do not constitute a model-independent measurement of the elemental
fractions at the source. The acceleration step is likewise idealised:
Eq.~\eqref{eq:electric_potential_energy} assigns an energy gain
$\eta Z e\mathcal{V}_{\max}$ without accounting for radiative or
photohadronic losses within the acceleration region. For protons in
magnetospheric current sheets around sub-Eddington SMBHs such as M87*, such
losses can restrict the attainable energy to at most a few EeV
\citep{Stathopoulos2024}. Heavier nuclei benefit from the larger charge
factor but remain subject to photodisintegration, so the present acceleration
prescription should be regarded as an upper envelope on the energy attainable
for a given $(\eta,Z,\mathcal{V}_{\max})$.

The inference uses the first two moments of the $X_{\max}$ distribution, which
provide complementary sensitivity to the mean mass and to the width of the
arriving composition. The UHECR energy spectrum and absolute flux are not
included in the likelihood, and the source luminosity function, source
evolution, magnetic transport, and anisotropy are not modelled.
The omission of the spectrum does not enter the present composition likelihood
directly, because $\langle X_{\max}\rangle$ and $\sigma(X_{\max})$ are
evaluated as per-energy-bin averages over whichever particles fall in each bin
and are therefore independent of the absolute number of particles per bin.
Because the acceleration deposits a charge-dependent energy
$E_0\simeq\eta Z e\mathcal{V}_{\max}$ with a relatively narrow
potential-sampling distribution, a single source at fixed distance and
$\mathcal{V}_{\max}$ tends to produce a more peaked arriving spectrum than the
observed broad UHECR spectrum, so reproducing the latter would likely require
a source population spanning a range of distances and maximum potentials,
together with the associated luminosity function and source evolution.
Such a joint fit is deferred to future work and could additionally help to
reduce the degeneracy between $\mu_\eta$ and $\mathcal{V}_{\max}$, since the
present fits constrain primarily the combination
$e^{\mu_\eta}\mathcal{V}_{\max}$.

The propagation model contains several simplifying approximations. Its
implementation is compared with a matched CMB-only CRPropa~3 calculation
for a representative Scenario~A configuration in
Appendix~\ref{app:validation}. Photodisintegration is calculated on the CMB
using the SS99 cross sections and a reduced PSB chain
\citep{SteckerSalamon1999,AhlersTaylor2010}, with multi-nucleon channels
incorporated through an effective mass-loss rate. Emitted nucleons are not
propagated as separate particles, so a more complete nuclear network could
modify the light-secondary population and the resulting mass spread.
Bethe--Heitler and adiabatic losses are applied as a single continuous step,
while photopion production is treated catastrophically for protons and
neglected for nuclei. The EBL is also omitted; its inclusion is expected to
matter most near the lower end of the fitted range, where infrared and optical
photons enhance the photodisintegration of intermediate-mass nuclei
\citep{Gilmore2012}, and could therefore refine the predicted composition
around $\log(E/\mathrm{eV})\simeq18.5$--$19$.

The conversion from propagated masses to shower observables uses the EPOS-LHC
generalised-Gumbel parametrisation throughout; alternative hadronic-interaction
models would modify the mass--$X_{\max}$ relation and constitute an additional
systematic uncertainty. The likelihood combines statistical and symmetrised
systematic uncertainties in quadrature and neglects correlations between
energy bins and between the two moments. The quadrature treatment broadens the
effective per-point uncertainties, while neglecting correlations simplifies the
likelihood structure; consequently, the reduced chi-square values below unity
for Scenario~A at $10$ and $30\,\mathrm{Mpc}$ should not be interpreted as
evidence of an unusually good fit, and the quoted $\Delta\chi^2$ values indicate
relative preference between the two templates under the adopted likelihood
rather than calibrated significances. The frozen Monte Carlo particle bank
makes the likelihood deterministic, and multi-seed tests at the initial
parameter vector indicate limited sensitivity to its particular realisation,
although complete reruns with independent banks would provide a stronger test
of finite-sample effects.

Future work will extend the framework with improved propagation and additional
observational constraints. Joint fits to the $X_{\max}$ moments and UHECR
spectrum could further constrain the injection composition and acceleration
conditions, while arrival-direction information could test whether a
population of nearby sources such as those discussed above is consistent with
the observed UHECR data.
\section{Summary}
\label{sec:summary}

We construct a direct composition-sensitive link between near-horizon
electric-field acceleration in AGN and the measured shower-maximum moments of
UHECRs. The fragmented current-sheet environment is represented by a
log-normal distribution of potential-sampling factors, so that, together with
the charge dependence of the energy gain, the model links the injected
composition to the nuclear population reaching UHECR energies. Coupling this
prescription to a semi-analytic propagation calculation -- tested against
CRPropa~3 in Appendix~\ref{app:validation} -- allows the acceleration
parameters to be constrained directly from the Auger DNN 2025 measurements.
The injected mass distributions are prescribed rather than predicted from
first principles; we test whether the proposed acceleration mechanism
reproduces the observed shower moments for these compositions and, where
successful, infer the compatible source conditions. The analysis therefore
goes beyond demonstrating that AGN satisfy the energetic requirements for
UHECR acceleration: it tests whether a specific potential-drop prescription
reproduces the observed composition and shower fluctuations after propagation.

Three results follow. First, both prescribed injection scenarios fit
substantially worse at $100\,\mathrm{Mpc}$ than at shorter distances; because
the $10$, $30$, and $100\,\mathrm{Mpc}$ cases were fitted independently, this
trend is an outcome of the propagation--acceleration analysis rather than an
assumption of identical source conditions. 
Second, the fits constrain mainly
the characteristic sampled potential
$e^{\mu_\eta}\mathcal{V}_{\max}$, reflecting the strong degeneracy between
the potential scale and the potential-sampling factor, while $\sigma_\eta$ is
better determined; for Scenario~A at $10$ and $30\,\mathrm{Mpc}$,
$\sigma_\eta\simeq0.6$, corresponding to a multiplicative dispersion of
approximately $1.8$. Third, both compositions remain viable at short
distances, although the more proton-poor template agrees better with the
shower moments. Within the compositions tested, this points toward a
relatively proton-poor accelerated population in nearby AGN
environments, while the viability of the hydrogen-richer template shows that
the inference is not tied to a single prescribed composition.

The inferred acceleration scale is compatible with nearby massive SMBHs
capable of supporting strongly magnetised, MAD-like conditions, with M87,
NGC~1407, and NGC~4261 providing useful benchmarks under the adopted
assumptions. The present fixed-distance analysis does not identify a unique
source; more generally, the results are compatible with contributions from a
local population of several nearby accelerators. Extending the framework to
freely varying injection fractions, improved propagation, the UHECR spectrum,
and arrival-direction information will allow this connection between
near-black-hole electrodynamics and the observed UHECR population to be tested
more comprehensively.

\begin{acknowledgements}

We used the generative AI tools ChatGPT (OpenAI; GPT-5.6 Sol) and Claude
(Anthropic; Claude Opus 4.8) for language editing and coding assistance; we
reviewed all output and take full responsibility for the content.
\end{acknowledgements}

\bibliographystyle{aa} 
\bibliography{biblio} 

@ARTICLE{tchekhovskoy2011,
       author = {{Tchekhovskoy}, Alexander and {Narayan}, Ramesh and {McKinney}, Jonathan C.},
        title = "{Efficient generation of jets from magnetically arrested accretion on a rapidly spinning black hole}",
      journal = {\mnras},
         year = 2011,
        month = nov,
       volume = {418},
       number = {1},
        pages = {L79-L83},
          doi = {10.1111/j.1745-3933.2011.01147.x},
archivePrefix = {arXiv},
       eprint = {1108.0412},
 primaryClass = {astro-ph.HE},
       adsurl = {https://ui.adsabs.harvard.edu/abs/2011MNRAS.418L..79T}
}

@ARTICLE{ripperda2022,
       author = {{Ripperda}, B. and {Liska}, M. and {Chatterjee}, K. and {Musoke}, G. and {Philippov}, A.~A. and {Markoff}, S.~B. and {Tchekhovskoy}, A. and {Younsi}, Z.},
        title = "{Black Hole Flares: Ejection of Accreted Magnetic Flux through 3D Plasmoid-mediated Reconnection}",
      journal = {\apjl},
         year = 2022,
        month = jan,
       volume = {924},
       number = {2},
          eid = {L32},
        pages = {L32},
          doi = {10.3847/2041-8213/ac46a1},
archivePrefix = {arXiv},
       eprint = {2109.15115},
 primaryClass = {astro-ph.HE},
       adsurl = {https://ui.adsabs.harvard.edu/abs/2022ApJ...924L..32R}
}

@ARTICLE{nathanail2025,
       author = {{Nathanail}, Antonios and {Mizuno}, Yosuke and {Contopoulos}, Ioannis and {Fromm}, Christian M. and {Cruz-Osorio}, Alejandro and {Moriyama}, Kotaro and {Rezzolla}, Luciano},
        title = "{The impact of resistivity on the variability of black hole accretion flows}",
      journal = {\aap},
         year = 2025,
        month = jan,
       volume = {693},
          eid = {A56},
        pages = {A56},
          doi = {10.1051/0004-6361/202451836},
archivePrefix = {arXiv},
       eprint = {2411.16684},
 primaryClass = {astro-ph.HE},
       adsurl = {https://ui.adsabs.harvard.edu/abs/2025A&A...693A..56N}
}

@ARTICLE{narayan2003,
       author = {{Narayan}, Ramesh and {Igumenshchev}, Igor V. and {Abramowicz}, Marek A.},
        title = "{Magnetically Arrested Disk: an Energetically Efficient Accretion Flow}",
      journal = {\pasj},
         year = 2003,
        month = dec,
       volume = {55},
        pages = {L69-L72},
          doi = {10.1093/pasj/55.6.L69},
archivePrefix = {arXiv},
       eprint = {astro-ph/0305029},
 primaryClass = {astro-ph},
       adsurl = {https://ui.adsabs.harvard.edu/abs/2003PASJ...55L..69N}
}

@ARTICLE{chatterjee2022,
       author = {{Chatterjee}, K. and {Narayan}, R.},
        title = "{Flux Eruption Events Drive Angular Momentum Transport in Magnetically Arrested Accretion Flows}",
      journal = {\apj},
         year = 2022,
        month = dec,
       volume = {941},
       number = {1},
          eid = {30},
        pages = {30},
          doi = {10.3847/1538-4357/ac9d97},
archivePrefix = {arXiv},
       eprint = {2210.08045},
 primaryClass = {astro-ph.HE},
       adsurl = {https://ui.adsabs.harvard.edu/abs/2022ApJ...941...30C}
}

@ARTICLE{Blandford1977,
       author = {{Blandford}, R.~D. and {Znajek}, R.~L.},
        title = "{Electromagnetic extraction of energy from Kerr black holes.}",
      journal = {\mnras},
         year = 1977,
        month = may,
       volume = {179},
        pages = {433-456},
          doi = {10.1093/mnras/179.3.433},
       adsurl = {https://ui.adsabs.harvard.edu/abs/1977MNRAS.179..433B}
}

@ARTICLE{ehtc2021,
       author = {{Event Horizon Telescope Collaboration} and {Akiyama}, Kazunori and {Algaba}, Juan Carlos and {Alberdi}, Antxon and {Alef}, Walter and {Anantua}, Richard and {Asada}, Keiichi and {Azulay}, Rebecca and {Baczko}, Anne-Kathrin and {Ball}, David and {Balokovi{\'c}}, Mislav and {Barrett}, John and {Benson}, Bradford A. and {Bintley}, Dan and {Blackburn}, Lindy and {Blundell}, Raymond and {Boland}, Wilfred and {Bouman}, Katherine L. and {Bower}, Geoffrey C. and {Boyce}, Hope and {Bremer}, Michael and {Brinkerink}, Christiaan D. and {Brissenden}, Roger and {Britzen}, Silke and {Broderick}, Avery E. and {Broguiere}, Dominique and {Bronzwaer}, Thomas and {Byun}, Do-Young and {Carlstrom}, John E. and {Chael}, Andrew and {Chan}, Chi-kwan and {Chatterjee}, Shami and {Chatterjee}, Koushik and {Chen}, Ming-Tang and {Chen}, Yongjun and {Chesler}, Paul M. and {Cho}, Ilje and {Christian}, Pierre and {Conway}, John E. and {Cordes}, James M. and {Crawford}, Thomas M. and {Crew}, Geoffrey B. and {Cruz-Osorio}, Alejandro and {Cui}, Yuzhu and {Davelaar}, Jordy and {De Laurentis}, Mariafelicia and {Deane}, Roger and {Dempsey}, Jessica and {Desvignes}, Gregory and {Dexter}, Jason and {Doeleman}, Sheperd S. and {Eatough}, Ralph P. and {Falcke}, Heino and {Farah}, Joseph and {Fish}, Vincent L. and {Fomalont}, Ed and {Ford}, H. Alyson and {Fraga-Encinas}, Raquel and {Friberg}, Per and {Fromm}, Christian M. and {Fuentes}, Antonio and {Galison}, Peter and {Gammie}, Charles F. and {Garc{\'\i}a}, Roberto and {Gelles}, Zachary and {Gentaz}, Olivier and {Georgiev}, Boris and {Goddi}, Ciriaco and {Gold}, Roman and {G{\'o}mez}, Jos{\'e} L. and {G{\'o}mez-Ruiz}, Arturo I. and {Gu}, Minfeng and {Gurwell}, Mark and {Hada}, Kazuhiro and {Haggard}, Daryl and {Hecht}, Michael H. and {Hesper}, Ronald and {Himwich}, Elizabeth and {Ho}, Luis C. and {Ho}, Paul and {Honma}, Mareki and {Huang}, Chih-Wei L. and {Huang}, Lei and {Hughes}, David H. and {Ikeda}, Shiro and {Inoue}, Makoto and {Issaoun}, Sara and {James}, David J. and {Jannuzi}, Buell T. and {Janssen}, Michael and {Jeter}, Britton and {Jiang}, Wu and {Jimenez-Rosales}, Alejandra and {Johnson}, Michael D. and {Jorstad}, Svetlana and {Jung}, Taehyun and {Karami}, Mansour and {Karuppusamy}, Ramesh and {Kawashima}, Tomohisa and {Keating}, Garrett K. and {Kettenis}, Mark and {Kim}, Dong-Jin and {Kim}, Jae-Young and {Kim}, Jongsoo and {Kim}, Junhan and {Kino}, Motoki and {Koay}, Jun Yi and {Kofuji}, Yutaro and {Koch}, Patrick M. and {Koyama}, Shoko and {Kramer}, Michael and {Kramer}, Carsten and {Krichbaum}, Thomas P. and {Kuo}, Cheng-Yu and {Lauer}, Tod R. and {Lee}, Sang-Sung and {Levis}, Aviad and {Li}, Yan-Rong and {Li}, Zhiyuan and {Lindqvist}, Michael and {Lico}, Rocco and {Lindahl}, Greg and {Liu}, Jun and {Liu}, Kuo and {Liuzzo}, Elisabetta and {Lo}, Wen-Ping and {Lobanov}, Andrei P. and {Loinard}, Laurent and {Lonsdale}, Colin and {Lu}, Ru-Sen and {MacDonald}, Nicholas R. and {Mao}, Jirong and {Marchili}, Nicola and {Markoff}, Sera and {Marrone}, Daniel P. and {Marscher}, Alan P. and {Mart{\'\i}-Vidal}, Iv{\'a}n and {Matsushita}, Satoki and {Matthews}, Lynn D. and {Medeiros}, Lia and {Menten}, Karl M. and {Mizuno}, Izumi and {Mizuno}, Yosuke and {Moran}, James M. and {Moriyama}, Kotaro and {Moscibrodzka}, Monika and {M{\"u}ller}, Cornelia and {Musoke}, Gibwa and {Mus Mej{\'\i}as}, Alejandro and {Michalik}, Daniel and {Nadolski}, Andrew and {Nagai}, Hiroshi and {Nagar}, Neil M. and {Nakamura}, Masanori and {Narayan}, Ramesh and {Narayanan}, Gopal and {Natarajan}, Iniyan and {Nathanail}, Antonios and {Neilsen}, Joey and {Neri}, Roberto and {Ni}, Chunchong and {Noutsos}, Aristeidis and {Nowak}, Michael A. and {Okino}, Hiroki and {Olivares}, H{\'e}ctor and {Ortiz-Le{\'o}n}, Gisela N. and {Oyama}, Tomoaki and {{\"O}zel}, Feryal and {Palumbo}, Daniel C.~M. and {Park}, Jongho and {Patel}, Nimesh and {Pen}, Ue-Li and {Pesce}, Dominic W. and {Pi{\'e}tu}, Vincent and {Plambeck}, Richard and {PopStefanija}, Aleksandar and {Porth}, Oliver and {P{\"o}tzl}, Felix M. and {Prather}, Ben and {Preciado-L{\'o}pez}, Jorge A. and {Psaltis}, Dimitrios and {Pu}, Hung-Yi and {Ramakrishnan}, Venkatessh and {Rao}, Ramprasad and {Rawlings}, Mark G. and {Raymond}, Alexander W. and {Rezzolla}, Luciano and {Ricarte}, Angelo and {Ripperda}, Bart and {Roelofs}, Freek and {Rogers}, Alan and {Ros}, Eduardo and {Rose}, Mel and {Roshanineshat}, Arash and {Rottmann}, Helge and {Roy}, Alan L. and {Ruszczyk}, Chet and {Rygl}, Kazi L.~J. and {S{\'a}nchez}, Salvador and {S{\'a}nchez-Arguelles}, David and {Sasada}, Mahito and {Savolainen}, Tuomas and {Schloerb}, F. Peter and {Schuster}, Karl-Friedrich and {Shao}, Lijing and {Shen}, Zhiqiang and {Small}, Des and {Sohn}, Bong Won and {SooHoo}, Jason and {Sun}, He and {Tazaki}, Fumie and {Tetarenko}, Alexandra J. and {Tiede}, Paul and {Tilanus}, Remo P.~J. and {Titus}, Michael and {Toma}, Kenji and {Torne}, Pablo and {Trent}, Tyler and {Traianou}, Efthalia and {Trippe}, Sascha and {van Bemmel}, Ilse and {van Langevelde}, Huib Jan and {van Rossum}, Daniel R. and {Wagner}, Jan and {Ward-Thompson}, Derek and {Wardle}, John and {Weintroub}, Jonathan and {Wex}, Norbert and {Wharton}, Robert and {Wielgus}, Maciek and {Wong}, George N. and {Wu}, Qingwen and {Yoon}, Doosoo and {Young}, Andr{\'e} and {Young}, Ken and {Younsi}, Ziri and {Yuan}, Feng and {Yuan}, Ye-Fei and {Zensus}, J. Anton and {Zhao}, Guang-Yao and {Zhao}, Shan-Shan},
        title = "{First M87 Event Horizon Telescope Results. VIII. Magnetic Field Structure near The Event Horizon}",
      journal = {\apjl},
         year = 2021,
        month = mar,
       volume = {910},
       number = {1},
          eid = {L13},
        pages = {L13},
          doi = {10.3847/2041-8213/abe4de},
archivePrefix = {arXiv},
       eprint = {2105.01173},
 primaryClass = {astro-ph.HE},
       adsurl = {https://ui.adsabs.harvard.edu/abs/2021ApJ...910L..13E}
}

@ARTICLE{ehtc2024,
       author = {{Event Horizon Telescope Collaboration} and {Akiyama}, Kazunori and {Alberdi}, Antxon and {Alef}, Walter and {Algaba}, Juan Carlos and {Anantua}, Richard and {Asada}, Keiichi and {Azulay}, Rebecca and {Bach}, Uwe and {Baczko}, Anne-Kathrin and {Ball}, David and {Balokovi{\'c}}, Mislav and {Bandyopadhyay}, Bidisha and {Barrett}, John and {Baub{\"o}ck}, Michi and {Benson}, Bradford A. and {Bintley}, Dan and {Blackburn}, Lindy and {Blundell}, Raymond and {Bouman}, Katherine L. and {Bower}, Geoffrey C. and {Boyce}, Hope and {Bremer}, Michael and {Brinkerink}, Christiaan D. and {Brissenden}, Roger and {Britzen}, Silke and {Broderick}, Avery E. and {Broguiere}, Dominique and {Bronzwaer}, Thomas and {Bustamante}, Sandra and {Byun}, Do-Young and {Carlstrom}, John E. and {Ceccobello}, Chiara and {Chael}, Andrew and {Chan}, Chi-kwan and {Chang}, Dominic O. and {Chatterjee}, Koushik and {Chatterjee}, Shami and {Chen}, Ming-Tang and {Chen}, Yongjun and {Cheng}, Xiaopeng and {Cho}, Ilje and {Christian}, Pierre and {Conroy}, Nicholas S. and {Conway}, John E. and {Cordes}, James M. and {Crawford}, Thomas M. and {Crew}, Geoffrey B. and {Cruz-Osorio}, Alejandro and {Cui}, Yuzhu and {Dahale}, Rohan and {Davelaar}, Jordy and {De Laurentis}, Mariafelicia and {Deane}, Roger and {Dempsey}, Jessica and {Desvignes}, Gregory and {Dexter}, Jason and {Dhruv}, Vedant and {Dihingia}, Indu K. and {Doeleman}, Sheperd S. and {Dougall}, Sean and {Dzib}, Sergio A. and {Eatough}, Ralph P. and {Emami}, Razieh and {Falcke}, Heino and {Farah}, Joseph and {Fish}, Vincent L. and {Fomalont}, Edward and {Ford}, H. Alyson and {Foschi}, Marianna and {Fraga-Encinas}, Raquel and {Freeman}, William T. and {Friberg}, Per and {Fromm}, Christian M. and {Fuentes}, Antonio and {Galison}, Peter and {Gammie}, Charles F. and {Garc{\'\i}a}, Roberto and {Gentaz}, Olivier and {Georgiev}, Boris and {Goddi}, Ciriaco and {Gold}, Roman and {G{\'o}mez-Ruiz}, Arturo I. and {G{\'o}mez}, Jos{\'e} L. and {Gu}, Minfeng and {Gurwell}, Mark and {Hada}, Kazuhiro and {Haggard}, Daryl and {Haworth}, Kari and {Hecht}, Michael H. and {Hesper}, Ronald and {Heumann}, Dirk and {Ho}, Luis C. and {Ho}, Paul and {Honma}, Mareki and {Huang}, Chih-Wei L. and {Huang}, Lei and {Hughes}, David H. and {Ikeda}, Shiro and {Impellizzeri}, C.~M. Violette and {Inoue}, Makoto and {Issaoun}, Sara and {James}, David J. and {Jannuzi}, Buell T. and {Janssen}, Michael and {Jeter}, Britton and {Jiang}, Wu and {Jim{\'e}nez-Rosales}, Alejandra and {Johnson}, Michael D. and {Jorstad}, Svetlana and {Joshi}, Abhishek V. and {Jung}, Taehyun and {Karami}, Mansour and {Karuppusamy}, Ramesh and {Kawashima}, Tomohisa and {Keating}, Garrett K. and {Kettenis}, Mark and {Kim}, Dong-Jin and {Kim}, Jae-Young and {Kim}, Jongsoo and {Kim}, Junhan and {Kino}, Motoki and {Koay}, Jun Yi and {Kocherlakota}, Prashant and {Kofuji}, Yutaro and {Koch}, Patrick M. and {Koyama}, Shoko and {Kramer}, Carsten and {Kramer}, Joana A. and {Kramer}, Michael and {Krichbaum}, Thomas P. and {Kuo}, Cheng-Yu and {La Bella}, Noemi and {Lauer}, Tod R. and {Lee}, Daeyoung and {Lee}, Sang-Sung and {Leung}, Po Kin and {Levis}, Aviad and {Li}, Zhiyuan and {Lico}, Rocco and {Lindahl}, Greg and {Lindqvist}, Michael and {Lisakov}, Mikhail and {Liu}, Jun and {Liu}, Kuo and {Liuzzo}, Elisabetta and {Lo}, Wen-Ping and {Lobanov}, Andrei P. and {Loinard}, Laurent and {Lonsdale}, Colin J. and {Lowitz}, Amy E. and {Lu}, Ru-Sen and {MacDonald}, Nicholas R. and {Mao}, Jirong and {Marchili}, Nicola and {Markoff}, Sera and {Marrone}, Daniel P. and {Marscher}, Alan P. and {Mart{\'\i}-Vidal}, Iv{\'a}n and {Matsushita}, Satoki and {Matthews}, Lynn D. and {Medeiros}, Lia and {Menten}, Karl M. and {Michalik}, Daniel and {Mizuno}, Izumi and {Mizuno}, Yosuke and {Moran}, James M. and {Moriyama}, Kotaro and {Moscibrodzka}, Monika and {Mulaudzi}, Wanga and {M{\"u}ller}, Cornelia and {M{\"u}ller}, Hendrik and {Mus}, Alejandro and {Musoke}, Gibwa and {Myserlis}, Ioannis and {Nadolski}, Andrew and {Nagai}, Hiroshi and {Nagar}, Neil M. and {Nakamura}, Masanori and {Narayanan}, Gopal and {Natarajan}, Iniyan and {Nathanail}, Antonios and {Fuentes}, Santiago Navarro and {Neilsen}, Joey and {Neri}, Roberto and {Ni}, Chunchong and {Noutsos}, Aristeidis and {Nowak}, Michael A. and {Oh}, Junghwan and {Okino}, Hiroki and {Olivares}, H{\'e}ctor and {Ortiz-Le{\'o}n}, Gisela N. and {Oyama}, Tomoaki and {{\"O}zel}, Feryal and {Palumbo}, Daniel C.~M. and {Paraschos}, Georgios Filippos and {Park}, Jongho and {Parsons}, Harriet and {Patel}, Nimesh and {Pen}, Ue-Li and {Pesce}, Dominic W. and {Pi{\'e}tu}, Vincent and {Plambeck}, Richard and {PopStefanija}, Aleksandar and {Porth}, Oliver and {P{\"o}tzl}, Felix M. and {Prather}, Ben and {Preciado-L{\'o}pez}, Jorge A. and {Psaltis}, Dimitrios and {Pu}, Hung-Yi and {Ramakrishnan}, Venkatessh and {Rao}, Ramprasad and {Rawlings}, Mark G. and {Raymond}, Alexander W. and {Rezzolla}, Luciano and {Ricarte}, Angelo and {Ripperda}, Bart and {Roelofs}, Freek and {Rogers}, Alan and {Romero-Ca{\~n}izales}, Cristina and {Ros}, Eduardo and {Roshanineshat}, Arash and {Rottmann}, Helge and {Roy}, Alan L. and {Ruiz}, Ignacio and {Ruszczyk}, Chet and {Rygl}, Kazi L.~J. and {S{\'a}nchez}, Salvador and {S{\'a}nchez-Arg{\"u}elles}, David and {S{\'a}nchez-Portal}, Miguel and {Sasada}, Mahito and {Satapathy}, Kaushik and {Savolainen}, Tuomas and {Schloerb}, F. Peter and {Schonfeld}, Jonathan and {Schuster}, Karl-Friedrich and {Shao}, Lijing and {Shen}, Zhiqiang and {Small}, Des and {Sohn}, Bong Won and {SooHoo}, Jason and {Sosapanta Salas}, Le{\'o}n David and {Souccar}, Kamal and {Stanway}, Joshua S. and {Sun}, He and {Tazaki}, Fumie and {Tetarenko}, Alexandra J. and {Tiede}, Paul and {Tilanus}, Remo P.~J. and {Titus}, Michael and {Torne}, Pablo and {Toscano}, Teresa and {Traianou}, Efthalia and {Trent}, Tyler and {Trippe}, Sascha and {Turk}, Matthew and {van Bemmel}, Ilse and {van Langevelde}, Huib Jan and {van Rossum}, Daniel R. and {Vos}, Jesse and {Wagner}, Jan and {Ward-Thompson}, Derek and {Wardle}, John and {Washington}, Jasmin E. and {Weintroub}, Jonathan and {Wharton}, Robert and {Wielgus}, Maciek and {Wiik}, Kaj and {Witzel}, Gunther and {Wondrak}, Michael F. and {Wong}, George N. and {Wu}, Qingwen and {Yadlapalli}, Nitika and {Yamaguchi}, Paul and {Yfantis}, Aristomenis and {Yoon}, Doosoo and {Young}, Andr{\'e} and {Young}, Ken and {Younsi}, Ziri and {Yu}, Wei and {Yuan}, Feng and {Yuan}, Ye-Fei and {Zensus}, J. Anton and {Zhang}, Shuo and {Zhao}, Guang-Yao and {Zhao}, Shan-Shan and {Najafi-Ziyazi}, Mahdi},
        title = "{First Sagittarius A* Event Horizon Telescope Results. VIII. Physical Interpretation of the Polarized Ring}",
      journal = {\apjl},
         year = 2024,
        month = apr,
       volume = {964},
       number = {2},
          eid = {L26},
        pages = {L26},
          doi = {10.3847/2041-8213/ad2df1},
       adsurl = {https://ui.adsabs.harvard.edu/abs/2024ApJ...964L..26E}
}

@ARTICLE{antonopoulou2025,
       author = {{Antonopoulou}, Eleni and {Loules}, Argyrios and {Nathanail}, Antonios},
        title = "{Magnetically arrested disk flux eruption events to describe SgrA* flares}",
      journal = {\aap},
         year = 2025,
        month = apr,
       volume = {696},
          eid = {A10},
        pages = {A10},
          doi = {10.1051/0004-6361/202453456},
archivePrefix = {arXiv},
       eprint = {2501.07521},
 primaryClass = {astro-ph.HE},
       adsurl = {https://ui.adsabs.harvard.edu/abs/2025A&A...696A..10A}
}

@ARTICLE{dexter2020,
       author = {{Dexter}, J. and {Tchekhovskoy}, A. and {Jim{\'e}nez-Rosales}, A. and {Ressler}, S.~M. and {Baub{\"o}ck}, M. and {Dallilar}, Y. and {de Zeeuw}, P.~T. and {Eisenhauer}, F. and {von Fellenberg}, S. and {Gao}, F. and {Genzel}, R. and {Gillessen}, S. and {Habibi}, M. and {Ott}, T. and {Stadler}, J. and {Straub}, O. and {Widmann}, F.},
        title = "{Sgr A* near-infrared flares from reconnection events in a magnetically arrested disc}",
      journal = {\mnras},
         year = 2020,
        month = oct,
       volume = {497},
       number = {4},
        pages = {4999-5007},
          doi = {10.1093/mnras/staa2288},
archivePrefix = {arXiv},
       eprint = {2006.03657},
 primaryClass = {astro-ph.HE},
       adsurl = {https://ui.adsabs.harvard.edu/abs/2020MNRAS.497.4999D}
}

@ARTICLE{igumenshchev2008,
       author = {{Igumenshchev}, Igor V.},
        title = "{Magnetically Arrested Disks and the Origin of Poynting Jets: A Numerical Study}",
      journal = {\apj},
         year = 2008,
        month = apr,
       volume = {677},
       number = {1},
        pages = {317-326},
          doi = {10.1086/529025},
archivePrefix = {arXiv},
       eprint = {0711.4391},
 primaryClass = {astro-ph},
       adsurl = {https://ui.adsabs.harvard.edu/abs/2008ApJ...677..317I}
}

@ARTICLE{vos2024,
       author = {{Vos}, Jesse and {Cerutti}, Benoit and {Moscibrodzka}, Monika and {Parfrey}, Kyle},
        title = "{Particle Acceleration in Collisionless Magnetically Arrested Disks}",
      journal = {arXiv e-prints},
         year = 2024,
        month = oct,
          eid = {arXiv:2410.19061},
        pages = {arXiv:2410.19061},
          doi = {10.48550/arXiv.2410.19061},
archivePrefix = {arXiv},
       eprint = {2410.19061},
 primaryClass = {astro-ph.HE},
       adsurl = {https://ui.adsabs.harvard.edu/abs/2024arXiv241019061V}
}

@ARTICLE{nathanail2022,
       author = {{Nathanail}, Antonios and {Mpisketzis}, Vasilis and {Porth}, Oliver and {Fromm}, Christian M. and {Rezzolla}, Luciano},
        title = "{Magnetic reconnection and plasmoid formation in three-dimensional accretion flows around black holes}",
      journal = {\mnras},
         year = 2022,
        month = jul,
       volume = {513},
       number = {3},
        pages = {4267-4277},
          doi = {10.1093/mnras/stac1118},
archivePrefix = {arXiv},
       eprint = {2111.03689},
 primaryClass = {astro-ph.HE},
       adsurl = {https://ui.adsabs.harvard.edu/abs/2022MNRAS.513.4267N}
}

@ARTICLE{ContopoulosKazanas1998,
       author = {{Contopoulos}, Ioannis and {Kazanas}, Demosthenes},
        title = "{A Cosmic Battery}",
      journal = {\apj},
         year = 1998,
        month = dec,
       volume = {508},
       number = {2},
        pages = {859-863},
          doi = {10.1086/306426},
archivePrefix = {arXiv},
       eprint = {astro-ph/9808223},
 primaryClass = {astro-ph},
       adsurl = {https://ui.adsabs.harvard.edu/abs/1998ApJ...508..859C}
}

@ARTICLE{Contopoulos2005,
       author = {{Contopoulos}, I.},
        title = "{The coughing pulsar magnetosphere}",
      journal = {\aap},
         year = 2005,
        month = nov,
       volume = {442},
       number = {2},
        pages = {579-586},
          doi = {10.1051/0004-6361:20053143},
archivePrefix = {arXiv},
       eprint = {astro-ph/0507201},
 primaryClass = {astro-ph},
       adsurl = {https://ui.adsabs.harvard.edu/abs/2005A&A...442..579C}
}

@article{Kampert2012,
         author = {{Kampert}, Karl-Heinz and {Unger}, Michael},
        title = "{Measurements of the cosmic ray composition with air shower experiments}",
      journal = {Astroparticle Physics},
         year = 2012,
        month = may,
       volume = {35},
       number = {10},
        pages = {660-678},
          doi = {10.1016/j.astropartphys.2012.02.004},
archivePrefix = {arXiv},
       eprint = {1201.0018},
 primaryClass = {astro-ph.HE},
       adsurl = {https://ui.adsabs.harvard.edu/abs/2012APh....35..660K}
}

@ARTICLE{Supanitsky2022,
       author = {{Supanitsky}, Alberto Daniel},
        title = "{Determination of the Cosmic-Ray Chemical Composition: Open Issues and Prospects}",
      journal = {Galaxies},
         year = 2022,
        month = jun,
       volume = {10},
       number = {3},
          eid = {75},
        pages = {75},
          doi = {10.3390/galaxies10030075},
archivePrefix = {arXiv},
       eprint = {2212.11695},
 primaryClass = {astro-ph.HE},
       adsurl = {https://ui.adsabs.harvard.edu/abs/2022Galax..10...75S}
}

@ARTICLE{Anchordoqui2019,
       author = {{Anchordoqui}, Luis A.},
        title = "{Ultra-high-energy cosmic rays}",
      journal = {\physrep},
         year = 2019,
        month = apr,
       volume = {801},
        pages = {1-93},
          doi = {10.1016/j.physrep.2019.01.002},
archivePrefix = {arXiv},
       eprint = {1807.09645},
 primaryClass = {astro-ph.HE},
       adsurl = {https://ui.adsabs.harvard.edu/abs/2019PhR...801....1A}
}

@ARTICLE{Ehlert2025,
       author = {{Ehlert}, Domenik and {Oikonomou}, Foteini and {Peretti}, Enrico},
        title = "{Ultra-high-energy cosmic rays from ultra-fast outflows of active galactic nuclei}",
      journal = {\mnras},
         year = 2025,
        month = may,
       volume = {539},
       number = {3},
        pages = {2435-2462},
          doi = {10.1093/mnras/staf457},
archivePrefix = {arXiv},
       eprint = {2411.05667},
 primaryClass = {astro-ph.HE},
       adsurl = {https://ui.adsabs.harvard.edu/abs/2025MNRAS.539.2435E}
}

@article{Bell2013,
    author = {Bell, A. R. and Schure, K. M. and Reville, B. and Giacinti, G.},
    title = {Cosmic-ray acceleration and escape from supernova remnants},
    journal = {Monthly Notices of the Royal Astronomical Society},
    volume = {431},
    number = {1},
    pages = {415-429},
    year = {2013},
    month = {02},
    issn = {0035-8711},
    doi = {10.1093/mnras/stt179},
    url = {https://doi.org/10.1093/mnras/stt179},
    eprint = {https://academic.oup.com/mnras/article-pdf/431/1/415/18243010/stt179.pdf},
}

@ARTICLE{Hillas1984,
       author = {{Hillas}, A.~M.},
        title = "{The Origin of Ultra-High-Energy Cosmic Rays}",
      journal = {\araa},
         year = 1984,
        month = jan,
       volume = {22},
        pages = {425-444},
          doi = {10.1146/annurev.aa.22.090184.002233},
       adsurl = {https://ui.adsabs.harvard.edu/abs/1984ARA&A..22..425H}
}

@article{Taylor2016,
  author  = {Taylor, Andrew M.},
  title   = {Cosmic rays beyond the knees},
  journal = {Nature},
  volume  = {531},
  number  = {7592},
  pages   = {43--44},
  year    = {2016},
  doi     = {10.1038/531043a}
}

@ARTICLE{Jermyn2022,
       author = {{Jermyn}, Adam S. and {Dittmann}, Alexander J. and {McKernan}, B. and {Ford}, K.~E.~S. and {Cantiello}, Matteo},
        title = "{Effects of an Immortal Stellar Population in AGN Disks}",
      journal = {\apj},
         year = 2022,
        month = apr,
       volume = {929},
       number = {2},
          eid = {133},
        pages = {133},
          doi = {10.3847/1538-4357/ac5d40},
archivePrefix = {arXiv},
       eprint = {2203.06187},
 primaryClass = {astro-ph.GA},
       adsurl = {https://ui.adsabs.harvard.edu/abs/2022ApJ...929..133J}
}

@ARTICLE{AliDib2023,
       author = {{Ali-Dib}, Mohamad and {Lin}, Douglas N.~C.},
        title = "{The impermanent fate of massive stars in AGN discs}",
      journal = {\mnras},
         year = 2023,
        month = dec,
       volume = {526},
       number = {4},
        pages = {5824-5838},
          doi = {10.1093/mnras/stad2774},
archivePrefix = {arXiv},
       eprint = {2309.04392},
 primaryClass = {astro-ph.GA},
       adsurl = {https://ui.adsabs.harvard.edu/abs/2023MNRAS.526.5824A}
}

@ARTICLE{Xu2026,
       author = {{Xu}, Zheng-Hao and {Chen}, Yi-Xian and {Lin}, Douglas N.~C.},
        title = "{Stellar Evolution with Radiative Feedback in AGN Disks}",
      journal = {\apj},
         year = 2026,
        month = feb,
       volume = {997},
       number = {2},
          eid = {206},
        pages = {206},
          doi = {10.3847/1538-4357/ae2271},
archivePrefix = {arXiv},
       eprint = {2511.03904},
 primaryClass = {astro-ph.GA},
       adsurl = {https://ui.adsabs.harvard.edu/abs/2026ApJ...997..206X}
}

@ARTICLE{Huang2023,
       author = {{Huang}, Jiamu and {Lin}, Douglas N.~C. and {Shields}, Gregory},
        title = "{Metal enrichment due to embedded stars in AGN discs}",
      journal = {\mnras},
         year = 2023,
        month = nov,
       volume = {525},
       number = {4},
        pages = {5702-5718},
          doi = {10.1093/mnras/stad2642},
archivePrefix = {arXiv},
       eprint = {2308.15761},
 primaryClass = {astro-ph.GA},
       adsurl = {https://ui.adsabs.harvard.edu/abs/2023MNRAS.525.5702H}
}

@article{Buitink2016,
  author    = {S. Buitink and A. Corstanje and H. Falcke and J. R. H\"orandel and T. Huege and A. Nelles and J. P. Rachen and L. Rossetto and P. Schellart and O. Scholten and S. ter Veen and S. Thoudam and T. N. G. Trinh and J. Anderson and A. Asgekar and I. M. Avruch and M. E. Bell and M. J. Bentum and G. Bernardi and P. Best and A. Bonafede and F. Breitling and J. W. Broderick and W. N. Brouw and M. Br\"uggen and H. R. Butcher and D. Carbone and B. Ciardi and J. E. Conway and F. de Gasperin and E. de Geus and A. Deller and R.-J. Dettmar and G. van Diepen and S. Duscha and J. Eisl\"offel and D. Engels and J. E. Enriquez and R. A. Fallows and R. Fender and C. Ferrari and W. Frieswijk and M. A. Garrett and J. M. Griessmeier and A. W. Gunst and M. P. van Haarlem and T. E. Hassall and G. Heald and J. W. T. Hessels and M. Hoeft and A. Horneffer and M. Iacobelli and H. Intema and E. Juette and A. Karastergiou and V. I. Kondratiev and M. Kramer and M. Kuniyoshi and G. Kuper and J. van Leeuwen and G. M. Loose and P. Maat and G. Mann and S. Markoff and R. McFadden and D. McKay-Bukowski and J. P. McKean and M. Mevius and D. D. Mulcahy and H. Munk and M. J. Norden and E. Orru and H. Paas and M. Pandey-Pommier and V. N. Pandey and M. Pietka and R. Pizzo and A. G. Polatidis and W. Reich and H. J. A. R\"ottgering and A. M. M. Scaife and D. J. Schwarz and M. Serylak and J. Sluman and O. Smirnov and B. W. Stappers and M. Steinmetz and A. Stewart and J. Swinbank and M. Tagger and Y. Tang and C. Tasse and M. C. Toribio and R. Vermeulen and C. Vocks and C. Vogt and R. J. van Weeren and R. A. M. J. Wijers and S. J. Wijnholds and M. W. Wise and O. Wucknitz and S. Yatawatta and P. Zarka and J. A. Zensus},
  title     = {A large light-mass component of cosmic rays at 1017–1017.5 electronvolts from radio observations},
  journal   = {Nature},
  volume    = {531},
  number    = {7592},
  pages     = {70--73},
  year      = {2016},
  doi       = {10.1038/nature16976},
  url       = {https://doi.org/10.1038/nature16976},
  issn      = {1476-4687}
}

@article{Kazanas1983,
  author    = {D. Kazanas and R. J. Protheroe},
  title     = {Active galaxies and the diffuse γ-ray background},
  journal   = {Nature},
  volume    = {302},
  number    = {5905},
  pages     = {228--230},
  year      = {1983},
  doi       = {10.1038/302228a0},
  url       = {https://doi.org/10.1038/302228a0},
  issn      = {1476-4687}
}

@ARTICLE{Kazanas1986,
       author = {{Kazanas}, D. and {Ellison}, D.~C.},
        title = "{The Central Engine of Quasars and Active Galactic Nuclei: Hadronic Interactions of Shock-accelerated Relativistic Protons}",
      journal = {\apj},
         year = 1986,
        month = may,
       volume = {304},
        pages = {178},
          doi = {10.1086/164152},
       adsurl = {https://ui.adsabs.harvard.edu/abs/1986ApJ...304..178K}
}

@ARTICLE{Aab2020b,
       author = {{Aab}, A. and {Abreu}, P. and {Aglietta}, M. and {Albury}, J.~M. and {Allekotte}, I. and {Almela}, A. and {Alvarez Castillo}, J. and {Alvarez-Mu{\~n}iz}, J. and {Alves Batista}, R. and {Anastasi}, G.~A. and {Anchordoqui}, L. and {Andrada}, B. and {Andringa}, S. and {Aramo}, C. and {Ara{\'u}jo Ferreira}, P.~R. and {Asorey}, H. and {Assis}, P. and {Avila}, G. and {Badescu}, A.~M. and {Bakalova}, A. and {Balaceanu}, A. and {Barbato}, F. and {Barreira Luz}, R.~J. and {Becker}, K.~H. and {Bellido}, J.~A. and {Berat}, C. and {Bertaina}, M.~E. and {Bertou}, X. and {Biermann}, P.~L. and {Bister}, T. and {Biteau}, J. and {Blanco}, A. and {Blazek}, J. and {Bleve}, C. and {Boh{\'a}{\v{c}}ov{\'a}}, M. and {Boncioli}, D. and {Bonifazi}, C. and {Bonneau Arbeletche}, L. and {Borodai}, N. and {Botti}, A.~M. and {Brack}, J. and {Bretz}, T. and {Briechle}, F.~L. and {Buchholz}, P. and {Bueno}, A. and {Buitink}, S. and {Buscemi}, M. and {Caballero-Mora}, K.~S. and {Caccianiga}, L. and {Calcagni}, L. and {Cancio}, A. and {Canfora}, F. and {Caracas}, I. and {Carceller}, J.~M. and {Caruso}, R. and {Castellina}, A. and {Catalani}, F. and {Cataldi}, G. and {Cazon}, L. and {Cerda}, M. and {Chinellato}, J.~A. and {Choi}, K. and {Chudoba}, J. and {Chytka}, L. and {Clay}, R.~W. and {Cobos Cerutti}, A.~C. and {Colalillo}, R. and {Coleman}, A. and {Coluccia}, M.~R. and {Concei{\c{c}}{\~a}o}, R. and {Condorelli}, A. and {Consolati}, G. and {Contreras}, F. and {Convenga}, F. and {Covault}, C.~E. and {Dasso}, S. and {Daumiller}, K. and {Dawson}, B.~R. and {Day}, J.~A. and {de Almeida}, R.~M. and {de Jes{\'u}s}, J. and {de Jong}, S.~J. and {De Mauro}, G. and {de Mello Neto}, J.~R.~T. and {De Mitri}, I. and {de Oliveira}, J. and {de Oliveira Franco}, D. and {de Souza}, V. and {De Vito}, E. and {Debatin}, J. and {del R{\'\i}o}, M. and {Deligny}, O. and {Dembinski}, H. and {Dhital}, N. and {Di Giulio}, C. and {Di Matteo}, A. and {D{\'\i}az Castro}, M.~L. and {Dobrigkeit}, C. and {D'Olivo}, J.~C. and {Dorosti}, Q. and {dos Anjos}, R.~C. and {Dova}, M.~T. and {Ebr}, J. and {Engel}, R. and {Epicoco}, I. and {Erdmann}, M. and {Escobar}, C.~O. and {Etchegoyen}, A. and {Falcke}, H. and {Farmer}, J. and {Farrar}, G. and {Fauth}, A.~C. and {Fazzini}, N. and {Feldbusch}, F. and {Fenu}, F. and {Fick}, B. and {Figueira}, J.~M. and {Filip{\v{c}}i{\v{c}}}, A. and {Fodran}, T. and {Freire}, M.~M. and {Fujii}, T. and {Fuster}, A. and {Galea}, C. and {Galelli}, C. and {Garc{\'\i}a}, B. and {Garcia Vegas}, A.~L. and {Gemmeke}, H. and {Gesualdi}, F. and {Gherghel-Lascu}, A. and {Ghia}, P.~L. and {Giaccari}, U. and {Giammarchi}, M. and {Giller}, M. and {Glombitza}, J. and {Gobbi}, F. and {Gollan}, F. and {Golup}, G. and {G{\'o}mez Berisso}, M. and {G{\'o}mez Vitale}, P.~F. and {Gongora}, J.~P. and {Gonz{\'a}lez}, N. and {Goos}, I. and {G{\'o}ra}, D. and {Gorgi}, A. and {Gottowik}, M. and {Grubb}, T.~D. and {Guarino}, F. and {Guedes}, G.~P. and {Guido}, E. and {Hahn}, S. and {Halliday}, R. and {Hampel}, M.~R. and {Hansen}, P. and {Harari}, D. and {Harvey}, V.~M. and {Haungs}, A. and {Hebbeker}, T. and {Heck}, D. and {Hill}, G.~C. and {Hojvat}, C. and {H{\"o}randel}, J.~R. and {Horvath}, P. and {Hrabovsk{\'y}}, M. and {Huege}, T. and {Hulsman}, J. and {Insolia}, A. and {Isar}, P.~G. and {Johnsen}, J.~A. and {Jurysek}, J. and {K{\"a}{\"a}p{\"a}}, A. and {Kampert}, K.~H. and {Keilhauer}, B. and {Kemp}, J. and {Klages}, H.~O. and {Kleifges}, M. and {Kleinfeller}, J. and {K{\"o}pke}, M. and {Kukec Mezek}, G. and {Lago}, B.~L. and {LaHurd}, D. and {Lang}, R.~G. and {Leigui de Oliveira}, M.~A. and {Lenok}, V. and {Letessier-Selvon}, A. and {Lhenry-Yvon}, I. and {Lo Presti}, D. and {Lopes}, L. and {L{\'o}pez}, R. and {Lorek}, R. and {Luce}, Q. and {Lucero}, A. and {Machado Payeras}, A. and {Malacari}, M. and {Mancarella}, G. and {Mandat}, D. and {Manning}, B.~C. and {Manshanden}, J. and {Mantsch}, P. and {Marafico}, S. and {Mariazzi}, A.~G.},
        title = "{Measurement of the cosmic-ray energy spectrum above 2.5 {\texttimes}{}10$^{18}$ eV using the Pierre Auger Observatory}",
      journal = {\prd},
         year = 2020,
        month = sep,
       volume = {102},
       number = {6},
          eid = {062005},
        pages = {062005},
          doi = {10.1103/PhysRevD.102.062005},
archivePrefix = {arXiv},
       eprint = {2008.06486},
 primaryClass = {astro-ph.HE},
       adsurl = {https://ui.adsabs.harvard.edu/abs/2020PhRvD.102f2005A}
}

@BOOK{Dermer2009,
       author = {{Dermer}, Charles D. and {Menon}, Govind},
        title = "{High Energy Radiation from Black Holes: Gamma Rays, Cosmic Rays, and Neutrinos}",
         year = 2009,
       adsurl = {https://ui.adsabs.harvard.edu/abs/2009herb.book.....D}
}

@ARTICLE{AlvesBatista2016,
       author = {{Alves Batista}, Rafael and {Dundovic}, Andrej and {Erdmann}, Martin and {Kampert}, Karl-Heinz and {Kuempel}, Daniel and {M{\"u}ller}, Gero and {Sigl}, Guenter and {van Vliet}, Arjen and {Walz}, David and {Winchen}, Tobias},
        title = "{CRPropa 3{\textemdash}a public astrophysical simulation framework for propagating extraterrestrial ultra-high energy particles}",
      journal = {\jcap},
         year = 2016,
        month = may,
       volume = {2016},
       number = {5},
          eid = {038},
        pages = {038},
          doi = {10.1088/1475-7516/2016/05/038},
archivePrefix = {arXiv},
       eprint = {1603.07142},
 primaryClass = {astro-ph.IM},
       adsurl = {https://ui.adsabs.harvard.edu/abs/2016JCAP...05..038A}
}

@ARTICLE{Aab2020,
       author = {{Aab}, A. and {Abreu}, P. and {Aglietta}, M. and {Albury}, J.~M. and {Allekotte}, I. and {Almela}, A. and {Alvarez Castillo}, J. and {Alvarez-Mu{\~n}iz}, J. and {Alves Batista}, R. and {Anastasi}, G.~A. and {Anchordoqui}, L. and {Andrada}, B. and {Andringa}, S. and {Aramo}, C. and {Ara{\'u}jo Ferreira}, P.~R. and {Asorey}, H. and {Assis}, P. and {Avila}, G. and {Badescu}, A.~M. and {Bakalova}, A. and {Balaceanu}, A. and {Barbato}, F. and {Barreira Luz}, R.~J. and {Becker}, K.~H. and {Bellido}, J.~A. and {Berat}, C. and {Bertaina}, M.~E. and {Bertou}, X. and {Biermann}, P.~L. and {Bister}, T. and {Biteau}, J. and {Blanco}, A. and {Blazek}, J. and {Bleve}, C. and {Boh{\'a}{\v{c}}ov{\'a}}, M. and {Boncioli}, D. and {Bonifazi}, C. and {Bonneau Arbeletche}, L. and {Borodai}, N. and {Botti}, A.~M. and {Brack}, J. and {Bretz}, T. and {Briechle}, F.~L. and {Buchholz}, P. and {Bueno}, A. and {Buitink}, S. and {Buscemi}, M. and {Caballero-Mora}, K.~S. and {Caccianiga}, L. and {Calcagni}, L. and {Cancio}, A. and {Canfora}, F. and {Caracas}, I. and {Carceller}, J.~M. and {Caruso}, R. and {Castellina}, A. and {Catalani}, F. and {Cataldi}, G. and {Cazon}, L. and {Cerda}, M. and {Chinellato}, J.~A. and {Choi}, K. and {Chudoba}, J. and {Chytka}, L. and {Clay}, R.~W. and {Cobos Cerutti}, A.~C. and {Colalillo}, R. and {Coleman}, A. and {Coluccia}, M.~R. and {Concei{\c{c}}{\~a}o}, R. and {Condorelli}, A. and {Consolati}, G. and {Contreras}, F. and {Convenga}, F. and {Covault}, C.~E. and {Dasso}, S. and {Daumiller}, K. and {Dawson}, B.~R. and {Day}, J.~A. and {de Almeida}, R.~M. and {de Jes{\'u}s}, J. and {de Jong}, S.~J. and {De Mauro}, G. and {de Mello Neto}, J.~R.~T. and {De Mitri}, I. and {de Oliveira}, J. and {de Oliveira Franco}, D. and {de Souza}, V. and {De Vito}, E. and {Debatin}, J. and {del R{\'\i}o}, M. and {Deligny}, O. and {Dembinski}, H. and {Dhital}, N. and {Di Giulio}, C. and {Di Matteo}, A. and {D{\'\i}az Castro}, M.~L. and {Dobrigkeit}, C. and {D'Olivo}, J.~C. and {Dorosti}, Q. and {dos Anjos}, R.~C. and {Dova}, M.~T. and {Ebr}, J. and {Engel}, R. and {Epicoco}, I. and {Erdmann}, M. and {Escobar}, C.~O. and {Etchegoyen}, A. and {Falcke}, H. and {Farmer}, J. and {Farrar}, G. and {Fauth}, A.~C. and {Fazzini}, N. and {Feldbusch}, F. and {Fenu}, F. and {Fick}, B. and {Figueira}, J.~M. and {Filip{\v{c}}i{\v{c}}}, A. and {Fodran}, T. and {Freire}, M.~M. and {Fujii}, T. and {Fuster}, A. and {Galea}, C. and {Galelli}, C. and {Garc{\'\i}a}, B. and {Garcia Vegas}, A.~L. and {Gemmeke}, H. and {Gesualdi}, F. and {Gherghel-Lascu}, A. and {Ghia}, P.~L. and {Giaccari}, U. and {Giammarchi}, M. and {Giller}, M. and {Glombitza}, J. and {Gobbi}, F. and {Gollan}, F. and {Golup}, G. and {G{\'o}mez Berisso}, M. and {G{\'o}mez Vitale}, P.~F. and {Gongora}, J.~P. and {Gonz{\'a}lez}, N. and {Goos}, I. and {G{\'o}ra}, D. and {Gorgi}, A. and {Gottowik}, M. and {Grubb}, T.~D. and {Guarino}, F. and {Guedes}, G.~P. and {Guido}, E. and {Hahn}, S. and {Halliday}, R. and {Hampel}, M.~R. and {Hansen}, P. and {Harari}, D. and {Harvey}, V.~M. and {Haungs}, A. and {Hebbeker}, T. and {Heck}, D. and {Hill}, G.~C. and {Hojvat}, C. and {H{\"o}randel}, J.~R. and {Horvath}, P. and {Hrabovsk{\'y}}, M. and {Huege}, T. and {Hulsman}, J. and {Insolia}, A. and {Isar}, P.~G. and {Johnsen}, J.~A. and {Jurysek}, J. and {K{\"a}{\"a}p{\"a}}, A. and {Kampert}, K.~H. and {Keilhauer}, B. and {Kemp}, J. and {Klages}, H.~O. and {Kleifges}, M. and {Kleinfeller}, J. and {K{\"o}pke}, M. and {Kukec Mezek}, G. and {Lago}, B.~L. and {LaHurd}, D. and {Lang}, R.~G. and {Leigui de Oliveira}, M.~A. and {Lenok}, V. and {Letessier-Selvon}, A. and {Lhenry-Yvon}, I. and {Lo Presti}, D. and {Lopes}, L. and {L{\'o}pez}, R. and {Lorek}, R. and {Luce}, Q. and {Lucero}, A. and {Machado Payeras}, A. and {Malacari}, M. and {Mancarella}, G. and {Mandat}, D. and {Manning}, B.~C. and {Manshanden}, J. and {Mantsch}, P. and {Marafico}, S. and {Mariazzi}, A.~G.},
        title = "{Features of the Energy Spectrum of Cosmic Rays above 2.5 {\texttimes}{}10$^{18}$ eV Using the Pierre Auger Observatory}",
      journal = {\prl},
         year = 2020,
        month = sep,
       volume = {125},
       number = {12},
          eid = {121106},
        pages = {121106},
          doi = {10.1103/PhysRevLett.125.121106},
archivePrefix = {arXiv},
       eprint = {2008.06488},
 primaryClass = {astro-ph.HE},
       adsurl = {https://ui.adsabs.harvard.edu/abs/2020PhRvL.125l1106A}
}

@ARTICLE{Loules2025,
       author = {{Loules}, Argyrios and {Nathanail}, Antonios and {Contopoulos}, Ioannis},
        title = "{The azimuthal structure of magnetically arrested disks during flux eruption events}",
      journal = {\aap},
         year = 2026,
        month = may,
       volume = {710},
          eid = {A37},
        pages = {A37},
          doi = {10.1051/0004-6361/202555642},
archivePrefix = {arXiv},
       eprint = {2604.14313},
 primaryClass = {astro-ph.HE},
       adsurl = {https://ui.adsabs.harvard.edu/abs/2026A&A...710A..37L}
}

@ARTICLE{Kotera2011,
       author = {{Kotera}, Kumiko and {Olinto}, Angela V.},
        title = "{The Astrophysics of Ultrahigh-Energy Cosmic Rays}",
      journal = {\araa},
         year = 2011,
        month = sep,
       volume = {49},
       number = {1},
        pages = {119-153},
          doi = {10.1146/annurev-astro-081710-102620},
archivePrefix = {arXiv},
       eprint = {1101.4256},
 primaryClass = {astro-ph.HE},
       adsurl = {https://ui.adsabs.harvard.edu/abs/2011ARA&A..49..119K}
}

@ARTICLE{Gilmore2012,
       author = {{Gilmore}, Rudy C. and {Somerville}, Rachel S. and {Primack}, Joel R. and {Dom{\'\i}nguez}, Alberto},
        title = "{Semi-analytic modelling of the extragalactic background light and consequences for extragalactic gamma-ray spectra}",
      journal = {\mnras},
         year = 2012,
        month = jun,
       volume = {422},
       number = {4},
        pages = {3189-3207},
          doi = {10.1111/j.1365-2966.2012.20841.x},
archivePrefix = {arXiv},
       eprint = {1104.0671},
 primaryClass = {astro-ph.CO},
       adsurl = {https://ui.adsabs.harvard.edu/abs/2012MNRAS.422.3189G}
}

@INPROCEEDINGS{Tchekhovskoy2015,
       author = {{Tchekhovskoy}, Alexander},
        title = "{Launching of Active Galactic Nuclei Jets}",
    booktitle = {The Formation and Disruption of Black Hole Jets},
         year = 2015,
       editor = {{Contopoulos}, Ioannis and {Gabuzda}, Denise and {Kylafis}, Nikolaos},
       series = {Astrophysics and Space Science Library},
       volume = {414},
        month = jan,
        pages = {45},
          doi = {10.1007/978-3-319-10356-3_3},
       adsurl = {https://ui.adsabs.harvard.edu/abs/2015ASSL..414...45T}
}

@article{Blumenthal1970,
  title = {Energy Loss of High-Energy Cosmic Rays in Pair-Producing Collisions with Ambient Photons},
  author = {Blumenthal, George R.},
  journal = {Phys. Rev. D},
  volume = {1},
  issue = {6},
  pages = {1596--1602},
  numpages = {0},
  year = {1970},
  month = {Mar},
  publisher = {American Physical Society},
  doi = {10.1103/PhysRevD.1.1596},
  url = {https://link.aps.org/doi/10.1103/PhysRevD.1.1596}
}

@ARTICLE{Chodorowski1992,
       author = {{Chodorowski}, Michal J. and {Zdziarski}, Andrzej A. and {Sikora}, Marek},
        title = "{Reaction Rate and Energy-Loss Rate for Photopair Production by Relativistic Nuclei}",
      journal = {\apj},
         year = 1992,
        month = nov,
       volume = {400},
        pages = {181},
          doi = {10.1086/171984},
       adsurl = {https://ui.adsabs.harvard.edu/abs/1992ApJ...400..181C}
}

@PHDTHESIS{Rachen1996,
       author = {{Rachen}, J{\"o}rg Paul},
        title = "{Interaction Processes and Statistical Properties of the Propagation of Cosmic Rays in Photon Backgrounds}",
       school = {Max-Planck-Institute for Radioastronomy, Bonn},
         year = 1996,
        month = sep,
       adsurl = {https://ui.adsabs.harvard.edu/abs/1996PhDT........59R}
}

@ARTICLE{Loules2025a,
       author = {{Loules}, Argyrios and {Vlahakis}, Nektarios},
        title = "{Modeling of resistive relativistic astrophysical jets: Semianalytic results following a paraxial formalism}",
      journal = {\aap},
         year = 2025,
        month = apr,
       volume = {696},
          eid = {A186},
        pages = {A186},
          doi = {10.1051/0004-6361/202452824},
archivePrefix = {arXiv},
       eprint = {2410.09826},
 primaryClass = {astro-ph.HE},
       adsurl = {https://ui.adsabs.harvard.edu/abs/2025A&A...696A.186L}
}

@ARTICLE{Matthews_2021,
       author = {{Matthews}, James H. and {Taylor}, Andrew M.},
        title = "{Particle acceleration in radio galaxies with flickering jets: GeV electrons to ultrahigh energy cosmic rays}",
      journal = {\mnras},
         year = 2021,
        month = jun,
       volume = {503},
       number = {4},
        pages = {5948-5964},
          doi = {10.1093/mnras/stab758},
archivePrefix = {arXiv},
       eprint = {2103.06900},
 primaryClass = {astro-ph.HE},
       adsurl = {https://ui.adsabs.harvard.edu/abs/2021MNRAS.503.5948M}
}

@ARTICLE{Sinha_2018,
       author = {{Sinha}, Atreyee and {Khatoon}, Rukaiya and {Misra}, Ranjeev and {Sahayanathan}, Sunder and {Mandal}, Soma and {Gogoi}, Rupjyoti and {Bhatt}, Nilay},
        title = "{The flux distribution of individual blazars as a key to understand the dynamics of particle acceleration}",
      journal = {\mnras},
         year = 2018,
        month = oct,
       volume = {480},
       number = {1},
        pages = {L116-L120},
          doi = {10.1093/mnrasl/sly136},
archivePrefix = {arXiv},
       eprint = {1807.09073},
 primaryClass = {astro-ph.HE},
       adsurl = {https://ui.adsabs.harvard.edu/abs/2018MNRAS.480L.116S}
}

@ARTICLE{Gaspari_2017,
       author = {{Gaspari}, Massimo and {S{\k{a}}dowski}, Aleksander},
        title = "{Unifying the Micro and Macro Properties of AGN Feeding and Feedback}",
      journal = {\apj},
         year = 2017,
        month = mar,
       volume = {837},
       number = {2},
          eid = {149},
        pages = {149},
          doi = {10.3847/1538-4357/aa61a3},
archivePrefix = {arXiv},
       eprint = {1701.07030},
 primaryClass = {astro-ph.HE},
       adsurl = {https://ui.adsabs.harvard.edu/abs/2017ApJ...837..149G}
}

@ARTICLE{Rieger_2010,
       author = {{Rieger}, F.~M. and {Volpe}, F.},
        title = "{Short-term VHE variability in blazars: PKS 2155-304}",
      journal = {\aap},
         year = 2010,
        month = sep,
       volume = {520},
          eid = {A23},
        pages = {A23},
          doi = {10.1051/0004-6361/201014273},
archivePrefix = {arXiv},
       eprint = {1007.4879},
 primaryClass = {astro-ph.HE},
       adsurl = {https://ui.adsabs.harvard.edu/abs/2010A&A...520A..23R}
}

@ARTICLE{Fang2012,
       author = {{Fang}, Ke and {Kotera}, Kumiko and {Olinto}, Angela V.},
        title = "{Newly Born Pulsars as Sources of Ultrahigh Energy Cosmic Rays}",
      journal = {\apj},
         year = 2012,
        month = may,
       volume = {750},
       number = {2},
          eid = {118},
        pages = {118},
          doi = {10.1088/0004-637X/750/2/118},
archivePrefix = {arXiv},
       eprint = {1201.5197},
 primaryClass = {astro-ph.HE},
       adsurl = {https://ui.adsabs.harvard.edu/abs/2012ApJ...750..118F}
}

@article{Fang2014,
  title = {Testing the newborn pulsar origin of ultrahigh energy cosmic rays with EeV neutrinos},
  author = {Fang, Ke and Kotera, Kumiko and Murase, Kohta and Olinto, Angela V.},
  journal = {Phys. Rev. D},
  volume = {90},
  issue = {10},
  pages = {103005},
  numpages = {7},
  year = {2014},
  month = {Nov},
  publisher = {American Physical Society},
  doi = {10.1103/PhysRevD.90.103005},
  url = {https://link.aps.org/doi/10.1103/PhysRevD.90.103005}
}

@ARTICLE{Guepin2018,
       author = {{Gu{\'e}pin}, Claire and {Kotera}, Kumiko and {Barausse}, Enrico and {Fang}, Ke and {Murase}, Kohta},
        title = "{Ultra-high-energy cosmic rays and neutrinos from tidal disruptions by massive black holes}",
      journal = {\aap},
         year = 2018,
        month = sep,
       volume = {616},
          eid = {A179},
        pages = {A179},
          doi = {10.1051/0004-6361/201732392},
archivePrefix = {arXiv},
       eprint = {1711.11274},
 primaryClass = {astro-ph.HE},
       adsurl = {https://ui.adsabs.harvard.edu/abs/2018A&A...616A.179G}
}

@ARTICLE{AbdulHalim2025,
       author = {{Abdul Halim}, A. and {Abreu}, P. and {Aglietta}, M. and {Allekotte}, I. and {Almeida Cheminant}, K. and {Almela}, A. and {Aloisio}, R. and {Alvarez-Mu{\~n}iz}, J. and {Ammerman Yebra}, J. and {Anastasi}, G.~A. and {Anchordoqui}, L. and {Andrada}, B. and {Andrade Dourado}, L. and {Andringa}, S. and {Apollonio}, L. and {Aramo}, C. and {Ara{\'u}jo Ferreira}, P.~R. and {Arnone}, E. and {Arteaga Vel{\'a}zquez}, J.~C. and {Assis}, P. and {Avila}, G. and {Avocone}, E. and {Bakalova}, A. and {Barbato}, F. and {Bartz Mocellin}, A. and {Berat}, C. and {Bertaina}, M.~E. and {Bhatta}, G. and {Bianciotto}, M. and {Biermann}, P.~L. and {Binet}, V. and {Bismark}, K. and {Bister}, T. and {Biteau}, J. and {Blazek}, J. and {Bleve}, C. and {Bl{\"u}mer}, J. and {Boh{\'a}{\v{c}}ov{\'a}}, M. and {Boncioli}, D. and {Bonifazi}, C. and {Bonneau Arbeletche}, L. and {Borodai}, N. and {Brack}, J. and {Brichetto Orchera}, P.~G. and {Briechle}, F.~L. and {Bueno}, A. and {Buitink}, S. and {Buscemi}, M. and {B{\"u}sken}, M. and {Bwembya}, A. and {Caballero-Mora}, K.~S. and {Cabana-Freire}, S. and {Caccianiga}, L. and {Campuzano}, F. and {Caruso}, R. and {Castellina}, A. and {Catalani}, F. and {Cataldi}, G. and {Cazon}, L. and {Cerda}, M. and {{\v{C}}erm{\'a}kov{\'a}}, B. and {Cermenati}, A. and {Chinellato}, J.~A. and {Chudoba}, J. and {Chytka}, L. and {Clay}, R.~W. and {Cobos Cerutti}, A.~C. and {Colalillo}, R. and {Coluccia}, M.~R. and {Concei{\c{c}}{\~a}o}, R. and {Condorelli}, A. and {Consolati}, G. and {Conte}, M. and {Convenga}, F. and {Correia dos Santos}, D. and {Costa}, P.~J. and {Covault}, C.~E. and {Cristinziani}, M. and {Cruz Sanchez}, C.~S. and {Dasso}, S. and {Daumiller}, K. and {Dawson}, B.~R. and {de Almeida}, R.~M. and {de Errico}, B. and {de Jes{\'u}s}, J. and {de Jong}, S.~J. and {de Mello Neto}, J.~R.~T. and {De Mitri}, I. and {de Oliveira}, J. and {de Oliveira Franco}, D. and {de Palma}, F. and {de Souza}, V. and {De Vito}, E. and {Del Popolo}, A. and {Deligny}, O. and {Denner}, N. and {Deval}, L. and {di Matteo}, A. and {do}, J.~A. and {Dobre}, M. and {Dobrigkeit}, C. and {D'Olivo}, J.~C. and {Domingues Mendes}, L.~M. and {Dorosti}, Q. and {dos Anjos}, J.~C. and {dos Anjos}, R.~C. and {Ebr}, J. and {Ellwanger}, F. and {Emam}, M. and {Engel}, R. and {Epicoco}, I. and {Erdmann}, M. and {Etchegoyen}, A. and {Evoli}, C. and {Falcke}, H. and {Farrar}, G. and {Fauth}, A.~C. and {Fehler}, T. and {Feldbusch}, F. and {Fenu}, F. and {Fernandes}, A. and {Fick}, B. and {Figueira}, J.~M. and {Filip}, P. and {Filip{\v{c}}i{\v{c}}}, A. and {Fitoussi}, T. and {Flaggs}, B. and {Fodran}, T. and {Fujii}, T. and {Fuster}, A. and {Galea}, C. and {Garc{\'\i}a}, B. and {Gaudu}, C. and {Gherghel-Lascu}, A. and {Ghia}, P.~L. and {Giaccari}, U. and {Glombitza}, J. and {Gobbi}, F. and {Gollan}, F. and {Golup}, G. and {G{\'o}mez Berisso}, M. and {G{\'o}mez Vitale}, P.~F. and {Gongora}, J.~P. and {Gonz{\'a}lez}, J.~M. and {Gonz{\'a}lez}, N. and {G{\'o}ra}, D. and {Gorgi}, A. and {Gottowik}, M. and {Guarino}, F. and {Guedes}, G.~P. and {Guido}, E. and {G{\"u}lzow}, L. and {Hahn}, S. and {Hamal}, P. and {Hampel}, M.~R. and {Hansen}, P. and {Harari}, D. and {Harvey}, V.~M. and {Haungs}, A. and {Hebbeker}, T. and {Hojvat}, C. and {H{\"o}randel}, J.~R. and {Horvath}, P. and {Hrabovsk{\'y}}, M. and {Huege}, T. and {Insolia}, A. and {Isar}, P.~G. and {Janecek}, P. and {Jilek}, V. and {Johnsen}, J.~A. and {Jurysek}, J. and {Kampert}, K. -H. and {Keilhauer}, B. and {Khakurdikar}, A. and {Kizakke Covilakam}, V.~V. and {Klages}, H.~O. and {Kleifges}, M. and {Knapp}, F. and {K{\"o}hler}, J. and {Krieger}, F. and {Kunka}, N. and {Lago}, B.~L. and {Langner}, N. and {Leigui de Oliveira}, M.~A. and {Lema-Capeans}, Y. and {Letessier-Selvon}, A. and {Lhenry-Yvon}, I. and {Lopes}, L. and {Lu}, L. and {Luce}, Q. and {Lundquist}, J.~P. and {Machado Payeras}, A. and {Majercakova}, M. and {Mandat}, D. and {Manning}, B.~C. and {Mantsch}, P. and {Mariani}, F.~M. and {Mariazzi}, A.~G. and {Mari{\c{s}}}, I.~C. and {Marsella}, G.},
        title = "{Measurement of the depth of maximum of air-shower profiles with energies between 1018.5 and 1020  eV using the surface detector of the Pierre Auger Observatory and deep learning}",
      journal = {\prd},
         year = 2025,
        month = jan,
       volume = {111},
       number = {2},
          eid = {022003},
        pages = {022003},
          doi = {10.1103/PhysRevD.111.022003},
archivePrefix = {arXiv},
       eprint = {2406.06319},
 primaryClass = {astro-ph.HE},
       adsurl = {https://ui.adsabs.harvard.edu/abs/2025PhRvD.111b2003A}
}

@ARTICLE{Globus,
       author = {{Globus}, No{\'e}mie and {Blandford}, Roger D.},
        title = "{Ultrahigh-Energy Cosmic Rays}",
      journal = {\araa},
         year = 2025,
        month = aug,
       volume = {63},
       number = {1},
        pages = {339-377},
          doi = {10.1146/annurev-astro-052622-033150},
archivePrefix = {arXiv},
       eprint = {2505.21846},
 primaryClass = {astro-ph.HE},
       adsurl = {https://ui.adsabs.harvard.edu/abs/2025ARA&A..63..339G}
}

@ARTICLE{Aabetal,
       author = {{Aab}, A. and {Abreu}, P. and {Aglietta}, M. and {Albuquerque}, I.~F.~M. and {Allekotte}, I. and {Almela}, A. and {Alvarez Castillo}, J. and {Alvarez-Mu{\~n}iz}, J. and {Anastasi}, G.~A. and {Anchordoqui}, L. and {Andrada}, B. and {Andringa}, S. and {Aramo}, C. and {Arsene}, N. and {Asorey}, H. and {Assis}, P. and {Avila}, G. and {Badescu}, A.~M. and {Balaceanu}, A. and {Barbato}, F. and {Barreira Luz}, R.~J. and {Beatty}, J.~J. and {Becker}, K.~H. and {Bellido}, J.~A. and {Berat}, C. and {Bertaina}, M.~E. and {Bertou}, X. and {Biermann}, P.~L. and {Biteau}, J. and {Blaess}, S.~G. and {Blanco}, A. and {Blazek}, J. and {Bleve}, C. and {Boh{\'a}{\v{c}}ov{\'a}}, M. and {Bonifazi}, C. and {Borodai}, N. and {Botti}, A.~M. and {Brack}, J. and {Brancus}, I. and {Bretz}, T. and {Bridgeman}, A. and {Briechle}, F.~L. and {Buchholz}, P. and {Bueno}, A. and {Buitink}, S. and {Buscemi}, M. and {Caballero-Mora}, K.~S. and {Caccianiga}, L. and {Cancio}, A. and {Canfora}, F. and {Caruso}, R. and {Castellina}, A. and {Catalani}, F. and {Cataldi}, G. and {Cazon}, L. and {Chavez}, A.~G. and {Chinellato}, J.~A. and {Chudoba}, J. and {Clay}, R.~W. and {Cobos Cerutti}, A.~C. and {Colalillo}, R. and {Coleman}, A. and {Collica}, L. and {Coluccia}, M.~R. and {Concei{\c{c}}{\~a}o}, R. and {Consolati}, G. and {Contreras}, F. and {Cooper}, M.~J. and {Coutu}, S. and {Covault}, C.~E. and {Cronin}, J. and {D'Amico}, S. and {Daniel}, B. and {Dasso}, S. and {Daumiller}, K. and {Dawson}, B.~R. and {de Almeida}, R.~M. and {de Jong}, S.~J. and {De Mauro}, G. and {de Mello Neto}, J.~R.~T. and {De Mitri}, I. and {de Oliveira}, J. and {de Souza}, V. and {Debatin}, J. and {Deligny}, O. and {D{\'\i}az Castro}, M.~L. and {Diogo}, F. and {Dobrigkeit}, C. and {D'Olivo}, J.~C. and {Dorosti}, Q. and {dos Anjos}, R.~C. and {Dova}, M.~T. and {Dundovic}, A. and {Ebr}, J. and {Engel}, R. and {Erdmann}, M. and {Erfani}, M. and {Escobar}, C.~O. and {Espadanal}, J. and {Etchegoyen}, A. and {Falcke}, H. and {Farmer}, J. and {Farrar}, G. and {Fauth}, A.~C. and {Fazzini}, N. and {Fenu}, F. and {Fick}, B. and {Figueira}, J.~M. and {Filip{\v{c}}i{\v{c}}}, A. and {Freire}, M.~M. and {Fujii}, T. and {Fuster}, A. and {Ga{\"\i}or}, R. and {Garc{\'\i}a}, B. and {Gat{\'e}}, F. and {Gemmeke}, H. and {Gherghel-Lascu}, A. and {Ghia}, P.~L. and {Giaccari}, U. and {Giammarchi}, M. and {Giller}, M. and {G{\l}as}, D. and {Glaser}, C. and {Golup}, G. and {G{\'o}mez Berisso}, M. and {G{\'o}mez Vitale}, P.~F. and {Gonz{\'a}lez}, N. and {Gorgi}, A. and {Grillo}, A.~F. and {Grubb}, T.~D. and {Guarino}, F. and {Guedes}, G.~P. and {Halliday}, R. and {Hampel}, M.~R. and {Hansen}, P. and {Harari}, D. and {Harrison}, T.~A. and {Haungs}, A. and {Hebbeker}, T. and {Heck}, D. and {Heimann}, P. and {Herve}, A.~E. and {Hill}, G.~C. and {Hojvat}, C. and {Holt}, E. and {Homola}, P. and {H{\"o}randel}, J.~R. and {Horvath}, P. and {Hrabovsk{\'y}}, M. and {Huege}, T. and {Hulsman}, J. and {Insolia}, A. and {Isar}, P.~G. and {Jandt}, I. and {Johnsen}, J.~A. and {Josebachuili}, M. and {Jurysek}, J. and {K{\"a}{\"a}p{\"a}}, A. and {Kambeitz}, O. and {Kampert}, K.~H. and {Keilhauer}, B. and {Kemmerich}, N. and {Kemp}, E. and {Kemp}, J. and {Kieckhafer}, R.~M. and {Klages}, H.~O. and {Kleifges}, M. and {Kleinfeller}, J. and {Krause}, R. and {Krohm}, N. and {Kuempel}, D. and {Kukec Mezek}, G. and {Kunka}, N. and {Kuotb Awad}, A. and {Lago}, B.~L. and {LaHurd}, D. and {Lang}, R.~G. and {Lauscher}, M. and {Legumina}, R. and {Leigui de Oliveira}, M.~A. and {Letessier-Selvon}, A. and {Lhenry-Yvon}, I. and {Link}, K. and {Lo Presti}, D. and {Lopes}, L. and {L{\'o}pez}, R. and {L{\'o}pez Casado}, A. and {Lorek}, R. and {Luce}, Q. and {Lucero}, A. and {Malacari}, M. and {Mallamaci}, M. and {Mandat}, D. and {Mantsch}, P. and {Mariazzi}, A.~G. and {Mari{\c{s}}}, I.~C. and {Marsella}, G. and {Martello}, D. and {Martinez}, H. and {Mart{\'\i}nez Bravo}, O.},
        title = "{An Indication of Anisotropy in Arrival Directions of Ultra-high-energy Cosmic Rays through Comparison to the Flux Pattern of Extragalactic Gamma-Ray Sources}",
      journal = {\apjl},
         year = 2018,
        month = feb,
       volume = {853},
       number = {2},
          eid = {L29},
        pages = {L29},
          doi = {10.3847/2041-8213/aaa66d},
archivePrefix = {arXiv},
       eprint = {1801.06160},
 primaryClass = {astro-ph.HE},
       adsurl = {https://ui.adsabs.harvard.edu/abs/2018ApJ...853L..29A}
}

@ARTICLE{Stathopoulos2024,
       author = {{Stathopoulos}, S.~I. and {Petropoulou}, M. and {Sironi}, L. and {Giannios}, D.},
        title = "{The role of magnetospheric current sheets in pair enrichment and ultra-high energy proton acceleration in M87*}",
      journal = {\jcap},
         year = 2024,
        month = dec,
       volume = {2024},
       number = {12},
          eid = {009},
        pages = {009},
          doi = {10.1088/1475-7516/2024/12/009},
archivePrefix = {arXiv},
       eprint = {2406.01211},
 primaryClass = {astro-ph.HE},
       adsurl = {https://ui.adsabs.harvard.edu/abs/2024JCAP...12..009S}
}

@ARTICLE{fornaxa,
       author = {{Nowak}, N. and {Saglia}, R.~P. and {Thomas}, J. and {Bender}, R. and {Davies}, R.~I. and {Gebhardt}, K.},
        title = "{The supermassive black hole of FornaxA}",
      journal = {\mnras},
         year = 2008,
        month = dec,
       volume = {391},
       number = {4},
        pages = {1629-1649},
          doi = {10.1111/j.1365-2966.2008.13960.x},
archivePrefix = {arXiv},
       eprint = {0809.0696},
 primaryClass = {astro-ph},
       adsurl = {https://ui.adsabs.harvard.edu/abs/2008MNRAS.391.1629N}
}

@article{ngc1275,
    author = {Scharwächter, J. and McGregor, P. J. and Dopita, M. A. and Beck, T. L.},
    title = {Kinematics and excitation of the molecular hydrogen accretion disc in NGC 1275},
    journal = {Monthly Notices of the Royal Astronomical Society},
    volume = {429},
    number = {3},
    pages = {2315-2332},
    year = {2013},
    month = {01},
    issn = {0035-8711},
    doi = {10.1093/mnras/sts502},
    url = {https://doi.org/10.1093/mnras/sts502},
    eprint = {https://academic.oup.com/mnras/article-pdf/429/3/2315/3248827/sts502.pdf},
}

@ARTICLE{ic4296,
       author = {{Graham}, Alister W.},
        title = "{Populating the Galaxy Velocity Dispersion - Supermassive Black Hole Mass Diagram: A Catalogue of (M$_{bh}$, {\ensuremath{\sigma}}) Values}",
      journal = {\pasa},
         year = 2008,
        month = nov,
       volume = {25},
       number = {4},
        pages = {167-175},
          doi = {10.1071/AS08013},
archivePrefix = {arXiv},
       eprint = {0807.2549},
 primaryClass = {astro-ph},
       adsurl = {https://ui.adsabs.harvard.edu/abs/2008PASA...25..167G}
}

@ARTICLE{HooperSarkarTaylor2008,
       author = {{Hooper}, Dan and {Sarkar}, Subir and {Taylor}, Andrew M.},
        title = "{Intergalactic propagation of ultrahigh energy cosmic ray nuclei: An analytic approach}",
      journal = {\prd},
         year = 2008,
        month = may,
       volume = {77},
       number = {10},
          eid = {103007},
        pages = {103007},
          doi = {10.1103/PhysRevD.77.103007},
archivePrefix = {arXiv},
       eprint = {0802.1538},
 primaryClass = {astro-ph},
       adsurl = {https://ui.adsabs.harvard.edu/abs/2008PhRvD..77j3007H}
}

@article{AhlersTaylor2010,
  title = {Analytic solutions of ultrahigh energy cosmic ray nuclei revisited},
  author = {Ahlers, Markus and Taylor, Andrew M.},
  journal = {Phys. Rev. D},
  volume = {82},
  issue = {12},
  pages = {123005},
  numpages = {15},
  year = {2010},
  month = {Dec},
  publisher = {American Physical Society},
  doi = {10.1103/PhysRevD.82.123005},
  url = {https://link.aps.org/doi/10.1103/PhysRevD.82.123005}
}

@article{SteckerSalamon1999,
doi = {10.1086/306816},
url = {https://doi.org/10.1086/306816},
year = {1999},
month = {feb},
publisher = {},
volume = {512},
number = {2},
pages = {521},
author = {Stecker, F. W. and Salamon, M. H.},
title = {Photodisintegration of Ultra-High-Energy Cosmic Rays: A New Determination},
journal = {The Astrophysical Journal}
}

@ARTICLE{PugetSteckerBredekamp1976,
       author = {{Puget}, J.~L. and {Stecker}, F.~W. and {Bredekamp}, J.~H.},
        title = "{Photonuclear interactions of ultrahigh energy cosmic rays and their astrophysical consequences.}",
      journal = {\apj},
         year = 1976,
        month = apr,
       volume = {205},
        pages = {638-654},
          doi = {10.1086/154321},
       adsurl = {https://ui.adsabs.harvard.edu/abs/1976ApJ...205..638P}
}

@article{AlvesBatista2022,
doi = {10.1088/1475-7516/2022/09/035},
url = {https://doi.org/10.1088/1475-7516/2022/09/035},
year = {2022},
month = {sep},
publisher = {IOP Publishing},
volume = {2022},
number = {09},
pages = {035},
author = {Alves Batista, Rafael and Becker Tjus, Julia and Dörner, Julien and Dundovic, Andrej and Eichmann, Björn and Frie, Antonius and Heiter, Christopher and Hoerbe, Mario R. and Kampert, Karl-Heinz and Merten, Lukas and Müller, Gero and Reichherzer, Patrick and Saveliev, Andrey and Schlegel, Leander and Sigl, Günter and van Vliet, Arjen and Winchen, Tobias},
title = {CRPropa 3.2 — an advanced framework for high-energy particle propagation in extragalactic and galactic spaces},
journal = {Journal of Cosmology and Astroparticle Physics}
}

@ARTICLE{Planck2020,
       author = {{Planck Collaboration} and {Aghanim}, N. and {Akrami}, Y. and {Ashdown}, M. and {Aumont}, J. and {Baccigalupi}, C. and {Ballardini}, M. and {Banday}, A.~J. and {Barreiro}, R.~B. and {Bartolo}, N. and {Basak}, S. and {Battye}, R. and {Benabed}, K. and {Bernard}, J.-P. and {Bersanelli}, M. and {Bielewicz}, P. and {Bock}, J.~J. and {Bond}, J.~R. and {Borrill}, J. and {Bouchet}, F.~R. and {Boulanger}, F. and {Bucher}, M. and {Burigana}, C. and {Butler}, R.~C. and {Calabrese}, E. and {Cardoso}, J.-F. and {Carron}, J. and {Challinor}, A. and {Chiang}, H.~C. and {Chluba}, J. and {Colombo}, L.~P.~L. and {Combet}, C. and {Contreras}, D. and {Crill}, B.~P. and {Cuttaia}, F. and {de Bernardis}, P. and {de Zotti}, G. and {Delabrouille}, J. and {Delouis}, J.-M. and {Di Valentino}, E. and {Diego}, J.~M. and {Dor{\'e}}, O. and {Douspis}, M. and {Ducout}, A. and {Dupac}, X. and {Dusini}, S. and {Efstathiou}, G. and {Elsner}, F. and {En{\ss}lin}, T.~A. and {Eriksen}, H.~K. and {Fantaye}, Y. and {Farhang}, M. and {Fergusson}, J. and {Fernandez-Cobos}, R. and {Finelli}, F. and {Forastieri}, F. and {Frailis}, M. and {Fraisse}, A.~A. and {Franceschi}, E. and {Frolov}, A. and {Galeotta}, S. and {Galli}, S. and {Ganga}, K. and {G{\'e}nova-Santos}, R.~T. and {Gerbino}, M. and {Ghosh}, T. and {Gonz{\'a}lez-Nuevo}, J. and {G{\'o}rski}, K.~M. and {Gratton}, S. and {Gruppuso}, A. and {Gudmundsson}, J.~E. and {Hamann}, J. and {Handley}, W. and {Hansen}, F.~K. and {Herranz}, D. and {Hildebrandt}, S.~R. and {Hivon}, E. and {Huang}, Z. and {Jaffe}, A.~H. and {Jones}, W.~C. and {Karakci}, A. and {Keih{\"a}nen}, E. and {Keskitalo}, R. and {Kiiveri}, K. and {Kim}, J. and {Kisner}, T.~S. and {Knox}, L. and {Krachmalnicoff}, N. and {Kunz}, M. and {Kurki-Suonio}, H. and {Lagache}, G. and {Lamarre}, J.-M. and {Lasenby}, A. and {Lattanzi}, M. and {Lawrence}, C.~R. and {Le Jeune}, M. and {Lemos}, P. and {Lesgourgues}, J. and {Levrier}, F. and {Lewis}, A. and {Liguori}, M. and {Lilje}, P.~B. and {Lilley}, M. and {Lindholm}, V. and {L{\'o}pez-Caniego}, M. and {Lubin}, P.~M. and {Ma}, Y.-Z. and {Mac{\'\i}as-P{\'e}rez}, J.~F. and {Maggio}, G. and {Maino}, D. and {Mandolesi}, N. and {Mangilli}, A. and {Marcos-Caballero}, A. and {Maris}, M. and {Martin}, P.~G. and {Martinelli}, M. and {Mart{\'\i}nez-Gonz{\'a}lez}, E. and {Matarrese}, S. and {Mauri}, N. and {McEwen}, J.~D. and {Meinhold}, P.~R. and {Melchiorri}, A. and {Mennella}, A. and {Migliaccio}, M. and {Millea}, M. and {Mitra}, S. and {Miville-Desch{\^e}nes}, M.-A. and {Molinari}, D. and {Montier}, L. and {Morgante}, G. and {Moss}, A. and {Natoli}, P. and {N{\o}rgaard-Nielsen}, H.~U. and {Pagano}, L. and {Paoletti}, D. and {Partridge}, B. and {Patanchon}, G. and {Peiris}, H.~V. and {Perrotta}, F. and {Pettorino}, V. and {Piacentini}, F. and {Polastri}, L. and {Polenta}, G. and {Puget}, J.-L. and {Rachen}, J.~P. and {Reinecke}, M. and {Remazeilles}, M. and {Renzi}, A. and {Rocha}, G. and {Rosset}, C. and {Roudier}, G. and {Rubi{\~n}o-Mart{\'\i}n}, J.~A. and {Ruiz-Granados}, B. and {Salvati}, L. and {Sandri}, M. and {Savelainen}, M. and {Scott}, D. and {Shellard}, E.~P.~S. and {Sirignano}, C. and {Sirri}, G. and {Spencer}, L.~D. and {Sunyaev}, R. and {Suur-Uski}, A.-S. and {Tauber}, J.~A. and {Tavagnacco}, D. and {Tenti}, M. and {Toffolatti}, L. and {Tomasi}, M. and {Trombetti}, T. and {Valenziano}, L. and {Valiviita}, J. and {Van Tent}, B. and {Vibert}, L. and {Vielva}, P. and {Villa}, F. and {Vittorio}, N. and {Wandelt}, B.~D. and {Wehus}, I.~K. and {White}, M. and {White}, S.~D.~M. and {Zacchei}, A. and {Zonca}, A.},
        title = "{Planck 2018 results. VI. Cosmological parameters}",
      journal = {\aap},
         year = 2020,
        month = sep,
       volume = {641},
          eid = {A6},
        pages = {A6},
          doi = {10.1051/0004-6361/201833910},
archivePrefix = {arXiv},
       eprint = {1807.06209},
 primaryClass = {astro-ph.CO},
       adsurl = {https://ui.adsabs.harvard.edu/abs/2020A&A...641A...6P}
}

@inbook{Palme2014,
title = "Solar System Abundances of the Elements",
author = "H. Palme and K. Lodders and A. Jones",
year = "2013",
month = nov,
doi = "10.1016/B978-0-08-095975-7.00118-2",
language = "English",
isbn = "9780080983004",
volume = "2",
pages = "15--36",
booktitle = "Planets, Asteriods, Comets and The Solar System",
publisher = "Elsevier Inc.",

}

@ARTICLE{Nagao2006b,
       author = {{Nagao}, T. and {Marconi}, A. and {Maiolino}, R.},
        title = "{The evolution of the broad-line region among SDSS quasars}",
      journal = {\aap},
         year = 2006,
        month = feb,
       volume = {447},
       number = {1},
        pages = {157-172},
          doi = {10.1051/0004-6361:20054024},
archivePrefix = {arXiv},
       eprint = {astro-ph/0510385},
 primaryClass = {astro-ph},
       adsurl = {https://ui.adsabs.harvard.edu/abs/2006A&A...447..157N}
}

@ARTICLE{Aab2017JCAP,
       author = {{Aab}, A. and {Abreu}, P. and {Aglietta}, M. and {Samarai}, I. Al and {Albuquerque}, I.~F.~M. and {Allekotte}, I. and {Almela}, A. and {Alvarez Castillo}, J. and {Alvarez-Mu{\~n}iz}, J. and {Anastasi}, G.~A. and {Anchordoqui}, L. and {Andrada}, B. and {Andringa}, S. and {Aramo}, C. and {Arqueros}, F. and {Arsene}, N. and {Asorey}, H. and {Assis}, P. and {Aublin}, J. and {Avila}, G. and {Badescu}, A.~M. and {Balaceanu}, A. and {Barreira Luz}, R.~J. and {Beatty}, J.~J. and {Becker}, K.~H. and {Bellido}, J.~A. and {Berat}, C. and {Bertaina}, M.~E. and {Bertou}, X. and {Biermann}, P.~L. and {Billoir}, P. and {Biteau}, J. and {Blaess}, S.~G. and {Blanco}, A. and {Blazek}, J. and {Bleve}, C. and {Boh{\'a}{\v{c}}ov{\'a}}, M. and {Boncioli}, D. and {Bonifazi}, C. and {Borodai}, N. and {Botti}, A.~M. and {Brack}, J. and {Brancus}, I. and {Bretz}, T. and {Bridgeman}, A. and {Briechle}, F.~L. and {Buchholz}, P. and {Bueno}, A. and {Buitink}, S. and {Buscemi}, M. and {Caballero-Mora}, K.~S. and {Caccianiga}, L. and {Cancio}, A. and {Canfora}, F. and {Caramete}, L. and {Caruso}, R. and {Castellina}, A. and {Cataldi}, G. and {Cazon}, L. and {Chavez}, A.~G. and {Chinellato}, J.~A. and {Chudoba}, J. and {Clay}, R.~W. and {Colalillo}, R. and {Coleman}, A. and {Collica}, L. and {Coluccia}, M.~R. and {Concei{\c{c}}{\~a}o}, R. and {Contreras}, F. and {Cooper}, M.~J. and {Coutu}, S. and {Covault}, C.~E. and {Cronin}, J. and {D'Amico}, S. and {Daniel}, B. and {Dasso}, S. and {Daumiller}, K. and {Dawson}, B.~R. and {de Almeida}, R.~M. and {de Jong}, S.~J. and {De Mauro}, G. and {de Mello Neto}, J.~R.~T. and {De Mitri}, I. and {de Oliveira}, J. and {de Souza}, V. and {Debatin}, J. and {Deligny}, O. and {Di Giulio}, C. and {di Matteo}, A. and {D{\'\i}az Castro}, M.~L. and {Diogo}, F. and {Dobrigkeit}, C. and {D'Olivo}, J.~C. and {Dorosti}, Q. and {dos Anjos}, R.~C. and {Dova}, M.~T. and {Dundovic}, A. and {Ebr}, J. and {Engel}, R. and {Erdmann}, M. and {Erfani}, M. and {Escobar}, C.~O. and {Espadanal}, J. and {Etchegoyen}, A. and {Falcke}, H. and {Farrar}, G. and {Fauth}, A.~C. and {Fazzini}, N. and {Fick}, B. and {Figueira}, J.~M. and {Filip{\v{c}}i{\v{c}}}, A. and {Fratu}, O. and {Freire}, M.~M. and {Fujii}, T. and {Fuster}, A. and {Gaior}, R. and {Garc{\'\i}a}, B. and {Garcia-Pinto}, D. and {Gat{\'e}}, F. and {Gemmeke}, H. and {Gherghel-Lascu}, A. and {Ghia}, P.~L. and {Giaccari}, U. and {Giammarchi}, M. and {Giller}, M. and {G{\l}as}, D. and {Glaser}, C. and {Golup}, G. and {G{\'o}mez Berisso}, M. and {G{\'o}mez Vitale}, P.~F. and {Gonz{\'a}lez}, N. and {Gorgi}, A. and {Gorham}, P. and {Grillo}, A.~F. and {Grubb}, T.~D. and {Guarino}, F. and {Guedes}, G.~P. and {Hampel}, M.~R. and {Hansen}, P. and {Harari}, D. and {Harrison}, T.~A. and {Harton}, J.~L. and {Haungs}, A. and {Hebbeker}, T. and {Heck}, D. and {Heimann}, P. and {Herve}, A.~E. and {Hill}, G.~C. and {Hojvat}, C. and {Holt}, E. and {Homola}, P. and {H{\"o}randel}, J.~R. and {Horvath}, P. and {Hrabovsk{\'y}}, M. and {Huege}, T. and {Hulsman}, J. and {Insolia}, A. and {Isar}, P.~G. and {Jandt}, I. and {Jansen}, S. and {Johnsen}, J.~A. and {Josebachuili}, M. and {K{\"a}{\"a}p{\"a}}, A. and {Kambeitz}, O. and {Kampert}, K.~H. and {Katkov}, I. and {Keilhauer}, B. and {Kemp}, E. and {Kemp}, J. and {Kieckhafer}, R.~M. and {Klages}, H.~O. and {Kleifges}, M. and {Kleinfeller}, J. and {Krause}, R. and {Krohm}, N. and {Kuempel}, D. and {Kukec Mezek}, G. and {Kunka}, N. and {Kuotb Awad}, A. and {LaHurd}, D. and {Lauscher}, M. and {Legumina}, R. and {Leigui de Oliveira}, M.~A. and {Letessier-Selvon}, A. and {Lhenry-Yvon}, I. and {Link}, K. and {Lopes}, L. and {L{\'o}pez}, R. and {L{\'o}pez Casado}, A. and {Luce}, Q. and {Lucero}, A. and {Malacari}, M. and {Mallamaci}, M. and {Mandat}, D. and {Mantsch}, P. and {Mariazzi}, A.~G. and {Mari{\textcommabelow s}}, I.~C. and {Marsella}, G. and {Martello}, D. and {Martinez}, H.},
        title = "{Combined fit of spectrum and composition data as measured by the Pierre Auger Observatory}",
      journal = {\jcap},
         year = 2017,
        month = apr,
       volume = {2017},
       number = {4},
          eid = {038},
        pages = {038},
          doi = {10.1088/1475-7516/2017/04/038},
archivePrefix = {arXiv},
       eprint = {1612.07155},
 primaryClass = {astro-ph.HE},
       adsurl = {https://ui.adsabs.harvard.edu/abs/2017JCAP...04..038A}
}

@ARTICLE{GallimoreImpellizzeri2023,
       author = {{Gallimore}, Jack F. and {Impellizzeri}, C.~M. Violette},
        title = "{High-sensitivity Observations of the H$_{2}$O Megamasers of NGC 1068: Precise Astrometry and Detailed Kinematics}",
      journal = {\apj},
         year = 2023,
        month = jul,
       volume = {951},
       number = {2},
          eid = {109},
        pages = {109},
          doi = {10.3847/1538-4357/acd846},
archivePrefix = {arXiv},
       eprint = {2305.13097},
 primaryClass = {astro-ph.GA},
       adsurl = {https://ui.adsabs.harvard.edu/abs/2023ApJ...951..109G}
}

@article{terBraak2008,
  author  = {{ter Braak}, C. J. F. and Vrugt, J. A.},
  title   = {Differential Evolution Markov Chain with snooker updater and fewer chains},
  journal = {Statistics and Computing},
  year    = {2008},
  volume  = {18},
  number  = {4},
  pages   = {435--446},
  doi     = {10.1007/s11222-008-9104-9}
}

@ARTICLE{IceCube2022NGC1068,
       author = {{IceCube Collaboration} and {Abbasi}, R. and {Ackermann}, M. and {Adams}, J. and {Aguilar}, J.~A. and {Ahlers}, M. and {Ahrens}, M. and {Alameddine}, J.~M. and {Alispach}, C. and {Alves}, Jr., A.~A. and {Amin}, N.~M. and {Andeen}, K. and {Anderson}, T. and {Anton}, G. and {Arg{\"u}elles}, C. and {Ashida}, Y. and {Axani}, S. and {Bai}, X. and {Balagopal}, A.~V. and {Barbano}, V.~A. and {Barwick}, S.~W. and {Bastian}, B. and {Basu}, V. and {Baur}, S. and {Bay}, R. and {Beatty}, J.~J. and {Becker}, K.-H. and {Becker Tjus}, J. and {Bellenghi}, C. and {Benzvi}, S. and {Berley}, D. and {Bernardini}, E. and {Besson}, D.~Z. and {Binder}, G. and {Bindig}, D. and {Blaufuss}, E. and {Blot}, S. and {Boddenberg}, M. and {Bontempo}, F. and {Borowka}, J. and {B{\"o}ser}, S. and {Botner}, O. and {B{\"o}ttcher}, J. and {Bourbeau}, E. and {Bradascio}, F. and {Braun}, J. and {Brinson}, B. and {Bron}, S. and {Brostean-Kaiser}, J. and {Browne}, S. and {Burgman}, A. and {Burley}, R.~T. and {Busse}, R.~S. and {Campana}, M.~A. and {Carnie-Bronca}, E.~G. and {Chen}, C. and {Chen}, Z. and {Chirkin}, D. and {Choi}, K. and {Clark}, B.~A. and {Clark}, K. and {Classen}, L. and {Coleman}, A. and {Collin}, G.~H. and {Conrad}, J.~M. and {Coppin}, P. and {Correa}, P. and {Cowen}, D.~F. and {Cross}, R. and {Dappen}, C. and {Dave}, P. and {de Clercq}, C. and {Delaunay}, J.~J. and {Delgado L{\'o}pez}, D. and {Dembinski}, H. and {Deoskar}, K. and {Desai}, A. and {Desiati}, P. and {de Vries}, K.~D. and {de Wasseige}, G. and {de With}, M. and {Deyoung}, T. and {Diaz}, A. and {D{\'\i}az-V{\'e}lez}, J.~C. and {Dittmer}, M. and {Dujmovic}, H. and {Dunkman}, M. and {Duvernois}, M.~A. and {Dvorak}, E. and {Ehrhardt}, T. and {Eller}, P. and {Engel}, R. and {Erpenbeck}, H. and {Evans}, J. and {Evenson}, P.~A. and {Fan}, K.~L. and {Fazely}, A.~R. and {Fedynitch}, A. and {Feigl}, N. and {Fiedlschuster}, S. and {Fienberg}, A.~T. and {Filimonov}, K. and {Finley}, C. and {Fischer}, L. and {Fox}, D. and {Franckowiak}, A. and {Friedman}, E. and {Fritz}, A. and {F{\"u}rst}, P. and {Gaisser}, T.~K. and {Gallagher}, J. and {Ganster}, E. and {Garcia}, A. and {Garrappa}, S. and {Gerhardt}, L. and {Ghadimi}, A. and {Glaser}, C. and {Glauch}, T. and {Gl{\"u}senkamp}, T. and {Goldschmidt}, A. and {Gonzalez}, J.~G. and {Goswami}, S. and {Grant}, D. and {Gr{\'e}goire}, T. and {Griswold}, S. and {G{\"u}nther}, C. and {Gutjahr}, P. and {Haack}, C. and {Hallgren}, A. and {Halliday}, R. and {Halve}, L. and {Halzen}, F. and {Hanson}, M. Ha Minh K. and {Hardin}, J. and {Harnisch}, A.~A. and {Haungs}, A. and {Hebecker}, D. and {Helbing}, K. and {Henningsen}, F. and {Hettinger}, E.~C. and {Hickford}, S. and {Hignight}, J. and {Hill}, C. and {Hill}, G.~C. and {Hoffman}, K.~D. and {Hoffmann}, R. and {Hokanson-Fasig}, B. and {Hoshina}, K. and {Huang}, F. and {Huber}, M. and {Huber}, T. and {Hultqvist}, K. and {H{\"u}nnefeld}, M. and {Hussain}, R. and {Hymon}, K. and {in}, S. and {Iovine}, N. and {Ishihara}, A. and {Jansson}, M. and {Japaridze}, G.~S. and {Jeong}, M. and {Jin}, M. and {Jones}, B.~J.~P. and {Kang}, D. and {Kang}, W. and {Kang}, X. and {Kappes}, A. and {Kappesser}, D. and {Kardum}, L. and {Karg}, T. and {Karl}, M. and {Karle}, A. and {Katz}, U. and {Kauer}, M. and {Kellermann}, M. and {Kelley}, J.~L. and {Kheirandish}, A. and {Kin}, K. and {Kintscher}, T. and {Kiryluk}, J. and {Klein}, S.~R. and {Koirala}, R. and {Kolanoski}, H. and {Kontrimas}, T. and {K{\"o}pke}, L. and {Kopper}, C. and {Kopper}, S. and {Koskinen}, D.~J. and {Koundal}, P. and {Kovacevich}, M. and {Kowalski}, M. and {Kozynets}, T. and {Kun}, E. and {Kurahashi}, N. and {Lad}, N. and {Lagunas Gualda}, C. and {Lanfranchi}, J.~L. and {Larson}, M.~J. and {Lauber}, F. and {Lazar}, J.~P.},
        title = "{Evidence for neutrino emission from the nearby active galaxy NGC 1068}",
      journal = {Science},
         year = 2022,
        month = nov,
       volume = {378},
       number = {6619},
        pages = {538-543},
          doi = {10.1126/science.abg3395},
archivePrefix = {arXiv},
       eprint = {2211.09972},
 primaryClass = {astro-ph.HE},
       adsurl = {https://ui.adsabs.harvard.edu/abs/2022Sci...378..538I}
}

@ARTICLE{Kimura2018,
       author = {{Kimura}, Shigeo S. and {Murase}, Kohta and {Zhang}, B. Theodore},
        title = "{Ultrahigh-energy cosmic-ray nuclei from black hole jets: Recycling galactic cosmic rays through shear acceleration}",
      journal = {\prd},
         year = 2018,
        month = jan,
       volume = {97},
       number = {2},
          eid = {023026},
        pages = {023026},
          doi = {10.1103/PhysRevD.97.023026},
archivePrefix = {arXiv},
       eprint = {1705.05027},
 primaryClass = {astro-ph.HE},
       adsurl = {https://ui.adsabs.harvard.edu/abs/2018PhRvD..97b3026K}
}

@ARTICLE{Colaco2026,
       author = {{Cola{\c{c}}o}, Andressa and {Azeredo}, Gabriel and {Parillo}, Isadora and {de Oliveira}, Cain{\~a} and {de Souza}, Vitor},
        title = "{Modeling individual nearby radio galaxies as ultra-high-energy cosmic-ray accelerators}",
      journal = {arXiv e-prints},
         year = 2026,
        month = feb,
          eid = {arXiv:2602.23458},
        pages = {arXiv:2602.23458},
          doi = {10.48550/arXiv.2602.23458},
archivePrefix = {arXiv},
       eprint = {2602.23458},
 primaryClass = {astro-ph.HE},
       adsurl = {https://ui.adsabs.harvard.edu/abs/2026arXiv260223458C}
}

@ARTICLE{Bergmann2007,
       author = {{Bergmann}, T. and {Engel}, R. and {Heck}, D. and {Kalmykov}, N.~N. and {Ostapchenko}, S. and {Pierog}, T. and {Thouw}, T. and {Werner}, K.},
        title = "{One-dimensional hybrid approach to extensive air shower simulation}",
      journal = {Astroparticle Physics},
         year = 2007,
        month = jan,
       volume = {26},
       number = {6},
        pages = {420-432},
          doi = {10.1016/j.astropartphys.2006.08.005},
archivePrefix = {arXiv},
       eprint = {astro-ph/0606564},
 primaryClass = {astro-ph},
       adsurl = {https://ui.adsabs.harvard.edu/abs/2007APh....26..420B}
}

@ARTICLE{2013Domenico,
       author = {{De Domenico}, Manlio and {Settimo}, Mariangela and {Riggi}, Simone and {Bertin}, Eric},
        title = "{Reinterpreting the development of extensive air showers initiated by nuclei and photons}",
      journal = {\jcap},
         year = 2013,
        month = jul,
       volume = {2013},
       number = {7},
          eid = {050},
        pages = {050},
          doi = {10.1088/1475-7516/2013/07/050},
archivePrefix = {arXiv},
       eprint = {1305.2331},
 primaryClass = {hep-ph},
       adsurl = {https://ui.adsabs.harvard.edu/abs/2013JCAP...07..050D}
}

@ARTICLE{2013Pierog,
       author = {{Pierog}, T. and {Karpenko}, Iu. and {Katzy}, J.~M. and {Yatsenko}, E. and {Werner}, K.},
        title = "{EPOS LHC : test of collective hadronization with LHC data}",
      journal = {arXiv e-prints},
         year = 2013,
        month = jun,
          eid = {arXiv:1306.0121},
        pages = {arXiv:1306.0121},
          doi = {10.48550/arXiv.1306.0121},
archivePrefix = {arXiv},
       eprint = {1306.0121},
 primaryClass = {hep-ph},
       adsurl = {https://ui.adsabs.harvard.edu/abs/2013arXiv1306.0121P}
}

@ARTICLE{emcee,
       author = {{Foreman-Mackey}, Daniel and {Farr}, Will and {Sinha}, Manodeep and {Archibald}, Anne and {Hogg}, David and {Sanders}, Jeremy and {Zuntz}, Joe and {Williams}, Peter and {Nelson}, Andrew and {de Val-Borro}, Miguel and {Erhardt}, Tobias and {Pashchenko}, Ilya and {Pla}, Oriol},
        title = "{emcee v3: A Python ensemble sampling toolkit for affine-invariant MCMC}",
      journal = {The Journal of Open Source Software},
         year = 2019,
        month = nov,
       volume = {4},
       number = {43},
          eid = {1864},
        pages = {1864},
          doi = {10.21105/joss.01864},
archivePrefix = {arXiv},
       eprint = {1911.07688},
 primaryClass = {astro-ph.IM},
       adsurl = {https://ui.adsabs.harvard.edu/abs/2019JOSS....4.1864F}
}

@article{terBraak,
author = {ter Braak, Cajo},
year = {2006},
month = {09},
pages = {239-249},
title = {A Markov Chain Monte Carlo version of the genetic algorithm Differential Evolution: Easy Bayesian computing for real parameter spaces},
volume = {16},
journal = {Statistics and Computing},
doi = {10.1007/s11222-006-8769-1}
}

@ARTICLE{AbdulHalim2023,
       author = {{Abdul Halim}, A. and {Abreu}, P. and {Aglietta}, M. and {Allekotte}, I. and {Almeida Cheminant}, K. and {Almela}, A. and {Alvarez-Mu{\~n}iz}, J. and {Ammerman Yebra}, J. and {Anastasi}, G.~A. and {Anchordoqui}, L. and {Andrada}, B. and {Andringa}, S. and {Aramo}, C. and {Ara{\'u}jo Ferreira}, P.~R. and {Arnone}, E. and {Arteaga Vel{\'a}zquez}, J.~C. and {Asorey}, H. and {Assis}, P. and {Avila}, G. and {Avocone}, E. and {Badescu}, A.~M. and {Bakalova}, A. and {Balaceanu}, A. and {Barbato}, F. and {Bellido}, J.~A. and {Berat}, C. and {Bertaina}, M.~E. and {Bhatta}, G. and {Biermann}, P.~L. and {Binet}, V. and {Bismark}, K. and {Bister}, T. and {Biteau}, J. and {Blazek}, J. and {Bleve}, C. and {Bl{\"u}mer}, J. and {Boh{\'a}{\v{c}}ov{\'a}}, M. and {Boncioli}, D. and {Bonifazi}, C. and {Bonneau Arbeletche}, L. and {Borodai}, N. and {Brack}, J. and {Bretz}, T. and {Brichetto Orchera}, P.~G. and {Briechle}, F.~L. and {Buchholz}, P. and {Bueno}, A. and {Buitink}, S. and {Buscemi}, M. and {B{\"u}sken}, M. and {Bwembya}, A. and {Caballero-Mora}, K.~S. and {Caccianiga}, L. and {Caracas}, I. and {Caruso}, R. and {Castellina}, A. and {Catalani}, F. and {Cataldi}, G. and {Cazon}, L. and {Cerda}, M. and {Chinellato}, J.~A. and {Chudoba}, J. and {Chytka}, L. and {Clay}, R.~W. and {Cobos Cerutti}, A.~C. and {Colalillo}, R. and {Coleman}, A. and {Coluccia}, M.~R. and {Concei{\c{c}}{\~a}o}, R. and {Condorelli}, A. and {Consolati}, G. and {Conte}, M. and {Contreras}, F. and {Convenga}, F. and {Correia dos Santos}, D. and {Covault}, C.~E. and {Cristinziani}, M. and {Cruz Sanchez}, C.~S. and {Dasso}, S. and {Daumiller}, K. and {Dawson}, B.~R. and {de Almeida}, R.~M. and {de Jes{\'u}s}, J. and {de Jong}, S.~J. and {de Mello Neto}, J.~R.~T. and {De Mitri}, I. and {de Oliveira}, J. and {de Oliveira Franco}, D. and {de Palma}, F. and {de Souza}, V. and {De Vito}, E. and {Del Popolo}, A. and {Deligny}, O. and {Deval}, L. and {di Matteo}, A. and {Dobre}, M. and {Dobrigkeit}, C. and {D'Olivo}, J.~C. and {Domingues Mendes}, L.~M. and {dos Anjos}, R.~C. and {Ebr}, J. and {Eman}, M. and {Engel}, R. and {Epicoco}, I. and {Erdmann}, M. and {Etchegoyen}, A. and {Falcke}, H. and {Farmer}, J. and {Farrar}, G. and {Fauth}, A.~C. and {Fazzini}, N. and {Feldbusch}, F. and {Fenu}, F. and {Fick}, B. and {Figueira}, J.~M. and {Filip{\v{c}}i{\v{c}}}, A. and {Fitoussi}, T. and {Flaggs}, B. and {Fodran}, T. and {Fujii}, T. and {Fuster}, A. and {Galea}, C. and {Galelli}, C. and {Garc{\'\i}a}, B. and {Gemmeke}, H. and {Gesualdi}, F. and {Gherghel-Lascu}, A. and {Ghia}, P.~L. and {Giaccari}, U. and {Giammarchi}, M. and {Glombitza}, J. and {Gobbi}, F. and {Gollan}, F. and {Golup}, G. and {G{\'o}mez Berisso}, M. and {G{\'o}mez Vitale}, P.~F. and {Gongora}, J.~P. and {Gonz{\'a}lez}, J.~M. and {Gonz{\'a}lez}, N. and {Goos}, I. and {G{\'o}ra}, D. and {Gorgi}, A. and {Gottowik}, M. and {Grubb}, T.~D. and {Guarino}, F. and {Guedes}, G.~P. and {Guido}, E. and {Hahn}, S. and {Hamal}, P. and {Hampel}, M.~R. and {Hansen}, P. and {Harari}, D. and {Harvey}, V.~M. and {Haungs}, A. and {Hebbeker}, T. and {Heck}, D. and {Hojvat}, C. and {H{\"o}randel}, J.~R. and {Horvath}, P. and {Hrabovsk{\'y}}, M. and {Huege}, T. and {Insolia}, A. and {Isar}, P.~G. and {Janecek}, P. and {Johnsen}, J.~A. and {Jurysek}, J. and {K{\"a}{\"a}p{\"a}}, A. and {Kampert}, K.~H. and {Keilhauer}, B. and {Khakurdikar}, A. and {Kizakke Covilakam}, V.~V. and {Klages}, H.~O. and {Kleifges}, M. and {Kleinfeller}, J. and {Knapp}, F. and {Kunka}, N. and {Lago}, B.~L. and {Langner}, N. and {Leigui de Oliveira}, M.~A. and {Lenok}, V. and {Letessier-Selvon}, A. and {Lhenry-Yvon}, I. and {Lo Presti}, D. and {Lopes}, L. and {L{\'o}pez}, R. and {Lu}, L. and {Luce}, Q. and {Lundquist}, J.~P. and {Machado Payeras}, A. and {Majercakova}, M. and {Mandat}, D. and {Manning}, B.~C. and {Manshanden}, J. and {Mantsch}, P. and {Marafico}, S. and {Mariani}, F.~M. and {Mariazzi}, A.~G. and {Mari{\textcommabelow s}}, I.~C. and {Marsella}, G. and {Martello}, D.},
        title = "{Constraining the sources of ultra-high-energy cosmic rays across and above the ankle with the spectrum and composition data measured at the Pierre Auger Observatory}",
      journal = {\jcap},
         year = 2023,
        month = may,
       volume = {2023},
       number = {5},
          eid = {024},
        pages = {024},
          doi = {10.1088/1475-7516/2023/05/024},
archivePrefix = {arXiv},
       eprint = {2211.02857},
 primaryClass = {astro-ph.HE},
       adsurl = {https://ui.adsabs.harvard.edu/abs/2023JCAP...05..024A}
}

@ARTICLE{ngc1407,
       author = {{Rusli}, S.~P. and {Thomas}, J. and {Saglia}, R.~P. and {Fabricius}, M. and {Erwin}, P. and {Bender}, R. and {Nowak}, N. and {Lee}, C.~H. and {Riffeser}, A. and {Sharp}, R.},
        title = "{The Influence of Dark Matter Halos on Dynamical Estimates of Black Hole Mass: 10 New Measurements for High-{\ensuremath{\sigma}} Early-type Galaxies}",
      journal = {\aj},
         year = 2013,
        month = sep,
       volume = {146},
       number = {3},
          eid = {45},
        pages = {45},
          doi = {10.1088/0004-6256/146/3/45},
archivePrefix = {arXiv},
       eprint = {1306.1124},
 primaryClass = {astro-ph.CO},
       adsurl = {https://ui.adsabs.harvard.edu/abs/2013AJ....146...45R}
}

@ARTICLE{cena,
       author = {{Cappellari}, Michele and {Neumayer}, N. and {Reunanen}, J. and {van der Werf}, P.~P. and {de Zeeuw}, P.~T. and {Rix}, H.-W.},
        title = "{The mass of the black hole in Centaurus A from SINFONI AO-assisted integral-field observations of stellar kinematics}",
      journal = {\mnras},
         year = 2009,
        month = apr,
       volume = {394},
       number = {2},
        pages = {660-674},
          doi = {10.1111/j.1365-2966.2008.14377.x},
archivePrefix = {arXiv},
       eprint = {0812.1000},
 primaryClass = {astro-ph},
       adsurl = {https://ui.adsabs.harvard.edu/abs/2009MNRAS.394..660C}
}

@ARTICLE{ngc1399,
       author = {{Gebhardt}, Karl and {Lauer}, Tod R. and {Pinkney}, Jason and {Bender}, Ralf and {Richstone}, Douglas and {Aller}, Monique and {Bower}, Gary and {Dressler}, Alan and {Faber}, S.~M. and {Filippenko}, Alexei V. and {Green}, Richard and {Ho}, Luis C. and {Kormendy}, John and {Siopis}, Christos and {Tremaine}, Scott},
        title = "{The Black Hole Mass and Extreme Orbital Structure in NGC 1399}",
      journal = {\apj},
         year = 2007,
        month = dec,
       volume = {671},
       number = {2},
        pages = {1321-1328},
          doi = {10.1086/522938},
archivePrefix = {arXiv},
       eprint = {0709.0585},
 primaryClass = {astro-ph},
       adsurl = {https://ui.adsabs.harvard.edu/abs/2007ApJ...671.1321G}
}

@ARTICLE{m87_vi,
       author = {{Event Horizon Telescope Collaboration} and {Akiyama}, Kazunori and {Alberdi}, Antxon and {Alef}, Walter and {Asada}, Keiichi and {Azulay}, Rebecca and {Baczko}, Anne-Kathrin and {Ball}, David and {Balokovi{\'c}}, Mislav and {Barrett}, John and {Bintley}, Dan and {Blackburn}, Lindy and {Boland}, Wilfred and {Bouman}, Katherine L. and {Bower}, Geoffrey C. and {Bremer}, Michael and {Brinkerink}, Christiaan D. and {Brissenden}, Roger and {Britzen}, Silke and {Broderick}, Avery E. and {Broguiere}, Dominique and {Bronzwaer}, Thomas and {Byun}, Do-Young and {Carlstrom}, John E. and {Chael}, Andrew and {Chan}, Chi-kwan and {Chatterjee}, Shami and {Chatterjee}, Koushik and {Chen}, Ming-Tang and {Chen}, Yongjun and {Cho}, Ilje and {Christian}, Pierre and {Conway}, John E. and {Cordes}, James M. and {Crew}, Geoffrey B. and {Cui}, Yuzhu and {Davelaar}, Jordy and {De Laurentis}, Mariafelicia and {Deane}, Roger and {Dempsey}, Jessica and {Desvignes}, Gregory and {Dexter}, Jason and {Doeleman}, Sheperd S. and {Eatough}, Ralph P. and {Falcke}, Heino and {Fish}, Vincent L. and {Fomalont}, Ed and {Fraga-Encinas}, Raquel and {Friberg}, Per and {Fromm}, Christian M. and {G{\'o}mez}, Jos{\'e} L. and {Galison}, Peter and {Gammie}, Charles F. and {Garc{\'\i}a}, Roberto and {Gentaz}, Olivier and {Georgiev}, Boris and {Goddi}, Ciriaco and {Gold}, Roman and {Gu}, Minfeng and {Gurwell}, Mark and {Hada}, Kazuhiro and {Hecht}, Michael H. and {Hesper}, Ronald and {Ho}, Luis C. and {Ho}, Paul and {Honma}, Mareki and {Huang}, Chih-Wei L. and {Huang}, Lei and {Hughes}, David H. and {Ikeda}, Shiro and {Inoue}, Makoto and {Issaoun}, Sara and {James}, David J. and {Jannuzi}, Buell T. and {Janssen}, Michael and {Jeter}, Britton and {Jiang}, Wu and {Johnson}, Michael D. and {Jorstad}, Svetlana and {Jung}, Taehyun and {Karami}, Mansour and {Karuppusamy}, Ramesh and {Kawashima}, Tomohisa and {Keating}, Garrett K. and {Kettenis}, Mark and {Kim}, Jae-Young and {Kim}, Junhan and {Kim}, Jongsoo and {Kino}, Motoki and {Koay}, Jun Yi and {Koch}, Patrick M. and {Koyama}, Shoko and {Kramer}, Michael and {Kramer}, Carsten and {Krichbaum}, Thomas P. and {Kuo}, Cheng-Yu and {Lauer}, Tod R. and {Lee}, Sang-Sung and {Li}, Yan-Rong and {Li}, Zhiyuan and {Lindqvist}, Michael and {Liu}, Kuo and {Liuzzo}, Elisabetta and {Lo}, Wen-Ping and {Lobanov}, Andrei P. and {Loinard}, Laurent and {Lonsdale}, Colin and {Lu}, Ru-Sen and {MacDonald}, Nicholas R. and {Mao}, Jirong and {Markoff}, Sera and {Marrone}, Daniel P. and {Marscher}, Alan P. and {Mart{\'\i}-Vidal}, Iv{\'a}n and {Matsushita}, Satoki and {Matthews}, Lynn D. and {Medeiros}, Lia and {Menten}, Karl M. and {Mizuno}, Yosuke and {Mizuno}, Izumi and {Moran}, James M. and {Moriyama}, Kotaro and {Moscibrodzka}, Monika and {M{\"u}ller}, Cornelia and {Nagai}, Hiroshi and {Nagar}, Neil M. and {Nakamura}, Masanori and {Narayan}, Ramesh and {Narayanan}, Gopal and {Natarajan}, Iniyan and {Neri}, Roberto and {Ni}, Chunchong and {Noutsos}, Aristeidis and {Okino}, Hiroki and {Olivares}, H{\'e}ctor and {Oyama}, Tomoaki and {{\"O}zel}, Feryal and {Palumbo}, Daniel C.~M. and {Patel}, Nimesh and {Pen}, Ue-Li and {Pesce}, Dominic W. and {Pi{\'e}tu}, Vincent and {Plambeck}, Richard and {PopStefanija}, Aleksandar and {Porth}, Oliver and {Prather}, Ben and {Preciado-L{\'o}pez}, Jorge A. and {Psaltis}, Dimitrios and {Pu}, Hung-Yi and {Ramakrishnan}, Venkatessh and {Rao}, Ramprasad and {Rawlings}, Mark G. and {Raymond}, Alexander W. and {Rezzolla}, Luciano and {Ripperda}, Bart and {Roelofs}, Freek and {Rogers}, Alan and {Ros}, Eduardo and {Rose}, Mel and {Roshanineshat}, Arash and {Rottmann}, Helge and {Roy}, Alan L. and {Ruszczyk}, Chet and {Ryan}, Benjamin R. and {Rygl}, Kazi L.~J. and {S{\'a}nchez}, Salvador and {S{\'a}nchez-Arguelles}, David and {Sasada}, Mahito and {Savolainen}, Tuomas and {Schloerb}, F. Peter and {Schuster}, Karl-Friedrich and {Shao}, Lijing and {Shen}, Zhiqiang and {Small}, Des and {Sohn}, Bong Won and {SooHoo}, Jason and {Tazaki}, Fumie and {Tiede}, Paul and {Tilanus}, Remo P.~J. and {Titus}, Michael and {Toma}, Kenji and {Torne}, Pablo and {Trent}, Tyler and {Trippe}, Sascha and {Tsuda}, Shuichiro and {van Bemmel}, Ilse and {van Langevelde}, Huib Jan and {van Rossum}, Daniel R. and {Wagner}, Jan and {Wardle}, John and {Weintroub}, Jonathan and {Wex}, Norbert and {Wharton}, Robert and {Wielgus}, Maciek and {Wong}, George N. and {Wu}, Qingwen and {Young}, Andr{\'e} and {Young}, Ken and {Younsi}, Ziri and {Yuan}, Feng},
        title = "{First M87 Event Horizon Telescope Results. VI. The Shadow and Mass of the Central Black Hole}",
      journal = {\apjl},
         year = 2019,
        month = apr,
       volume = {875},
       number = {1},
          eid = {L6},
        pages = {L6},
          doi = {10.3847/2041-8213/ab1141},
archivePrefix = {arXiv},
       eprint = {1906.11243},
 primaryClass = {astro-ph.GA},
       adsurl = {https://ui.adsabs.harvard.edu/abs/2019ApJ...875L...6E}
}

@ARTICLE{NGC315,
       author = {{Boizelle}, Benjamin D. and {Walsh}, Jonelle L. and {Barth}, Aaron J. and {Buote}, David A. and {Baker}, Andrew J. and {Darling}, Jeremy and {Ho}, Luis C. and {Cohn}, Jonathan and {Kabasares}, Kyle M.},
        title = "{Black Hole Mass Measurements of Radio Galaxies NGC 315 and NGC 4261 Using ALMA CO Observations}",
      journal = {\apj},
         year = 2021,
        month = feb,
       volume = {908},
       number = {1},
          eid = {19},
        pages = {19},
          doi = {10.3847/1538-4357/abd24d},
archivePrefix = {arXiv},
       eprint = {2012.04669},
 primaryClass = {astro-ph.GA},
       adsurl = {https://ui.adsabs.harvard.edu/abs/2021ApJ...908...19B}
}

@ARTICLE{cygnusa,
       author = {{Tadhunter}, C. and {Marconi}, A. and {Axon}, D. and {Wills}, K. and {Robinson}, T.~G. and {Jackson}, N.},
        title = "{Spectroscopy of the near-nuclear regions of Cygnus A: estimating the mass of the supermassive black hole}",
      journal = {\mnras},
         year = 2003,
        month = jul,
       volume = {342},
       number = {3},
        pages = {861-875},
          doi = {10.1046/j.1365-8711.2003.06588.x},
archivePrefix = {arXiv},
       eprint = {astro-ph/0302513},
 primaryClass = {astro-ph},
       adsurl = {https://ui.adsabs.harvard.edu/abs/2003MNRAS.342..861T}
}

\begin{appendix}
\onecolumn

\section{Posterior distributions and MCMC convergence}
\label{app:posteriors}

\begin{figure*}[htb]
    \centering
    \subfigure[A, $d=10\,\mathrm{Mpc}$]{
        \includegraphics[width=0.32\textwidth]
        {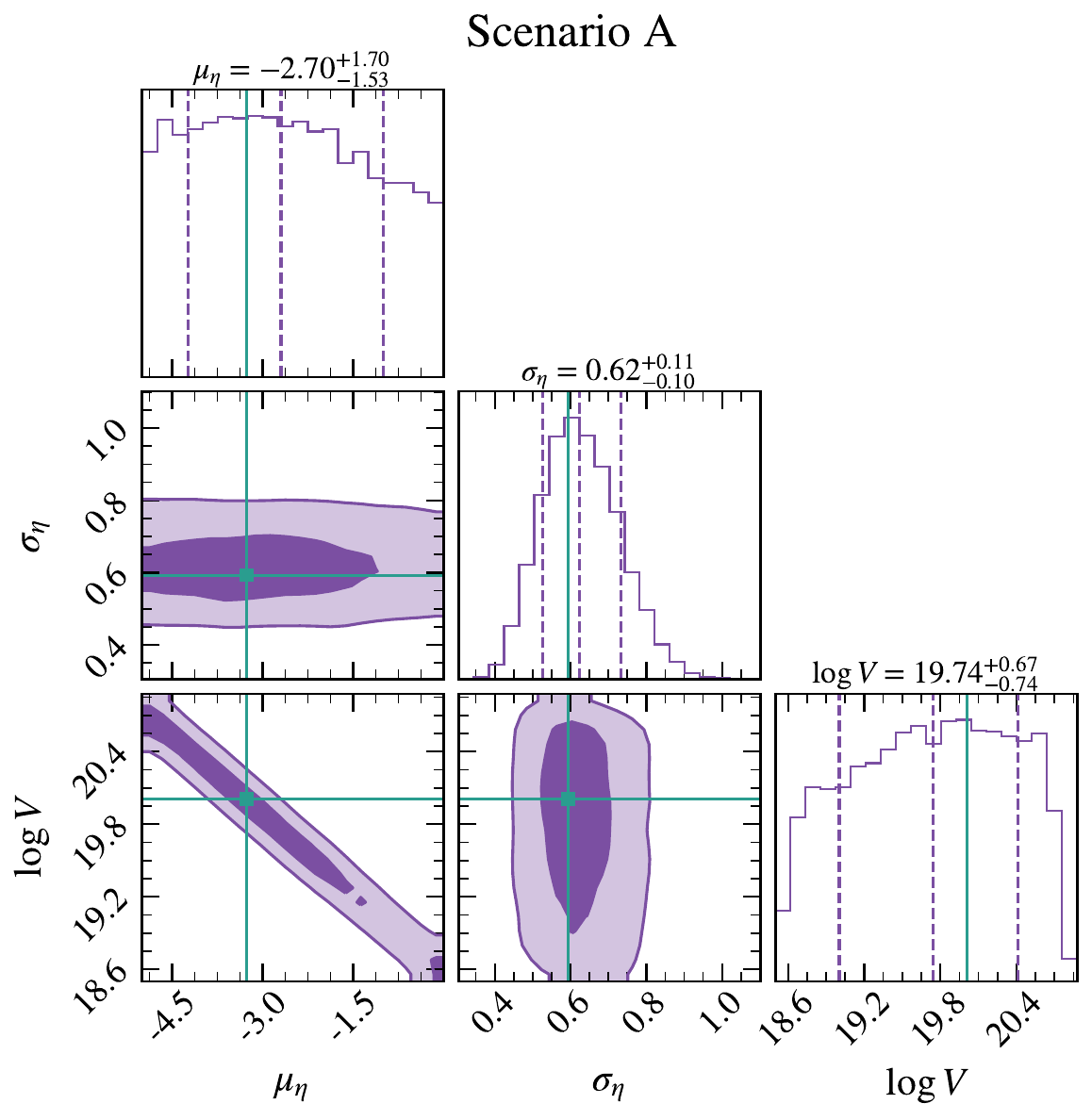}
    }\hfill
    \subfigure[A, $d=30\,\mathrm{Mpc}$]{
        \includegraphics[width=0.32\textwidth]
        {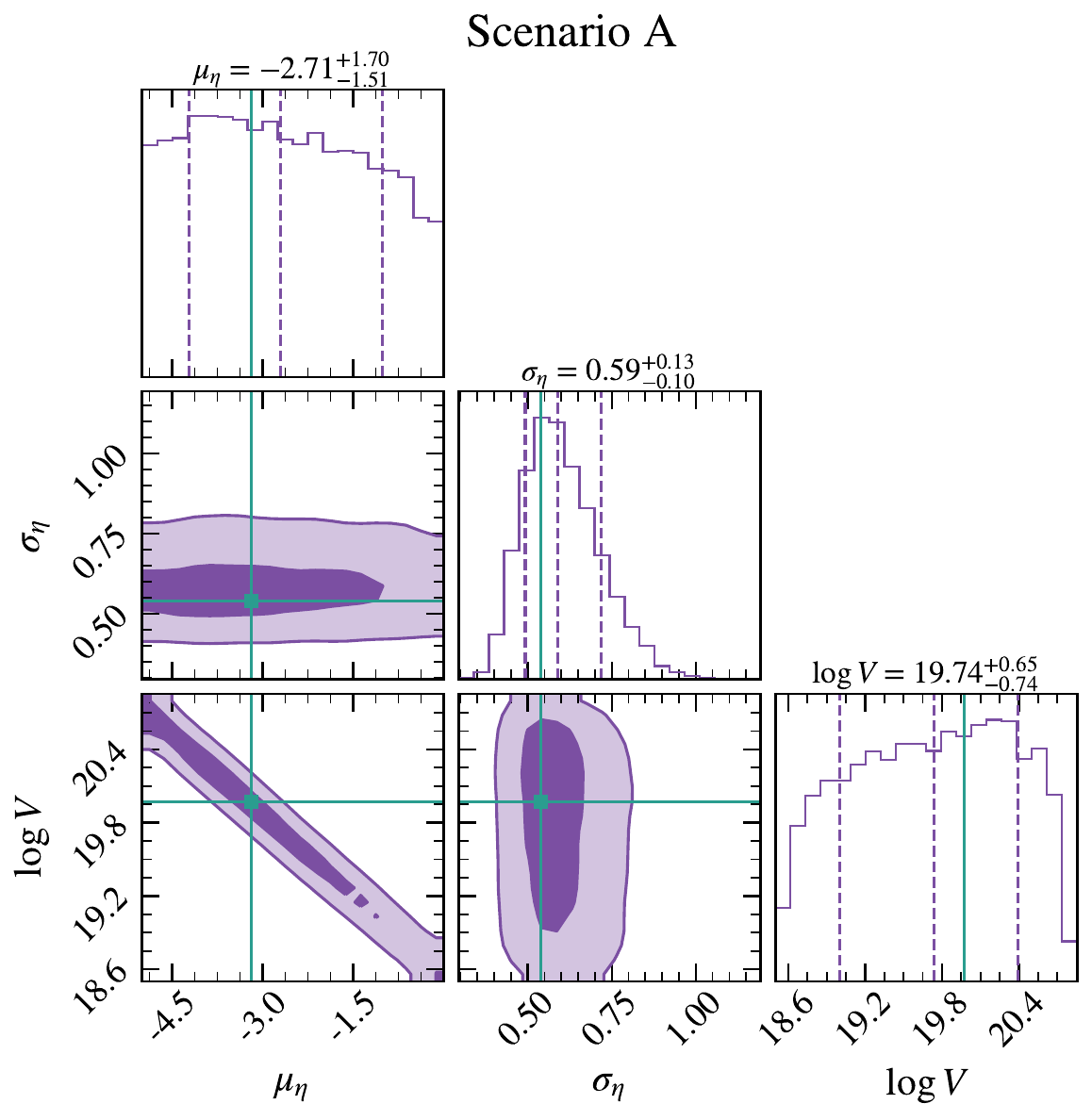}
    }\hfill
    \subfigure[A, $d=100\,\mathrm{Mpc}$]{
        \includegraphics[width=0.32\textwidth]
        {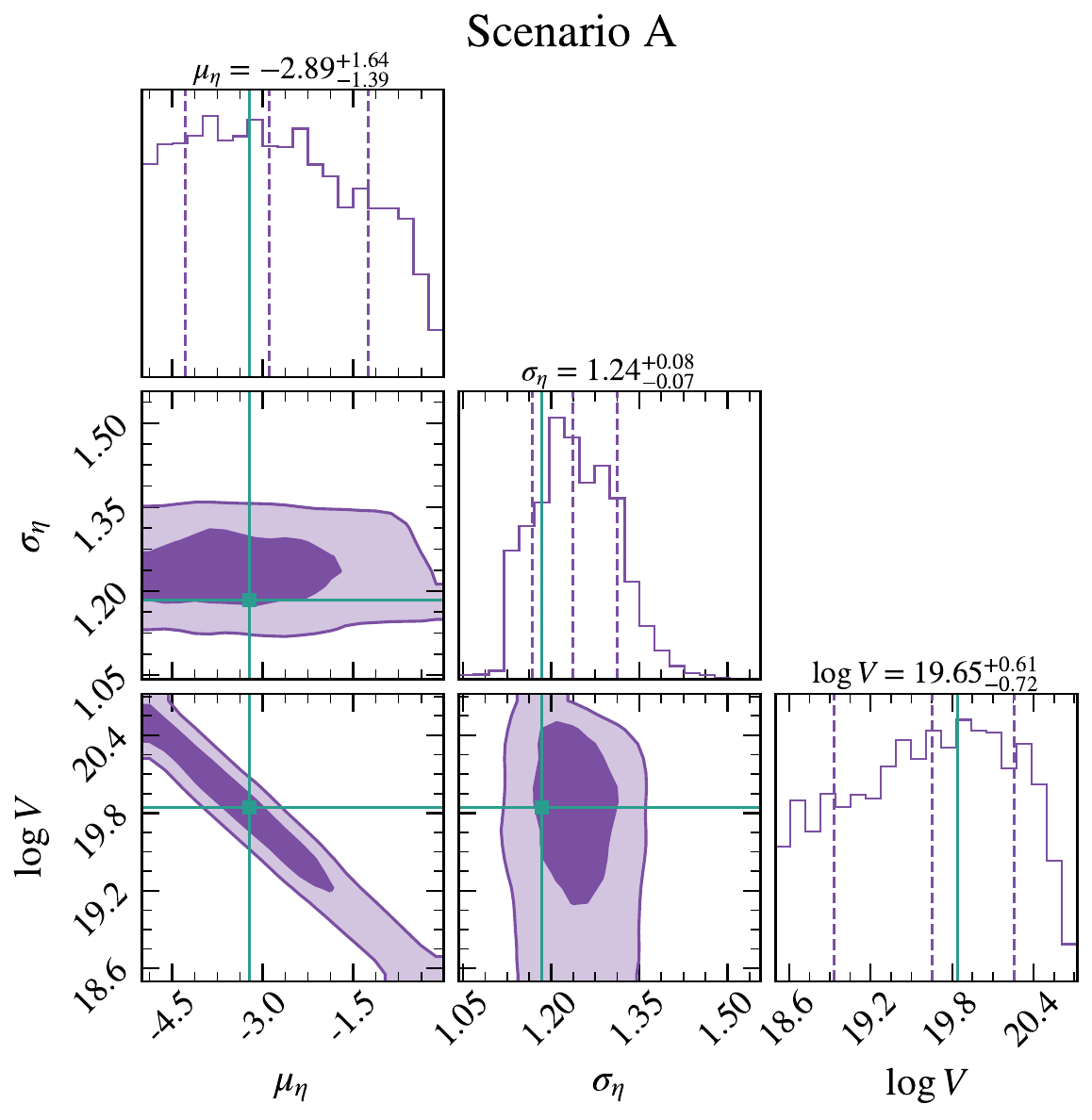}
    }\\[4pt]
    \subfigure[B, $d=10\,\mathrm{Mpc}$]{
        \includegraphics[width=0.32\textwidth]
        {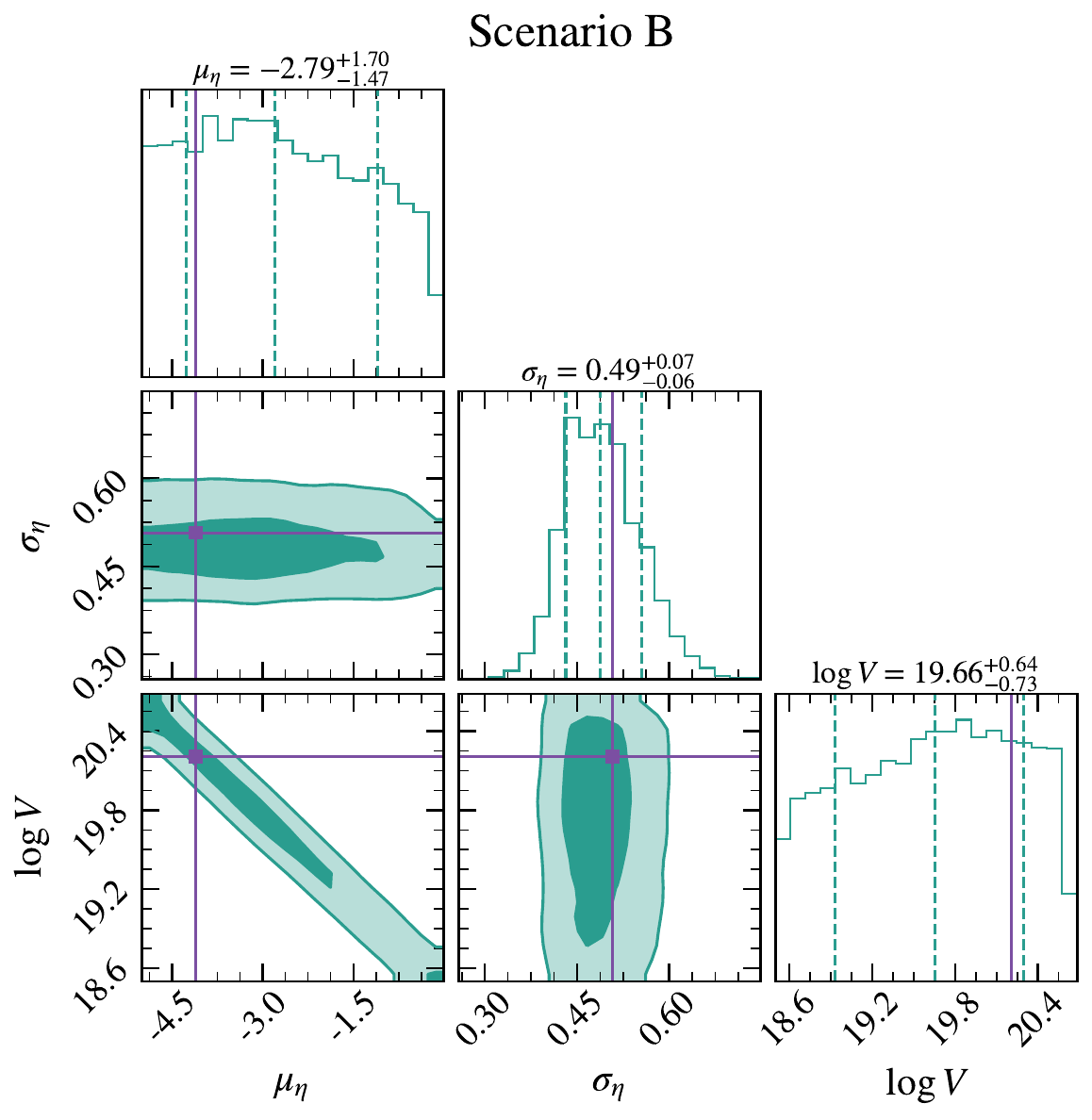}
    }\hfill
    \subfigure[B, $d=30\,\mathrm{Mpc}$]{
        \includegraphics[width=0.32\textwidth]
        {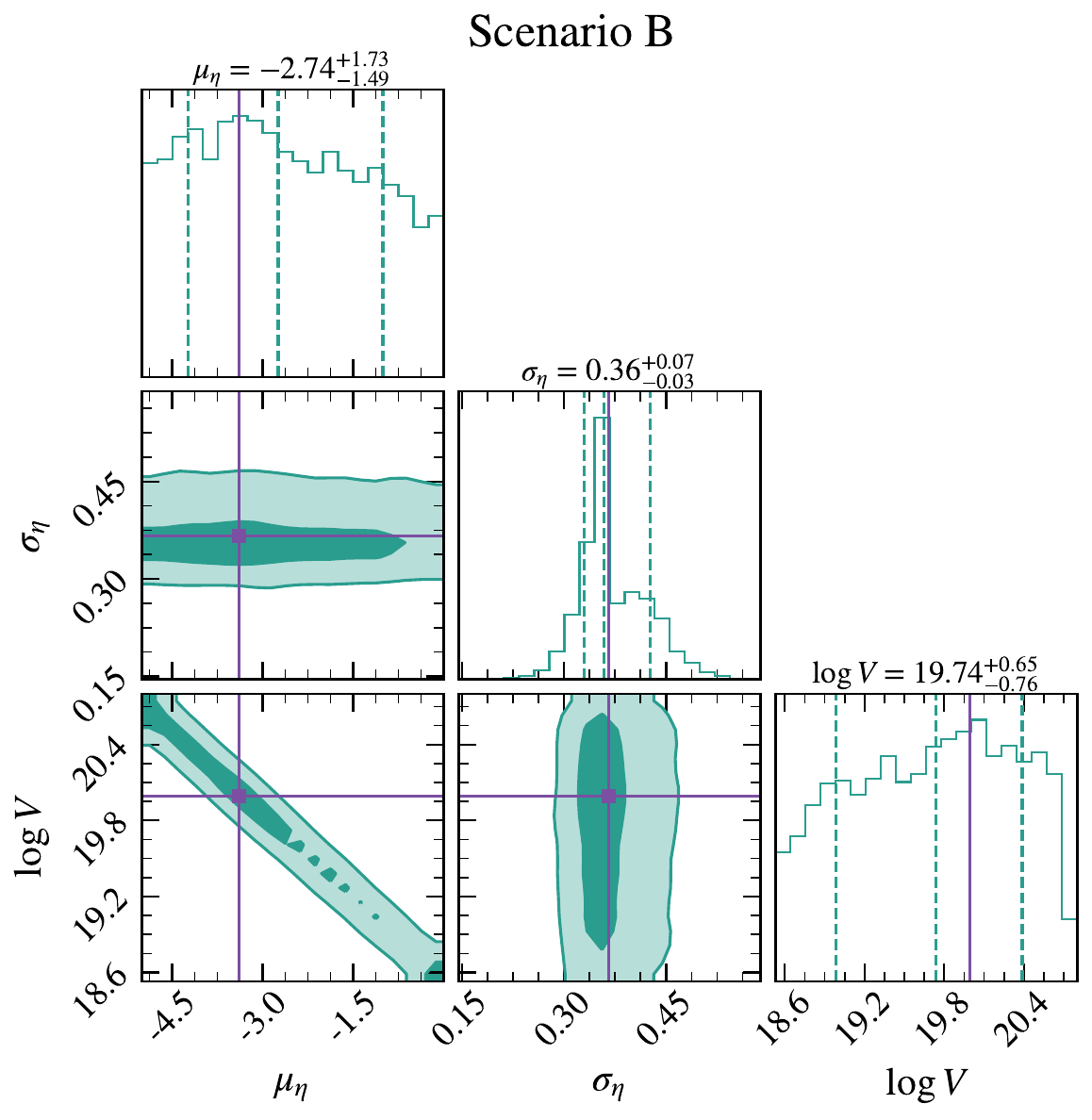}
    }\hfill
    \subfigure[B, $d=100\,\mathrm{Mpc}$]{
        \includegraphics[width=0.32\textwidth]
        {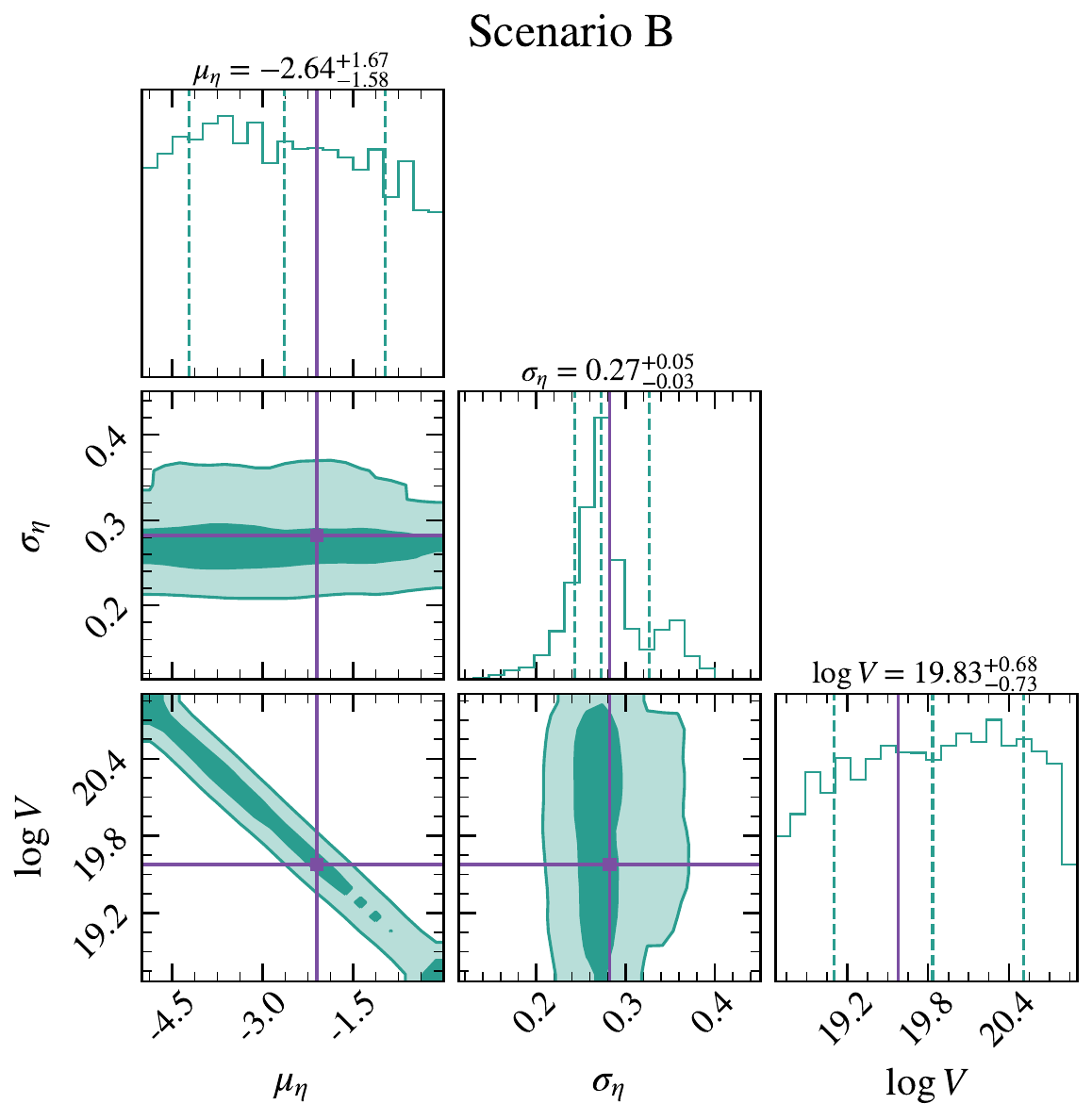}
    }
    \caption{
    Marginalised posterior distributions of
    $(\mu_\eta,\sigma_\eta,\ell_V)$, where
    $\ell_V=\log(\mathcal{V}_{\max}/\mathrm{V})$, for Scenario~A
    (top row, purple) and Scenario~B (bottom row, teal) at $d=10$, 30,
    and $100\,\mathrm{Mpc}$. Off-diagonal panels show the joint $1\sigma$
    and $2\sigma$ credible regions, enclosing 39.3\% and 86.5\% of the
    posterior mass. On the diagonal, dashed vertical lines mark the 16th,
    50th, and 84th percentiles of each marginal distribution, with the
    corresponding median and $68\%$ interval quoted above each panel. The
    solid lines and square markers in the contrasting colour (teal for
    Scenario~A, purple for Scenario~B) indicate the maximum-posterior
    point. The strong $\mu_\eta$--$\ell_V$ anticorrelation appears as a
    diagonal ridge in every fit, whereas $\sigma_\eta$ is comparatively
    well localised. Scenario~A shifts to a substantially larger
    $\sigma_\eta$ at $d=100\,\mathrm{Mpc}$, while Scenario~B favours
    progressively smaller values as the distance increases.
    }
    \label{fig:corners_all}
\end{figure*}

\begin{figure*}[htb]
    \centering
    \subfigure[A, $d=10\,\mathrm{Mpc}$]{
        \includegraphics[width=0.32\textwidth]
        {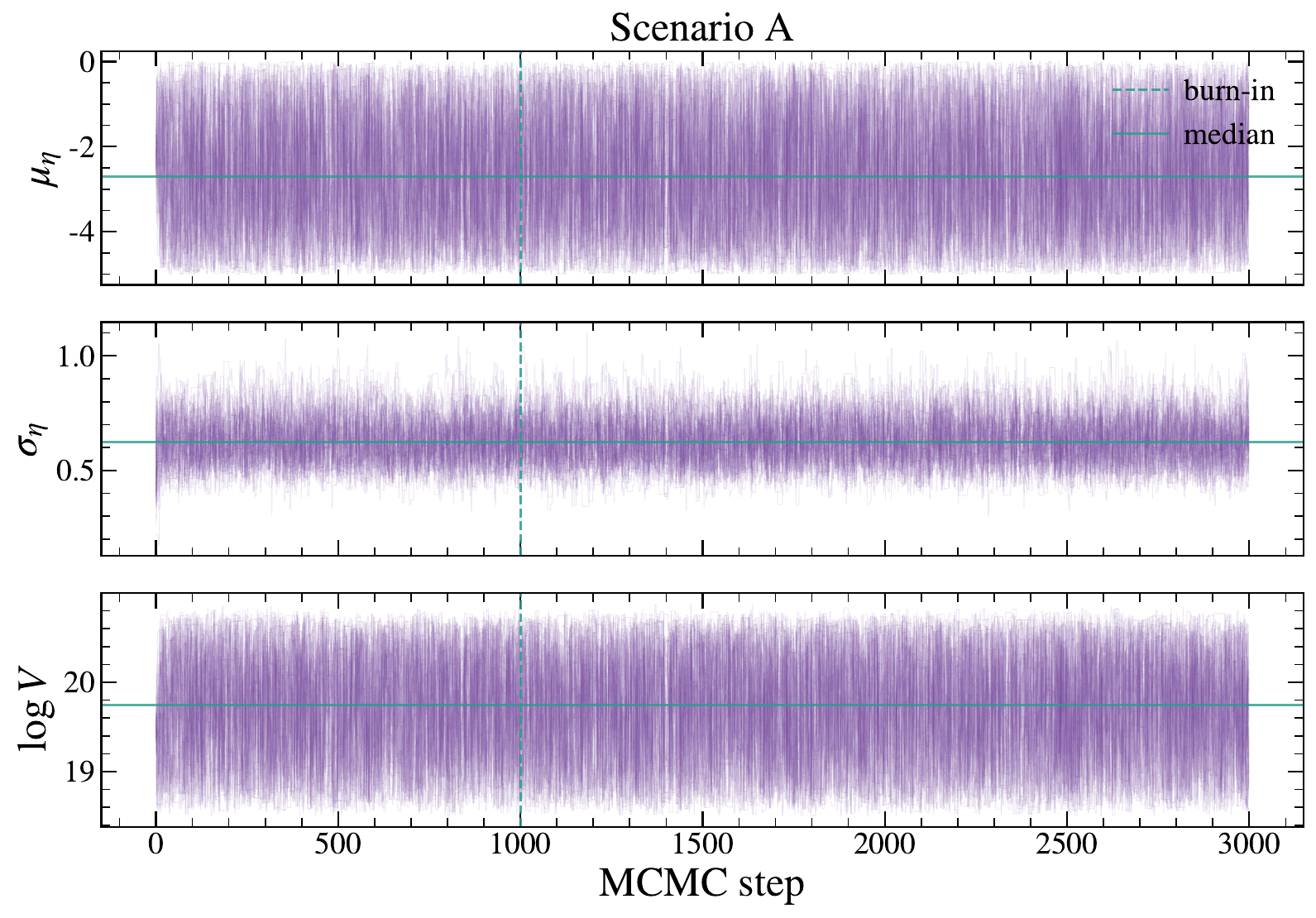}
    }\hfill
    \subfigure[A, $d=30\,\mathrm{Mpc}$]{
        \includegraphics[width=0.32\textwidth]
        {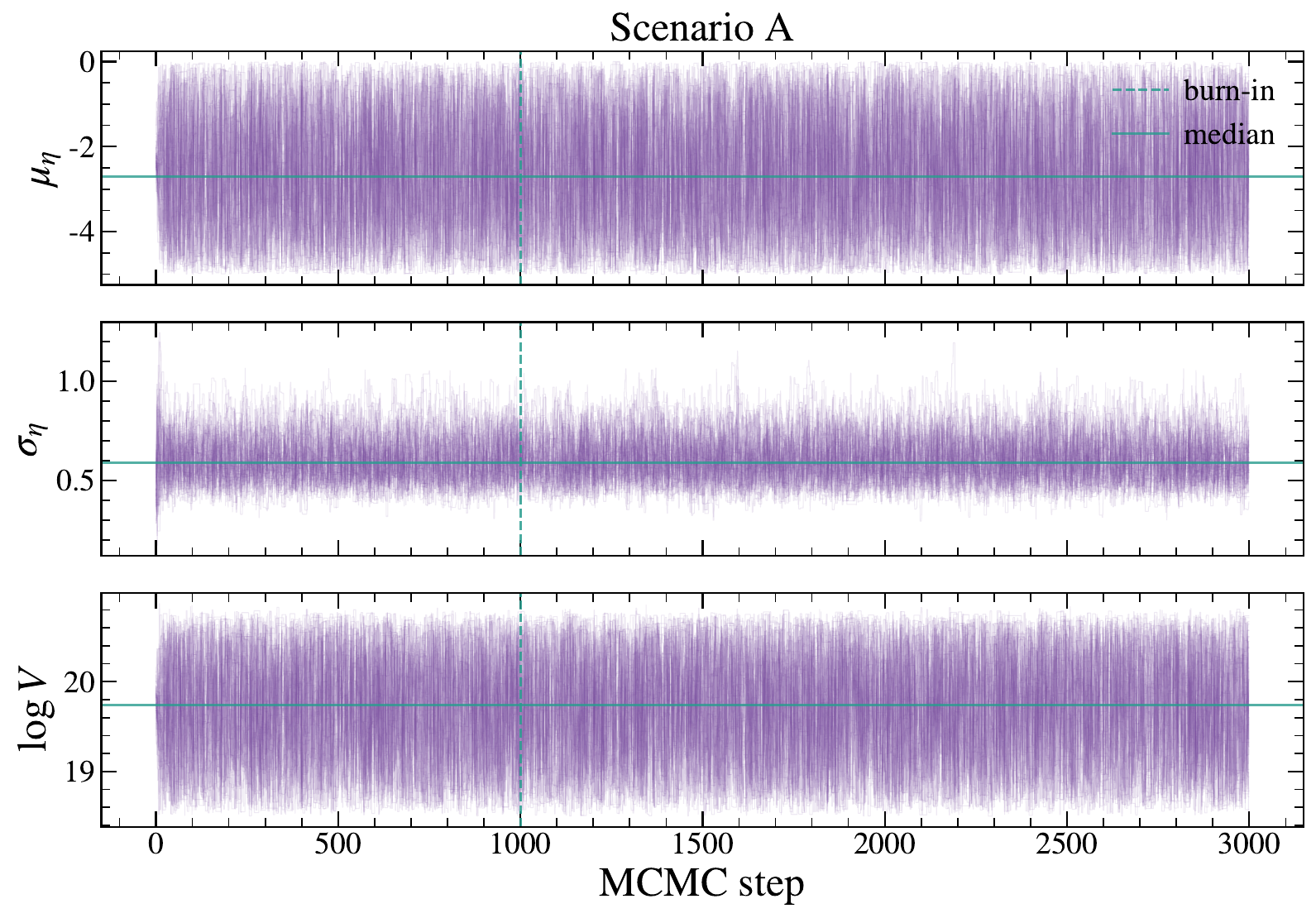}
    }\hfill
    \subfigure[A, $d=100\,\mathrm{Mpc}$]{
        \includegraphics[width=0.32\textwidth]
        {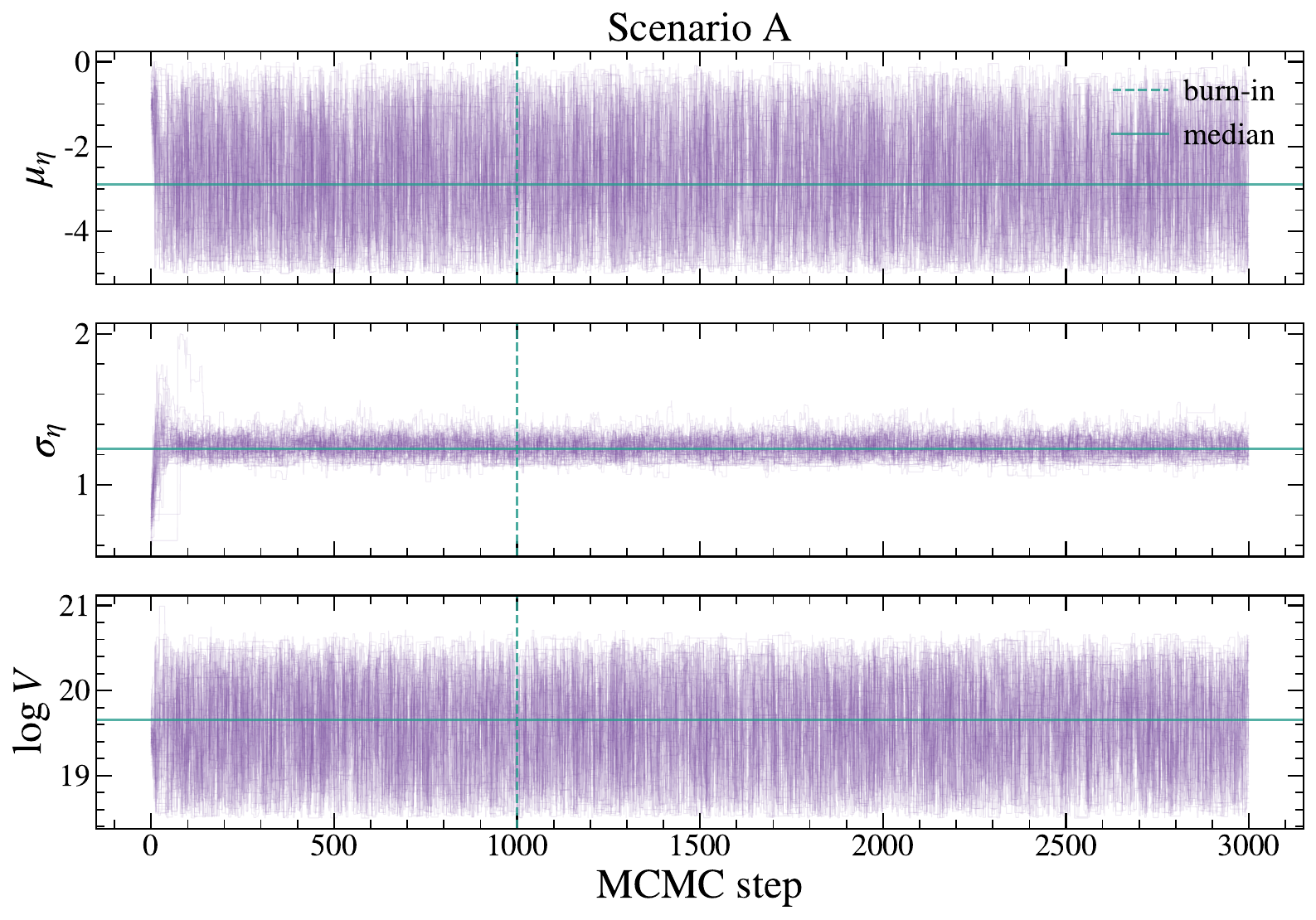}
    }\\[4pt]
    \subfigure[B, $d=10\,\mathrm{Mpc}$]{
        \includegraphics[width=0.32\textwidth]
        {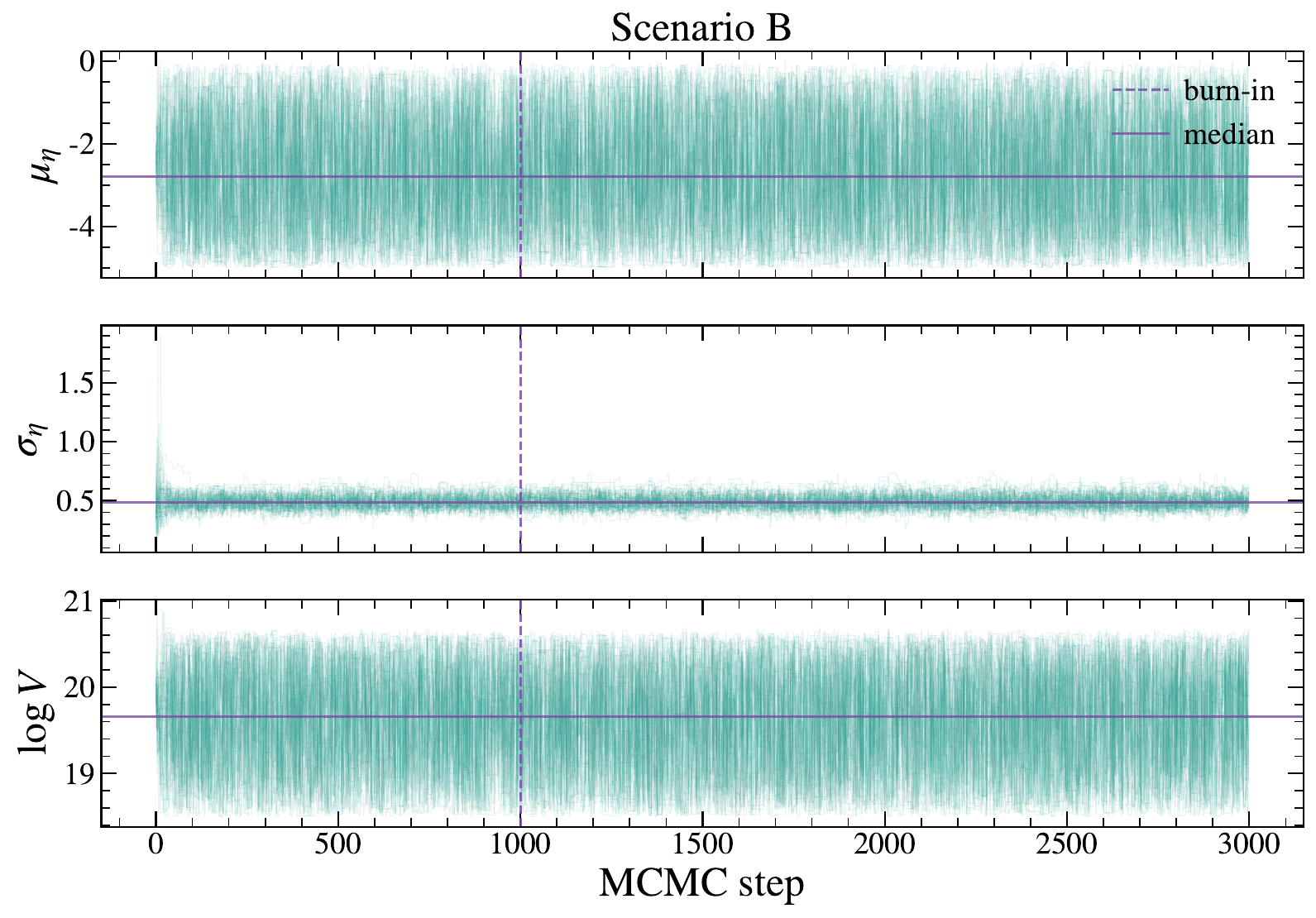}
    }\hfill
    \subfigure[B, $d=30\,\mathrm{Mpc}$]{
        \includegraphics[width=0.32\textwidth]
        {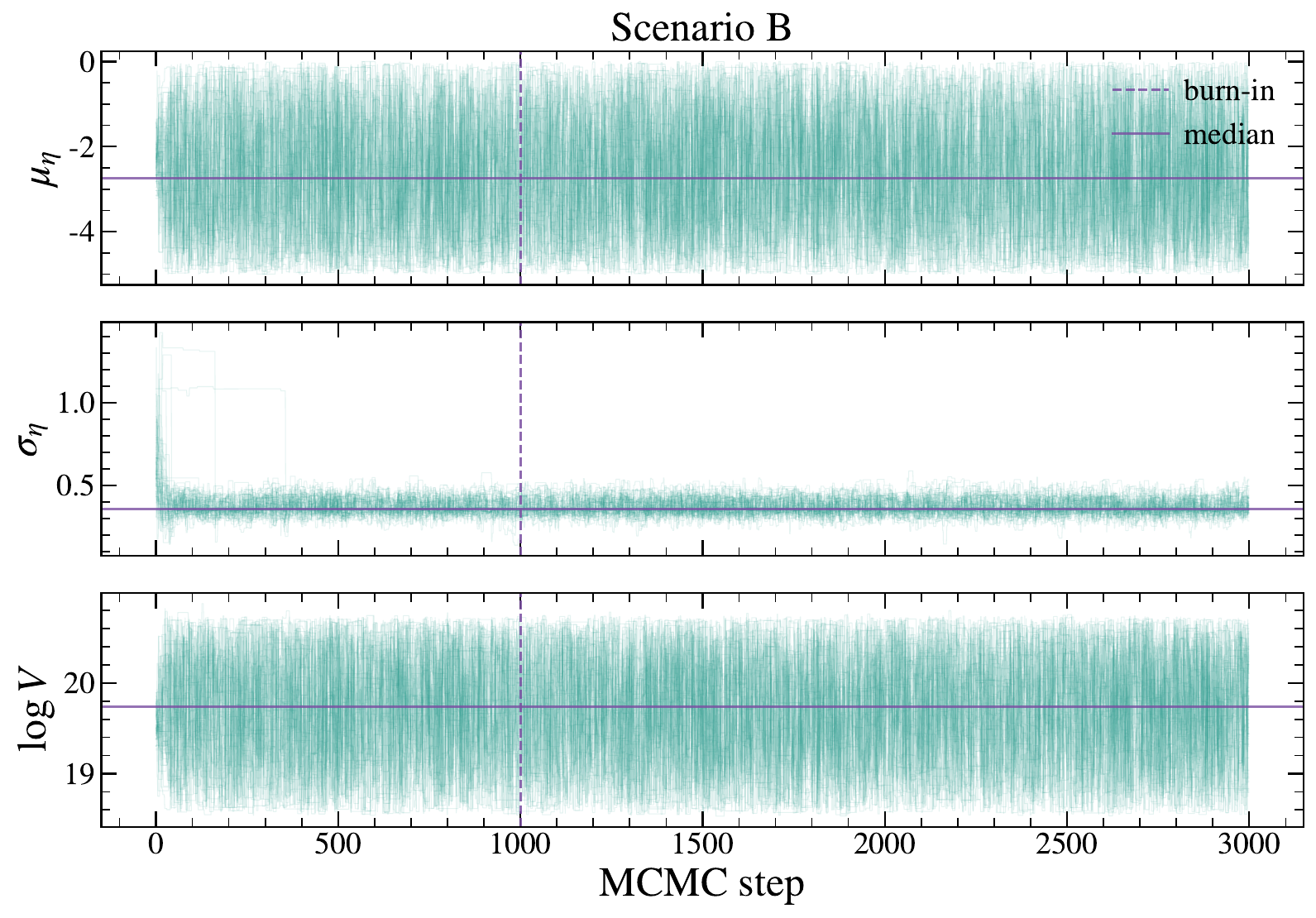}
    }\hfill
    \subfigure[B, $d=100\,\mathrm{Mpc}$]{
        \includegraphics[width=0.32\textwidth]
        {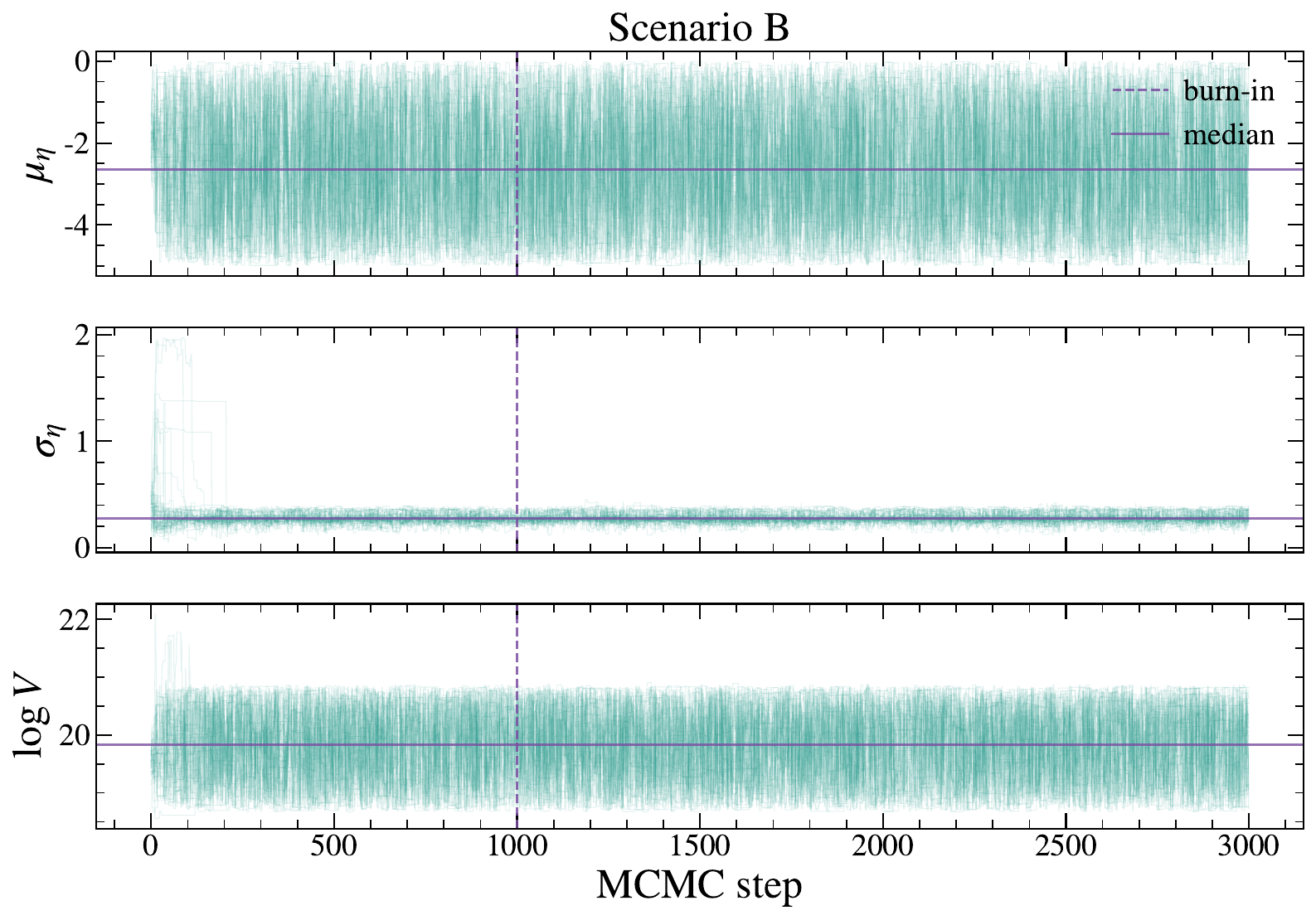}
    }
    \caption{
    MCMC traces of $\mu_\eta$, $\sigma_\eta$, and
    $\ell_V=\log(\mathcal{V}_{\max}/\mathrm{V})$ for Scenario~A
    (top row) and Scenario~B (bottom row) at $d=10$, 30, and
    $100\,\mathrm{Mpc}$. Within each panel, the three parameters are
    displayed from top to bottom. Dashed vertical lines mark the end of
    the 1000-step burn-in, and solid horizontal lines show the posterior
    medians. The retained portions of the chains are stationary, with no
    obvious persistent drift or stuck walkers.
    }
    \label{fig:traces_all}
\end{figure*}

Figure~\ref{fig:corners_all} presents the marginalised posterior distributions
for the six fixed-distance fits, while Fig.~\ref{fig:traces_all} shows the
corresponding MCMC traces. After the adopted burn-in, the traces are stationary,
with no obvious persistent drift or stuck walkers, and each $\sigma_\eta$
marginal exhibits a single dominant mode. The mean walker acceptance fraction
lies between $0.17$ and $0.28$ across the six fits, and each retained chain is
2000 steps long, corresponding to between $64$ and $161$ times the largest
integrated autocorrelation time $\tau_{\max}$ among the three parameters and
therefore exceeding the $50\,\tau_{\max}$ usually required for reliable
autocorrelation estimates. 

The strong $\mu_\eta$--$\ell_V$ anticorrelation discussed in
Sect.~\ref{sec:results_mcmc} appears as a diagonal ridge in every corner plot,
reflecting the fact that both parameters control the characteristic
acceleration-energy scale. By comparison, $\sigma_\eta$ is more tightly
localised: its shift toward a larger value for Scenario~A at
$d=100\,\mathrm{Mpc}$, and its progressive decrease with distance for
Scenario~B, are clearly visible.

\clearpage

\section{Candidate sources}

\begin{table*}[htb]
\centering
\caption{
AGN frequently discussed as possible UHECR sources. Representative
distances $D$ are taken from the \textit{NASA/IPAC Extragalactic
Database (NED)}.  Black-hole masses are representative published
dynamical measurements or literature estimates adopted from the cited
studies; the measurement methods and uncertainties are heterogeneous,
and the masses have not been rescaled to the distances listed here. Estimates of the maximum electric potential $\mathcal{V}_{\max}$ follow Eq.~\eqref{eq:Vmax}, adopting
$v_r/c=1$ and $h/r=0.05$. We adopt $k=1$, corresponding to an
approximately horizon-scale potential drop. More localized acceleration
regions with $k<1$ reduce all potential estimates linearly with $k$.
The two columns correspond to illustrative low-accretion
($L/L_{\rm Edd}=10^{-6}$) and Eddington-limited
($L/L_{\rm Edd}=1$) cases and do not represent measured Eddington
ratios for the individual sources.
}
\label{tab:source_table}

\begin{tabular}{lcccc}
\hline
\hline
\textbf{Source} &
$\boldsymbol{D}^{\,a}$ [Mpc] &
$\boldsymbol{M_{\rm BH}}$ [$10^9\,M_\odot$] &
\multicolumn{2}{c}{
$\boldsymbol{\mathcal{V}_{\max}}$ [$10^{19}$\,V]
} \\
\cline{4-5}
& & &
$L/L_{\rm Edd}=10^{-6}$ &
$L/L_{\rm Edd}=1$ \\
\hline

Centaurus A (NGC~5128) &
$\simeq3.8$ &
$0.055$ \citep{cena} &
0.066&
66\\

NGC~1068 &
$\simeq14.4$ &
$0.0172$ \citep{GallimoreImpellizzeri2023} &
$0.037$ &
$36.9$ \\

Virgo A (M87) &
$\simeq16.8$ &
$6.5$ \citep{m87_vi}&
0.718&
$718$ \\

NGC~1399 &
$\simeq17.7$ &
$0.51$ \citep{ngc1399} &
$0.201$&
$201$\\
Fornax A (NGC~1316) &
$\simeq19.2$ &
$0.15$ \citep{fornaxa} &
$0.109$&
109\\

NGC~1407 &
$\simeq24.1$ &
$4.5$ \citep{ngc1407} &
$0.597$&
$597$ \\
NGC~4261 (3C\,270) &
$\simeq32.0$ &
$1.67$ \citep{NGC315} &
$0.364$&
$364$\\

IC~4296 &
$\simeq46.9$ &
$1.34$ \citep{ic4296} &
$0.326$&
$326$\\

3C\,84 (NGC~1275) &
$\simeq69.0$ &
$0.80^{\,b}$ \citep{ngc1275} &
$0.252$&
$252$\\

Cygnus A &
$\simeq232$ &
$2.5$ \citep{cygnusa} &
$0.445$&
$445$ \\

\hline
\end{tabular}

\vspace{0.15cm}
\parbox{0.97\textwidth}{
\footnotesize
$^{a}$ Representative distances from NED. The published dynamical
masses generally depend on the distances adopted in the respective
studies and are used here without distance rescaling.
$^{b}$ \cite{ngc1275} report an enclosed central mass of
$8^{+7}_{-2}\times10^8\,M_\odot$. Because the molecular-gas mass
inside the modelled region may be non-negligible, this value may
represent an upper limit on $M_{\rm BH}$. Potential values are
calculated using the listed central mass estimates; mass uncertainties
are not propagated.
}
\end{table*}

\section{Test of the propagation treatment}
\label{app:validation}

The semi-analytic transport adopted in Sect.~\ref{sec:losses} is used for a specific
reason. The Bayesian inference of Sect.~\ref{sec:bayes} evaluates the forward model of
order $10^{5}$ times per fit (48 walkers over 3000 steps), each time propagating
$N_{\rm mc}=10^{6}$ particles, so a full Monte Carlo transport code
would be computationally prohibitive within the present likelihood implementation. Our
treatment instead applies tabulated loss lengths to a frozen particle bank, which is fast
enough for repeated evaluation while retaining the dominant physics. We assess this
approximation by comparing both the characteristic propagation scales and the end-to-end
shower-moment predictions with the public Monte Carlo framework CRPropa~3
\citep{AlvesBatista2016,AlvesBatista2022}.

\begin{figure*}
  \centering
  \includegraphics[width=\textwidth]{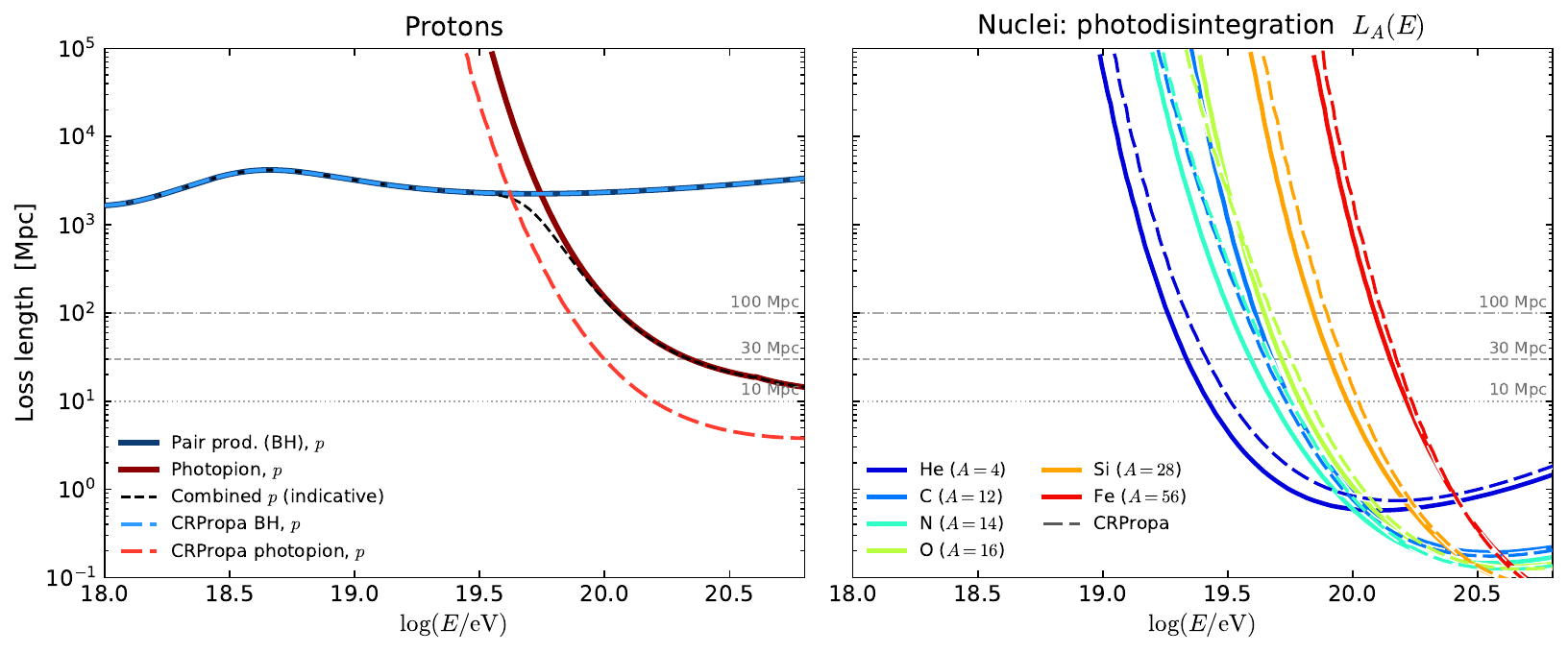}
  \caption{Comparison of characteristic propagation scales. \emph{Left:}
  the proton Bethe--Heitler energy-loss length,
Eq.~\eqref{eq:lambda_BH_nucleus}, and the proton photopion interaction
length, Eq.~\eqref{eq:lambda_pion}, on the present-day CMB.
  \emph{Right:} effective photodisintegration length $L_A(E)$ of
  Eq.~\eqref{eq:effective_nucleon_loss} for representative nuclei. Horizontal guides
  mark the three propagation distances used in the fits (10, 30, 100~Mpc). Solid curves
  show the quantities adopted in this work, while dashed curves show the corresponding
  CRPropa~3 CMB-only quantities. For photodisintegration, our $L_A$ is a
  multiplicity-weighted effective nucleon-loss length, whereas the CRPropa curves show
  the total interaction length $\lambda_{\rm dis}=1/\Gamma_{\rm dis}$; the two therefore
  are not expected to coincide point by point.}
  \label{fig:val_losslengths}
\end{figure*}

We first examine the three energy-loss processes that drive the propagation.
Figure~\ref{fig:val_losslengths} shows the Bethe--Heitler (pair-production) energy-loss
length for protons, Eq.~\eqref{eq:lambda_BH_nucleus}, the proton photopion interaction
length, Eq.~\eqref{eq:lambda_pion}, and the effective photodisintegration length
$L_A(E)$ of Eq.~\eqref{eq:effective_nucleon_loss} for representative nuclei, all
evaluated on the present-day CMB. The proton curves show the
expected behaviour: the pair-production scale remains of order Gpc over the fitted
range, while the photopion scale falls to a few tens of Mpc above the GZK threshold.
For nuclei, the effective photodisintegration length decreases rapidly once the
giant-dipole resonance becomes important.

The dashed curves in Fig.~\ref{fig:val_losslengths} show the corresponding
CRPropa~3 CMB-only quantities. For Bethe--Heitler and photopion production these are
taken from the tabulated CRPropa loss/rate tables. For photodisintegration, however,
our $L_A$ is a multiplicity-weighted effective nucleon-loss length, whereas the
CRPropa curve shows the total interaction length
$\lambda_{\rm dis}=1/\Gamma_{\rm dis}$. The two nuclear quantities therefore should
not be expected to coincide point by point. Figure~\ref{fig:val_losslengths} is thus
used primarily to compare the characteristic propagation scales and their energy
dependence.

We then compare the full pipeline against CRPropa~3 end to end. For the representative
case of the Scenario~A composition at $d=30$~Mpc, we inject the identical
accelerated population -- the same particles with the same $(E_0,Z,A)$ produced by the
acceleration prescription of Eq.~\eqref{eq:electric_potential_energy} -- into both our
transport and a one-dimensional CRPropa configuration with CMB-only interactions
(electron-pair production, photopion production, and photodisintegration), and convert
the arriving particles of both engines to $\langle X_{\max}\rangle$ and
$\sigma(X_{\max})$ using the same generalised Gumbel parametrisation and moment
calculation of Sect.~\ref{sec:xmax}. Because the acceleration step and the mapping from
$(E,A)$ to $X_{\max}$ are shared, any difference between the two predictions is
attributable to the propagation treatment.

\begin{figure*}
  \centering
  \includegraphics[width=\textwidth]{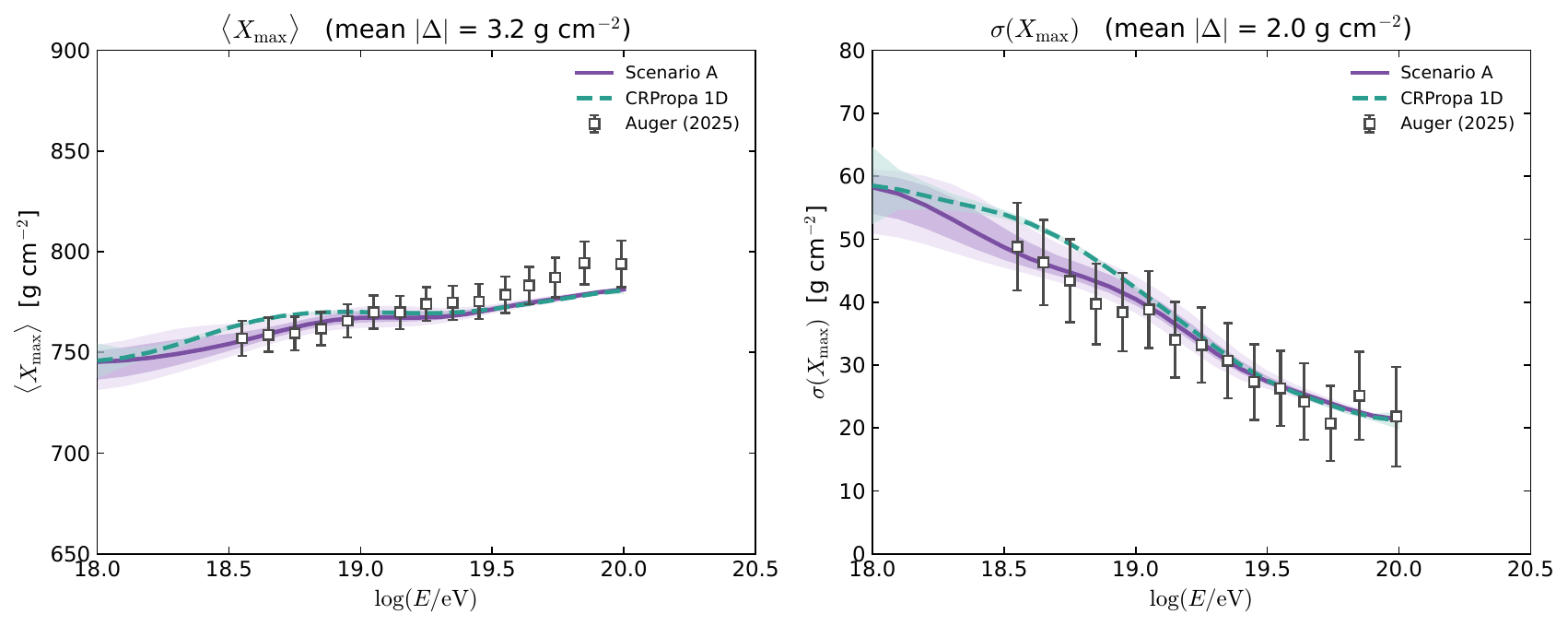}
  \caption{End-to-end comparison against a CMB-only CRPropa configuration for
  the Scenario~A composition at $d=30$~Mpc, evaluated at the posterior-median
  parameters of Table~\ref{tab:posterior_summary}. The identical accelerated
  population, generated with Eq.~\eqref{eq:electric_potential_energy}, is propagated
  by our semi-analytic transport (purple) and by a one-dimensional CRPropa simulation
  with CMB-only interactions (teal dashed; the band shows the per-bin Monte Carlo
  uncertainty), and both are converted to shower moments using the same generalised
  Gumbel parametrisation and moment calculation of Sect.~\ref{sec:xmax}. Light and dark
  shaded regions show the 95\% and 68\% posterior-predictive intervals of
  Figs.~\ref{fig:combined_xmax} and~\ref{fig:combined_sigma} for context; square markers
  show the Auger DNN 2025 measurements \citep{AbdulHalim2025}.
  \emph{Left:} $\langle X_{\max}\rangle$. \emph{Right:} $\sigma(X_{\max})$. As in
  Figs.~\ref{fig:combined_xmax} and~\ref{fig:combined_sigma}, the curves are smoothed
  along the energy axis with a Gaussian filter of standard deviation 1.5~bins for
  visualisation only; the quoted mean absolute differences are computed from the
  unsmoothed predictions. The two treatments agree to $3.2$ and
  $2.0$~g\,cm$^{-2}$, respectively, for this representative matched CMB-only
  configuration.}
  \label{fig:val_crpropa}
\end{figure*}

As shown in Fig.~\ref{fig:val_crpropa}, the two agree to a mean absolute difference of
$3.2$~g\,cm$^{-2}$ in $\langle X_{\max}\rangle$ and $2.0$~g\,cm$^{-2}$ in
$\sigma(X_{\max})$ across the fitted energy range. The residuals are largest toward the
lowest energies and in $\sigma(X_{\max})$, where the shower moments are most sensitive
to differences in the propagated mass distribution. For this representative matched
CMB-only configuration, the semi-analytic forward model therefore reproduces the
CRPropa shower moments to within a few g\,cm$^{-2}$ over the fitted range. This test
does not validate every propagation approximation or the omitted EBL contribution, but
provides an end-to-end check of the transport treatment used in the likelihood.

\end{appendix}
\end{document}